\documentclass[12pt,a4paper]{article}
\usepackage[T2A]{fontenc}
\usepackage[cp1252]{inputenc}
\usepackage[english]{babel}
\usepackage{bm,epsfig,amssymb,amsfonts,amsmath,cite,xcolor,adjustbox}
\newcommand{\be}{\begin{equation}}
\newcommand{\ee}{\end{equation}}
\newcommand{\beq}{\begin{eqnarray}}
\newcommand{\eeq}{\end{eqnarray}}
\newcommand{\bea}{\begin{eqnarray}}
\newcommand{\eea}{\end{eqnarray}}
\newcommand{\ds}{\displaystyle}
\newcommand{\vep}{{\bm p}}
\newcommand{\veq}{{\bm q}}
\newcommand{\vek}{{\bm k}}
\newcommand{\ves}{{\bm s}}
\newcommand{\veY}{{\bm Y}}
\newcommand{\ven}{{\bm n}}
\newcommand{\X}{X(3872)}
\newcommand{\Jp}{\gamma J/\psi}
\newcommand{\p}{\gamma\psi'}
\newcommand{\pp}{\gamma\psi}
\newcommand{\veg}{{\bm g}}
\DeclareMathSymbol{\varGamma}{\mathord}{letters}{"00}
\newcommand{\Br}{\mbox{Br}}
\newcommand{\pipi}{\pi^+\pi^-}
\newcommand{\pipipi}{\pi^+\pi^-\pi^0}

\def\bbletal{et~al.}
\def\bblin{in}

\graphicspath{{Figs.dir/}}

\title{\bf $X(3872)$ in the molecular model}
\author{Yu. S. Kalashnikova$^{\rm a,b}$, A. V. Nefediev$^{\rm b}$\\[5mm]
${\rm ^a}$ {\small\it Institute for Theoretical and Experimental Physics,}\\
{\small\it 117218, B.Cheremushkinskaya 25, Moscow, Russia}\\
${\rm ^b}$ {\small\it P.N. Lebedev Physical Institute of the Russian Academy of Sciences,}\\
{\small\it 119991, Leninskiy Prospect 53, Moscow, Russia}
}
\date{}

\begin{document}

\maketitle

\begin{abstract}
We discuss methods and approaches to the description of molecular states in the spectrum of heavy quarks and investigate in detail various properties of the exotic charmonium-like state X(3872) in 
the 
framework of the mesonic molecule model.
\end{abstract}

\tableofcontents

\section{Introduction}\label{Xexp}

In the end of the 20-th and at the very beginning of this century, charmonium spectroscopy was
considered as a respectable but a bit dull subject. Indeed, since the charmonium revolution of 1974, all $\bar{c}c$ excitations below
the open-charm threshold were discovered, together with a fair amount of higher vector states directly accessible in the $e^+e^-$ annihilation.
With the advent of $B$-factories, everybody anticipated an observation of other higher $\bar{c}c$ charmonia, however, no surprises were expected from the theoretical point of view: 
nonrelativistic quark models of the Cornell-like type \cite{Eichten:1978tg} seemed to be quite adequate in description of
the spectra and radiative transitions of the observed charmonia. It 
appeared, however, that it was premature to assume that we had fully understood the physics of charmonium: experimental developments at
$B$-factories revealed a bunch of states which, on the one hand, definitely contained a $\bar{c}c$ pair but, on the other hand,
could not be described with a simple quark-antiquark assignment. In the literature, such states are conventionally referred to as exotic. 
To see a considerable progress achieved recently in studies of exotic chamonium-like states it is sufficient to check the review paper  \cite{Pakhlova:2010zza} which contains a brilliant description of the situation with exotic hadrons in the end of the first decade of the 20-th century. 
In particular, the number of exotic states known at that time was less than ten while now the total number claimed exotic states in the spectrum of charmonium exeeds 20, and approximately half of them are regarded as well established.

The first and the most well-studied representative of this family is the state $\X$ observed in 2003 by the Belle Collaboration in the reaction 
$B^+\to K^+ X\to K^+(\pi^+\pi^-J/\psi)$ \cite{Choi:2003ue}. To begin with, the state lies fantastically close to the $D^0\bar{D}^{*0}$ threshold --- for the average value of the mass one has 
\cite{Tanabashi:2018oca}
\be 
M_X=3871.69\pm 0.17~\mbox{MeV}, 
\label{Xmass} 
\ee 
that is, within just 1 MeV from the $D^0\bar{D}^{*0}$ threshold \cite{Tanabashi:2018oca},
\be
E_B=M_{D^0}+M_{\bar{D}^{*0}}-M_X=-(1.1^{+0.6+0.1}_{-0.4-0.3})~\mbox{MeV}.
\label{EBexp} 
\ee 
Only the upper limit on the $\X$ total width exists \cite{Tanabashi:2018oca}, 
\be 
\varGamma_X<1.2~\mbox{MeV}.
\label{widthX} 
\ee 

Similarly, only an upper limit is imposed by the BABAR Collaboration on the total branching fraction of the $\X$ production in weak decays of the $B$-meson \cite{Kato:2017gfv},
\be 
\Br(B \to KX) < 2.6 \times 10^{-4}.
\label{babartotal} 
\ee 
Later on, this state was found in the $\pi^+\pi^-\pi^0 J/\psi$ mode \cite{Abe:2005ix}, and it appeared that the two-pion and three-pion branchings were of the same order of magnitude, 
\be 
\frac{\Br(X \to \pi^+\pi^-\pi^0J/\psi)}{\Br(X \to \pi^+\pi^-J/\psi)}=1.0 \pm 0.4 \pm 0.3. 
\label{omega} 
\ee 
The latest results for the branching fractions are 
\bea
&&\Br(B^+\to K^+\X)\Br(\X\to\pi^+\pi^-J/\psi)\nonumber\\
&&\hspace*{60mm}=[8.63\pm 0.82({\rm stat})\pm 0.52({\rm syst})]\times 10^{-6},\nonumber\\[-2mm]
\\[-2mm]
&&\Br(B^0\to K^0\X)\Br(\X\to\pi^+\pi^-J/\psi)\nonumber\\
&&\hspace*{60mm}=[4.3\pm 1.2({\rm stat})\pm 0.4({\rm syst})]\times
10^{-6},
\nonumber 
\eea 
for the dipion mode \cite{Choi:2011fc}, and 
\bea
&&\Br(B^+\to K^+\X)\Br(\X\to\pi^+\pi^-\pi^0J/\psi)\nonumber\\
&&\hspace*{60mm}=[0.6\pm 0.2({\rm stat})\pm 0.1({\rm syst})]\times 10^{-5},\nonumber\\[-2mm]
\\[-2mm]
&&\Br(B^0\to K^0\X)\Br(\X\to\pi^+\pi^-\pi^0J/\psi)\nonumber\\
&&\hspace*{60mm}=[0.6\pm 0.3({\rm stat})\pm 0.1({\rm syst})]\times
10^{-5},
\nonumber 
\eea 
for the tripion mode \cite{delAmoSanchez:2010jr}. Studies of the pion spectra have shown that the dipion in the $\pi^+\pi^-J/\psi$ mode
originates from the $\rho$-meson while the tripion in the $\pi^+\pi^-\pi^0 J/\psi$ mode originates from the $\omega$-meson.
Because of this, in what follows the two- and three-pion mode will be referred to as the $\rho J/\psi$ and $\omega J/\psi$ mode, respectively.

The threshold proximity of the $\X$ raises a burning question of the $D^0\bar{D}^{*0}$ mode. Indeed, such a mode was found in the $B$-meson
decay, $B^+\to K^+D^0\bar{D}^0\pi^0$ \cite{Gokhroo:2006bt}, with a rather large branching fraction,\footnote{Data on the open-charm modes
are presented in different ways by different collaborations. Belle presents the data on the $D\bar{D}\pi$ channel while BABAR sums the data
obtained for the $D\bar{D}\pi$ and $D\bar{D}\gamma$ final states and refers to the corresponding mode as $D\bar{D}^*$. To simplify
notations, in what follows, open-charm channel will be labelled as $D\bar{D}^*$ irrespectively of the particular collaboration under consideration.}
\be 
\Br(B^+\to K^+D^0\bar{D}^0\pi^0)=(1.02\pm 0.31^{+0.21}_{-0.29})\times 10^{-4}. 
\label{BrB} 
\ee 
First measurements gave a bit higher mass than that found in the inelastic modes with hidden charm, namely the value 
\be 
M_X=3875.2\pm 0.7^{+0.3}_{-1.6}\pm 0.8~\mbox{MeV} 
\label{3875v1} 
\ee 
was presented in Ref.~\cite{Gokhroo:2006bt}. A more recent Belle result for the $\X$ mass in the $D\bar{D}^*$ mode reads \cite{Adachi:2008sua}
\be 
M_X= 3872.9^{+0.6+0.4}_{-0.4-0.5}~\mbox{MeV}. 
\ee 
Although the BABAR Collaboration also gives a bit higher value for the mass in this mode in comparison with the dipion mode \cite{Aubert:2007rva}, it is commonly accepted that this is one and the 
same state.

The value (\ref{BrB}) implies that the $\X$ is produced in $B$-meson decays with the branching fractions of the same order of magnitude as ordinary 
$\bar{c}c$ charmonia \cite{Tanabashi:2018oca},
\be 
\Br(B\to K J/\psi,K\psi',K\chi_{c1})\sim 10^{-4}.
\label{ccbarbr}
\ee

The given state is also observed in the radiative modes. The BABAR Collaboration gives the following data for the $\gamma J/\psi$ and $\gamma\psi'$ ones \cite{Aubert:2008ae},
\begin{eqnarray}
&&\Br(B^{\pm} \to K^{\pm} X)\Br(X \to \gamma J/\psi)=(2.8 \pm 0.8 \pm 0.2) \times 10^{-6},\nonumber\\[-2mm]
\label{Aubert:2008ae}\\[-2mm]
&&\Br(B^{\pm} \to K^{\pm} X)\Br(X \to \gamma \psi')=(9.5 \pm 2.9
\pm 0.6) \times 10^{-6},\nonumber
\end{eqnarray}
and for the ratios,
\begin{eqnarray}
\frac{\Br(X\to \gamma
J/\psi)}{\Br(X\to \pi^+\pi^- J/\psi)}&=&0.33\pm 0.12,\label{raddec}\\
\frac{\Br(X\to \gamma \psi')}{\Br(X\to \pi^+\pi^-J/\psi)}&=&1.1\pm
0.4.\label{raddecprime}
\end{eqnarray}
The data from the Belle Collaboration read \cite{Bhardwaj:2011dj}
\begin{eqnarray}
&&\Br(B^{\pm} \to K^{\pm} X)\Br(X \to \gamma
J/\psi)=(1.78^{+0.48}_{-0.44}\pm 0.12)
\times 10^{-6},\nonumber\\[-2mm]
\label{bellegamma}\\[-2mm]
&&\Br(B^{\pm} \to K^{\pm} X)\Br(X \to \gamma \psi')<3.45 \times
10^{-6},\nonumber
\end{eqnarray}
so that there is a discrepancy between the two collaborations on the $\gamma\psi'$ mode. Meanwhile, there exists also a recent measurement of the
ratio of the branching fractions for the given modes, performed by the LHCb Collaboration \cite{Aaij:2014ala}, 
\be 
R\equiv \frac{\Br(X\to\gamma \psi')}{\Br(X\to \gamma J/\psi)}=2.46\pm0.64\pm0.29.
\label{RLHCb} 
\ee

The story of the $\X$ quantum numbers is rather dramatic. Studies of the dipion mode performed by CDF at Tevatron \cite{Abulencia:2006ma} allowed one to restrict the quantum numbers of the $\X$ to 
only two options: $1^{++}$ or $2^{-+}$. There are good phenomenological reasons
to assign the $\X$ $1^{++}$ quantum numbers, so the
$1^{++}$ option became a more popular version. However, the BABAR
analysis \cite{delAmoSanchez:2010jr} resulted in the conclusion that the
$\pi^+\pi^-\pi^0$ mass distribution was described better with the angular momentum $L=1$ in the $\omega J/\psi$ system than with the
angular momentum $L=0$. The former implies a negative $P$-parity
for the $\X$ and, as a consequence, entails the $2^{-+}$ assignment. Meanwhile it is shown in Ref.~\cite{Hanhart:2011tn}
that, if one performs a combined analysis of the data obtained at $B$-factories for both the two-pion and three-pion mode, then the $1^{++}$ option
is preferable. It was only in 2013 when it became possible to reject
(at the 8$\sigma$ level) the $2^{-+}$
hypothesis, due to high-statistics measurements of the two-pion mode performed in the LHCb
experiment at CERN \cite{Aaij:2013zoa,Aaij:2015eva}.

With the $1^{++}$ quantum numbers the most natural possibility for the $\X$ is the genuine $2^3P_1$ $\bar{c}c$ charmonium, however this
possibility seems to be ruled out by the mass argument: the quark model calculations give for the mass of the $2^3P_1$ state a higher value ---
see, for example, Refs.~\cite{Badalian:1999fe,Barnes:2003vb,Barnes:2005pb}. Besides that, the ratio (\ref{omega}) of the branchings for the modes 
$\rho J/\psi$ and $\omega J/\psi$ indicates a strong isospin symmetry violation which is difficult to explain in the framework of a simple $\bar{c}c$ model.

\begin{figure}[t]
\begin{center}
\begin{tabular}{ccccccc}
\includegraphics[width=0.1\textwidth]{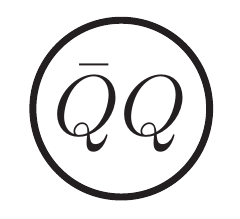}&\hspace*{0.05\textwidth}&
\includegraphics[width=0.1\textwidth]{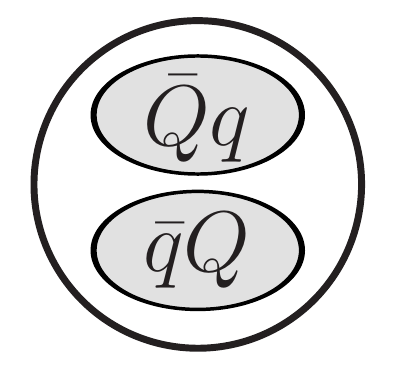}&\hspace*{0.05\textwidth}&
\includegraphics[width=0.1\textwidth]{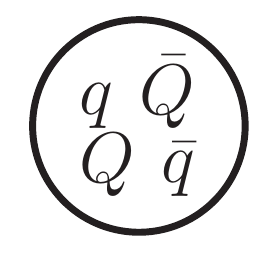}&\hspace*{0.05\textwidth}&
\includegraphics[width=0.1\textwidth]{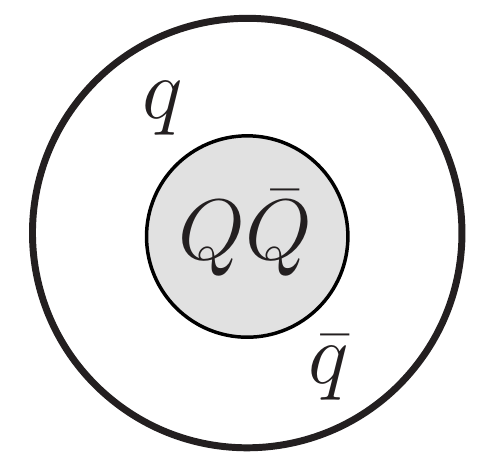}\\
Quarkonium&& Molecule && Tetraquark && Hadrocharmonium
\end{tabular}
\end{center}
\caption{Models for the $X(3872)$ compatible with the $1^{++}$ quantum numbers.}\label{fig:Xmodels}
\end{figure}

There is no deficit in models for the $\X$ --- see Fig.~\ref{fig:Xmodels}, where models discussed in the literature are depicted schematically.
One of them, the hadronic molecule model, had had a history long enough (see, for example, papers \cite{Voloshin:1976ap,DeRujula:1976zlg}),
and the discovery of the $X(3872)$ state at the $D\bar{D}^*$ threshold made this model quite popular again
\cite{Tornqvist:2004qy,Tornqvist:1991ks,Swanson:2003tb,Wong:2003xk,Wang:2013daa,Kang:2016jxw}.
It should be noticed that, in the molecular model, the aforementioned isospin symmetry violation in the $\X$ decays appears in a natural way
\cite{Suzuki:2005ha,Gamermann:2007fi}. Another popular model for the $\X$ is the tetraquark model (see papers
\cite{Maiani:2004vq,Ebert:2005nc,Ebert:2007rn,Esposito:2014rxa}). Finally, a recent arrival is the hadrocharmonium model
\cite{Dubynskiy:2008mq,Dubynskiy:2008di} which allows one to explain in a natural way the large probability for the exotic state to decay into
a lower charmonium plus light hadrons.

Generally speaking, the wave function of the $\X$ should contain, in addition to the genuine $\bar{c}c$ component, extra components like
the $D \bar D^*$ molecular ones, a tetraquark, a $\bar{c}c g$ hybrid, and what not, whose nature is unclear. Relative weights of
all these components are to be determined dynamically and, in the absence of an analytical solution of Quantum Chromodynamics, are very
model-dependent. However, for a near-threshold resonance, it is possible to define the admixture of the hadronic molecule in the wave function of the resonance in a model-independent 
way \cite{Weinberg:1962hj,Weinberg:1963zza,Weinberg:1965zz}. In the present review we argue that in case of the $\X$ state one can
indeed estimate this admixture using the aforementioned approach and demonstrate that the molecular $D \bar D^*$ component dominates
the wave function of the $\X$. We also discuss various scenarios of the formation of such a mesonic molecule and the role of the molecular component played in the pionic and radiative decays of the 
$\X$.

It is worthwhile noting that papers on the $\X$ are yet the most cited publications of the Belle Collaboration since it was established. 
This is not only because the state $\X$ is interesting by itself. It was already mentioned in the Introduction that the spectroscopy of heavy quark flavours experiences a renaissance because the 
experiment constantly brings information on new states which do not fit into the quark model scheme. In the meantime, the molecular model is one of the most successful approaches which allows one to 
explain qualitatively, as well as quantitatively in many cases, the properties of such exotic states in the spectrum of heavy quarks and to predict new resonances. The interested reader can find a 
detailed discussion of this model applied to hadronic spectroscopy in a recent review paper \cite{Guo:2017jvc}. Below we will only mention two such applications closely related to the $\X$.

The first application deals with the so-called $Z$-resonances which are charged and, therefore, do not admit a simple quark-antiquark interpretation --- obviously, their minimal quark contents is 
four-quark. The $Z$-states in the spectrum of charmonium, conventionally denoted as $Z_c$, lie near strong thresholds and, therefore, have good reasons 
to be interpreted as hadronic molecules. In particular, the state $Z_c(3900)$ observed by BESSIII \cite{Ablicim:2013mio} and Belle \cite{Liu:2013dau} is often discussed in the literature as an 
isovector $D \bar {D}^*$ molecule with the quantum numbers $J^{PC}=1^{+-}$, that is, as a spin partner of the $\X$. Remarkably, analogous charged 
$Z$-states are found recently in the spectrum of bottomonium --- the so-called $Z_b(10610)$ and $Z_b(10650)$ lying near the $B\bar{B}^*$ and $B^*\bar{B}^*$ thresholds, respectively 
\cite{Belle:2011aa,Garmash:2015rfd}. A molecular interpretation of these states is suggested in paper \cite{Bondar:2011ev} and further developed 
in a series of later works. 

The other application to be mentioned is related to transitions between molecular states, in particular, involving exotic vector charmonia, traditionally denoted as $Y$'s. One of the most well-studied 
$Y$-resonances is $Y(4260)$ observed by BABAR in the channel $J/\psi\pi^+\pi^-$ \cite{Aubert:2005rm}. This state is not seen in the inclusive cross section $e^+e^- \to $ hadrons that has no 
satisfactory explanation for a genuine $\bar{c}c$ charmonium, and the proximity of its mass to the $D_1\bar{D}$ threshold hints a large 
admixture of the hadronic component in its wave function, that is, its possible molecular nature
\cite{Wang:2013cya,Wang:2013kra,Cleven:2013mka,Wu:2013onz}. The assumption that $Y(4260)$ is a $D_1\bar{D}$ molecule while $Z_c(3900)$ is a $D\bar{D^*}$ one allows one to develop a nice 
model for the decay $Y(4260) \to \pi Z_c(3900)$ --- see paper \cite{Wang:2013cya}. The radiative transition $Y(4260) \to \gamma \X$ can be described in an analogous manner \cite{Guo:2013nza}. 

We might witness the beginning of a new, molecular, spectroscopy. The state $\X$ is an ideal (at least the best possible so far) laboratory to study exotic molecules. On the one hand, there exists a 
vast experimental material on it. On the other hand, this is a typical near-threshold resonance containing heavy quarks. As a result, the $\X$ allows one to employ the full power of the theoretical 
and phenomenological approaches developed so far to extract from the experimental data all possible information on the nature and properties of this state as well as the interactions responsible 
for its formation. This way, this paper is devoted to a review of various methods and approaches used in the molecular model and an illustration of their practical application at the 
example of the most well studied state $\X$. Systematisation of our knowledge on exotic near-threshold states is particularly actual and timely now in view of the launch of the upgraded $B$-factory 
Belle-II as well as plans of the super-$c$-$\tau$ factory construction in Novosibirsk. The physical programme of both experiments contain investigations in hadronic spectroscopy. 

In order to make reading this review easier, we collect all definitions and conventions used in it in Section~\ref{sec:def}. The latter also contains a justification of a dedicated study of 
near-threshold states. If appropriate, each section containing a large amount of technical details and formulae begins with a short introduction to the physical results arrived at in this section. If 
the reader is aimed at a most general acquainting with the subject of the review with technical details being unimportant, such a reader may want to skip most of the formulae in this section; to this 
end we provide brief instructions what can and what cannot be skipped at the first of sketchy reading. We also spotlight important conclusions which cannot be skipped and should, therefore, be paid 
attention to. 

\section{Definitions and conventions}
\label{sec:def}

We start from explaining some notions used in the review. The word ``resonance'' is used as a synonym for ``state''. Such an identification is motivated by the fact that each state has decay channels 
and, therefore, the corresponding pole will always reside on an unphysical sheet of the Riemann surface, away from the real axis (the Schwarz reflection principle requires that a symmetric ``mirror'' 
pole also exists, its imaginary part having the opposite sign). In other words, such a state cannot be regarded as bound or virtual in the proper sense of the word. Nevertheless, it proves 
convenient to use the terminology of bound and virtual states from the point of view of the pole location near a given threshold and on the corresponding Riemann Sheet with respect to this 
threshold. Consider a simple example. For $n$ decay channels, the Riemann surface for the state under study is multi-sheeted, with the number of sheets equal to $2^n$. However, if only a limited 
region near the given threshold is considered while all other thresholds can be viewed as remote, it is sufficient to confine oneself to a 
double-sheeted Riemann surface for the nearest threshold and regard one of the sheets as physical (it turns to a genuine physical sheet as the coupling to all other channels is switched off) and the 
other sheet as unphysical. With such a definition, a below-threshold pole lying on the physical sheet of such a double-sheeted Riemann surface will be conventionally called a bound state while a 
similar pole on the unphysical sheet will be called a virtual state in spite of the fact that in either case the pole will be somewhat shifted away from the real axis. It is important to note that, 
from the point of view of the full multi-sheeted Riemann surface, both aforementioned sheets are unphysical, so that shifting poles to the complex plane does not result in a contradiction with the 
probability conservation or other basic principles of quantum mechanics. 

All decay channels of the studied resonance will be discriminated as ``elastic'' and ``inelastic'' which are strong decays into an open-flavour and hidden-flavour final 
state, respectively; the corresponding 
thresholds are also called elastic and inelastic. All elastic channels considered in the review are assumed to be $S$-wave while there is no constraint on the relative momentum in the 
inelastic channels. A state residing at a strong threshold will be referred to as a near-threshold state and the energy region around this threshold will be called a near-threshold region. The size 
of 
such a region can be defined for each particular problem under study, however in any case it is bounded by the nearest elastic and inelastic thresholds.

We will denote as ``elementary'' or ``bare'' a state whose structure does not affect the effect under study, for example, the line shape of a near-threshold resonance. In many cases, the elementary 
state can be understood as a compact quark compound, for example, a genuine quarkonium, tetraquark, hybrid, and so on. Then the hard scale which governs the behaviour of the transition form factors 
between different channels of the reaction and which is defined by the binding forces of the corresponding state is called the (inverse) range of force. 
For example, the inverse range of force $\beta$ for the coupling of an elementary state to a hadronic channel is defined by the inter-quark interaction responsible for the formation of this 
elementary 
state. If not stated otherwise, it is assumed that $\beta\simeq 1~\mbox{GeV}$ and that it exceeds all typical momenta in the problem. The corresponding effective field theory is built in the leading 
order in the expansion in the inverse range of force. Finally, we will mention as ``molecular'' a near-threshold state whose wave function is dominated 
by the hadronic component. The nature of such a molecule is not specified --- a full variety of options is allowed for it, from a bound or virtual state to an above-threshold resonance. In a similar 
way, possible binding forces in the molecule are not limited to the $t$-channel exchanges; they can have different origins --- establishing the nature of such forces is one of the most important 
problems 
in the phenomenology of near-threshold states. 

Finally, let us comment on the applicability of the terms ``mass'' and ``width'' for near-threshold states. To begin with, a fundamental characteristic of a resonance is the location of its pole in 
the complex plane. In practice, the location of this pole may not be possible to determine in case of a multi-sheet Riemann surface and in presence of strong coupled channels effects. On the other 
hand, the mass and the width are the standard parameters of the Breit-Wigner distribution which is quite successful in description of isolated resonances, that is, the states in the spectrum lying 
fairly 
far from other states and from thresholds of the channels with the given quantum numbers. 

Consider, for example, the $S$-matrix element describing scattering of heavy-light mesons off each other,
$(\bar{Q}q)+(\bar{q}Q)\to (\bar{Q}q)+(\bar{q}Q)$ (here $Q$ and $q$ are the heavy and light quark, respectively), and express it through the amplitude $A$  as
\be
S=1+2iA,
\ee
so that unitarity requires the condition
\be
AA^{\dagger}=\frac{1}{2i}(A-A^{\dagger})
\label{unitarity}
\ee
to hold, which is fulfilled automatically if a real quantity $K$ (in a multichannel case, $K$ is a hermitian matrix) is introduced as
\be
A=K(1-iK)^{-1}.
\label{Aampl}
\ee

If the scattering process proceeds through the formation of a resonance, the function $K$ can be written in the form
\be
K=
G(s)\frac{1}{M^2-s}G(s)=\frac{\Gamma(s)\sqrt{s}}{M^2-s},\quad \Gamma(s)=\frac{G^2(s)}{\sqrt{s}},
\label{K1}
\ee
where $s$ is the invariant energy, and we introduced the vertex function $G(s)$ and the parameter $M$ which define the coupling of the resonance to the hadronic channel and its propagation function, 
respectively. Then, with the help of Eq.~(\ref{Aampl}) for the amplitude $A$, one can arrive at the expression
\be
A=\frac{\Gamma(s)\sqrt{s}}{M^2-s-i\Gamma(s)\sqrt{s}}.
\label{BW1}
\ee

Consider a region in the vicinity of the resonance, $s\sim M^2$. If the vertex function is nearly a constant in this region, that is, one can set
$\Gamma(s)\approx\Gamma(M^2)\equiv\Gamma$, and $\Gamma\ll M$,
then the amplitude (\ref{BW1}) can be expanded retaining only the leading term,
\be
A\approx\frac{\Gamma/2}{M-\sqrt{s}-i\Gamma/2}.
\label{BW1nr}
\ee
The distribution just arrived at is the celebrated Breit--Wigner distribution, and the quantities $M$ and $\Gamma$ are known as the mass and the width of the resonance. 

Meanwhile, if the resonance resides near threshold, that is, $M\simeq 2M_{\bar{Q}q}\equiv M_{\rm th}$, then the dependence of the vertex function on the energy is important and cannot be neglected. 
As 
a 
result, the line shape of a near-threshold resonance is not described by the Breit-Wigner distribution. Furthermore, the notions of the mass and width cannot be defined for it. Indeed, near threshold,
the width of a two-body $S$-wave decay behaves as $\Gamma\propto\sqrt{E}$, where the energy $E$ is counted from the threshold $M_{\rm th}$, that is, 
$\sqrt{s}=M_{\rm th}+E$. Then the amplitude (\ref{BW1nr}) takes the form
\be
A\propto \frac{1}{E-E_0+i\times\mbox{const}\times\sqrt{E}},
\label{BW1nr2}
\ee
where, for convenience, the quantity $E_0=M-M_{\rm th}$ was introduced. It is easy to see that the distribution (\ref{BW1nr2}) possesses few features not inherent for the Breit--Wigner 
distribution (\ref{BW1nr}), namely
\begin{itemize} 
\item the line shape is asymmetric around $E=E_0$;
\item the maximum of the curve does not necessarily lie at $E_0$ since the function $\sqrt{E}$ needs to be analytically continued below threshold, that is, for  
$E<0$, where it will contribute to the real part of the denominator and will move the pole;
\item for a particular relation between the parameters, the peak resides strictly at threshold, that is, at $E=0$, irrespectively of the value 
taken by the parameter $E_0$;
\item the visible width of the peak is not given by a single parameter, like the width $\Gamma$ in the formula (\ref{BW1nr}), which would define the imaginary part of the pole position in 
the energy complex plane. 
\item derivative in the energy from the amplitude turns to infinity at the threshold.
\end{itemize}

The aforementioned properties of the distribution (\ref{BW1nr2}) belong to thresholds phenomena which constitute the subject of the present review.
In particular, in chapter~\ref{sec:2chan} we will consider the form of the distributions for near-threshold states resulted from the interplay of different dynamics, including a multi-channel case. 

Therefore, as was mentioned above, it is not possible as a matter of principle to introduce the notions of the mass and width for a near-threshold state. 
It should be noted then that the mass of the $\X$ quoted in Eq.~(\ref{Xmass}) should not be taken literally since it was obtained from a Breit-Wigner fit for the line shape. The most important 
information encoded in Eq.~(\ref{EBexp}) is that the $\X$ resides very close to the neutral two-body threshold $D\bar{D}^*$, so that taking it into account is critically important for understanding 
the 
nature of this state.

Also, it proves convenient to deal with the notion of the binding energy which, therefore, should be properly defined. It follows from the discussion above that a simple way used in formula 
(\ref{EBexp}) is not adequate given the uncertainty in the definition of the mass $M_X$. In what follows, as the energy of a near-threshold state we will understand the difference between the 
zero of the real part of the denominator of the distribution and the threshold position (in other words, this is the real part of the pole of the amplitude relative to the threshold). Meanwhile, 
convenient 
and universal parameters describing the imaginary part of the denominator are the coupling constants of the resonance to its various decay channels.

It should also be noted that the Breit-Wigner amplitude (\ref{BW1nr}) possesses only one pole in the complex energy plane which lies at $E=E_0-i\Gamma/2$. 
Formally, this is at odds with the Schwarz reflection principle and, for this reason, violates analyticity of the amplitude. 
However, this violation is not large if one considers the energy region near the resonance since the path of the mirror pole, located at $E_0+i\Gamma/2$, to this region is much longer 
than that for the pole at $E_0-i\Gamma/2$. At the same time, when working near threshold, one has to retain both symmetric poles since none of them can be regarded 
as dominating in this region. Obviously, the amplitude (\ref{BW1nr}) does not meet this requirement and needs to be generalised.

\section{Elementary or compound state?}\label{weinapp}

As mentioned in the Introduction, the wave function of a near-threshold resonance may contain both a compact, ``elementary'', and molecular component. 
Investigation of the nature of such a resonance implies \emph{inter alia} establishing the relative weights of these components in a model-independent way. The cornerstone of the method suggested by 
Weinberg in his works \cite{Weinberg:1962hj,Weinberg:1963zza,Weinberg:1965zz} is the analysis of low-energy observables which, indeed, allows one to make conclusions whether the given state is 
``elementary'' or compound. The method is based on the idea of extending the basis of the coupled channels and including into it the ``bare'' resonance whose wave function gets renormalised because 
of the interaction with the ``compound'' channels. Ideologically this method is reminiscent of the standard field theoretical approach where interaction of fields results in a decrease of the 
probability to observe the ``bare'' field from unity to a smaller value, $0\leqslant Z\leqslant 1$. An important achievement gained in the Weinberg approach is establishing a one-to-one correspondence 
between such a $Z$-factor and experimentally observable quantities that allows one to come to a model-independent conclusion on the structure of the wave function of the resonance. In paper 
\cite{Weinberg:1965zz}, this approach was applied to the case of a below-threshold state --- the deuteron. In work \cite{Baru:2003qq}, a natural generalisation of the given approach to the 
case of an arbitrary near-threshold resonance was suggested, equally applicable to both below- and above-threshold resonances; it also allowed to incorporate inelasticity. It is important to note 
that the most relevant formalism to be used with the Weinberg approach is the $t$-matrix (or scattering operator) approach, brilliantly described in book 
\cite{BZP}, after it is extended to include the ``elementary'' state. Details of such a generalised formalism considered in this section can be found in Refs.~\cite{Baru:2010ww,Hanhart:2011jz}

In Subsection~\ref{general} one should pay attention to the formulation of the coupled-channel problem (Eqs.~(\ref{state})-(\ref{Hmol})), to the different types of solution of such a problem 
(Eqs.~(\ref{bound}) and (\ref{chi0c})), and to the definition of the $Z$-factor (Eq.~(\ref{Zf})) and the spectral density as its convenient generalisation (Eq.~(\ref{wE})) which will be used below in 
the analysis of the experimental data for the $\X$. In Subsection \ref{weinbergformula}, important information is contained in the Weinberg formulae (\ref{a})
and (\ref{re}), as well as in Eq.~(\ref{tmat}) which gives the most general form of the energy distribution in case of only one hadronic channel with a direct interaction between mesons in it. In 
Subsection \ref{sec:2chan} we describe generalisation of the results from Subsection \ref{weinbergformula} to a two-channel case. The main results of this subsection are illustrated 
in Figs.~\ref{brqFig}-\ref{brh2Fig}, the most important conclusion being the observation that quite a nontrivial line shape of a near-threshold resonance can be provided by only two parameters
defining the position of the zero of the scattering amplitude and the contribution of the nonlinear term $k_1k_2$, respectively, where $k_1$ and $k_2$ are the momenta in the mesonic channels. Given 
that the existing experimental data on the $\X$ do not allow one to identify such irregularities of the line shape, Subsection \ref{sec:2chan} provides 
a surplus to be used in the future, when new experiments will be able to provide us with accurate information on near-threshold states containing heavy quarks. This part can be skipped during 
cursory reading of the review or if the reader is only interested in the state-of-the-art of the experimental situation with the $\X$.

\subsection{Coupled-channel scheme}
\label{general}

Consider a physical resonance which is a mixture of a bare elementary state $|\psi_0\rangle$ (for example, $\bar{c}c$ or tetraquark) and
multiple two-body components (labelled as $i=1,2,\ldots$), and represent its wave function as
\be
|\Psi\rangle=\left(
\begin{array}{c}
c|\psi_0\rangle\\
\chi_1(\vep)|M_{11}M_{12}\rangle\\
\chi_2(\vep)|M_{21}M_{22}\rangle\\
\cdots
\end{array}
\right),
\label{state}
\ee
where $c$ is the probability amplitude for the bare state while $\chi_i(\vep)$ describes the relative motion in the system of two mesons $\{M_{i1}M_{i2}\}$ with the relative momentum $\vep$.

The system is described by the Schr{\"o}dinger equation
\be 
{\cal H}|\Psi\rangle=M|\Psi\rangle
\label{Sheq}
\ee
with the Hamiltonian
\be
\hat{\cal H}=
\left(
\begin{array}{cccc}
H_0&V_{01}&V_{02}&\cdots\\
V_{10}&H_{h_1}&V_{12}&\cdots\\
V_{20}&V_{21}&H_{h_2}&\cdots\\
\cdots&\cdots&\cdots&\cdots
\end{array}
\right).
\label{Hmol}
\ee

Here 
$$
H_0|\psi_0\rangle=M_0|\psi_0\rangle,
$$
where $M_0$ is the mass of the bare state. In what follows it will be convenient to work in terms of the energy counted from the lowest threshold with $i=1$, that is,
$$
M=m_{11}+m_{12}+E,\quad M_0=m_{11}+m_{12}+E_0.
$$

We assume the direct interactions between mesonic channels $i$ and $j$ to exist and encode them in the potentials $V_{ij}(\vep,\vep')$ (including the diagonal terms with $i=j$), so that
$$
H_{h_i}(\vep,\vep')=\left(m_{i1}+m_{i2}+\frac{p^2}{2\mu_i}\right)\delta^{(3)}(\vep-\vep')+
V_{ii}(\vep,\vep'),
$$
where $m_{i1}$ and $m_{i2}$ are the meson masses in the $i$-th channel, with the reduced mass
$$
\mu_i=\frac{m_{i1}m_{i2}}{m_{i1}+m_{i2}}.
$$
The $i$-th mesonic channel and the bare state communicate via the off-diagonal quark-meson potential $V_{0i}$ which is given by the transition vertex $f_i(\vep)$.

The $t$-matrix for the coupled-channel problem satisfies a Lippmann-Schwinger equation (written schematically),
\be
t=V-VSt,
\ee
where $S$ is a diagonal matrix of free propagators,
\be
S_0(E)=\frac{1}{E_0-E-i0},
\ee
\be
S_i(\vep)=\frac{1}{p^2/(2\mu_i)-E+\Delta_i-i0},\quad \Delta_i=(m_{i1}+m_{i2})-(m_{11}+m_{12}).
\ee

The formalism of the $t$-matrix is, of course, fully equivalent to a more ``standard'' method based on the Schr{\"o}dinger equation for the wave function of the system --- see, for example, the 
textbook \cite{LL3}. Nevertheless, the $t$-matrix (or the scattering matrix as it is sometimes called in the literature --- see, for example, the book \cite{BZP}) grants some advantages. In 
particular, it allows one to consider the scattering problem for a 
off-mass-shell particle, that is, to formally ``disentangle'' the momentum and the energy which for on-mass-shell particles are 
linked to each other through the dispersion relation. 

As the first step, it is convenient to define the $t$-matrix for the potential problem (that is, the $t$-matrix of the direct interaction in the mesonic channel $t^V$) which obeys the following 
equation:
\be
t_{ij}^V(\vep,\vep')=V_{ij}(\vep,\vep')-\sum_k\int V_{ik}(\vep,\veq)S_k(\veq)t_{kj}^V(\veq,\vep') d^3q,
\label{tVdef}
\ee
and introduce the dressed vertex functions,
\begin{eqnarray}
\ds\phi_i(\vep)&=&f_i(\vep)-\sum_k\int t_{ik}^V(\vep,\veq)S_k(\veq)f_k(\veq)d^3q,\label{sol11}\\
\ds\bar{\phi}_i(\vep')&=&f_i(\vep)-\sum_k\int S_k(\veq)f_k(\veq)t_{ki}^V(\veq,\vep')d^3q.
\label{dvf2}
\end{eqnarray}
In terms of these quantities the full $t$-matrix can be expressed as
\cite{Hanhart:2011jz}
\bea
&&t_{00}(E)=-\frac{(E-E_0){\cal G}(E)}{E-E_0+{\cal G}(E)},\label{t00sol}\\
&&t_{0i}(\vep,E)=\frac{E-E_0}{E-E_0+{\cal G}(E)}\bar{\phi}_i(\vep),\label{t0isol}\\
&&t_{i0}(\vep,E)=\frac{E-E_0}{E-E_0+{\cal G}(E)}\phi_i(\vep),\label{ti0sol}\\
&&t_{ij}(\vep,\vep',E)=t_{ij}^V(\vep,\vep')+\frac{\phi_i(\vep)\bar{\phi}_j(\vep')}{E-E_0+{\cal G}(E)},\label{tijsol}
\eea
where
\be
{\cal G}(E)=\sum_i\int f_i^2(\veq)S_i(\veq)d^3q
-\sum_{i,j}\int f_i(\vek)S_i(\vek)t_{ij}^V(\vek,\veq)S_j(\veq)f_j(\veq)d^3k d^3q.
\label{Gsol}
\ee

The system under consideration can possess bound states
(generally, more than one). For the $\alpha$-th bound state with the binding
energy $E_B^{(\alpha)}$, the solution of equation
(\ref{Sheq}) can be written as
\be
c_B^{(\alpha)}=\cos\theta_{\alpha},\quad
\chi_B^{(i\alpha)}(\vep)=S_i(\vep)\phi_i(\vep)\sin\theta_{\alpha}.
\label{bound}
\ee
The wave function of this bound state is normalised,
$$
{}_B^{(\alpha)}\langle\Psi|\Psi\rangle_B^{(\alpha)}=1,
$$
which gives, for the mixing angle $\theta_{\alpha}$ the ralation
\be
\tan^2\theta_{\alpha}=\sum_i\int S^2_i(\vep)\phi^2_i(\vep)d^3p.
\label{tan2theta}
\ee

The quantity
\be
Z_{\alpha}=|\langle\psi_0|\Psi\rangle_B^{(\alpha)}|^2=\cos^2\theta_{\alpha}
\label{Zf}
\ee
introduced in Ref.~\cite{Weinberg:1965zz} gives the probability to find a bare state
in the wave function of the $\alpha$-th bound state.

The solution of equation (\ref{Sheq}) with the free asymptotic in the hadronic channel $i$ takes the form
\bea
\chi^{(i)}_{j;\vek_i}(\vep)&=&\delta_{ij}\delta^{(3)}(\vep-\vek_i)-S_j(\vep)
t_{ji}(\vep,\vek_i,E),\nonumber\\[-2mm]
\label{chi0c}\\[-2mm]
c^{(i)}_{\vek_i}(E)&=&-\frac{t_{0i}(\vek_i,E)}{E_0-E},\nonumber
\eea
where $k_i=\sqrt{2\mu_i(E-\Delta_i)}$.
The solution for the coefficient $c$ (see Eq.~(\ref{chi0c})) allows one to define a continuum counterpart of the quantity (\ref{Zf}) --- the
spectral density $w(E)$ --- which gives the probability to find the bare
state in the continuum wave function \cite{Bogdanova:1991zz}. It reads
\be
w(E)=\sum_i \mu_i k_i\Theta(E-\Delta_i)\int \left|c^{(i)}_{\vek_i}(E)\right|^2d o_{\vek_i},
\label{wE}
\ee
where $\theta$ is the Heaviside function.
As shown in Ref.~\cite{Bogdanova:1991zz}, the normalisation condition for the distribution $w(E)$ is
\be
\int_0^{\infty} w(E)dE=1-\sum_{\alpha}
Z_{\alpha},
\label{wnorm}
\ee
where the sum goes over all bound states present in the system.

\begin{figure}[t]
\centerline{\raisebox{5mm}{\epsfig{file=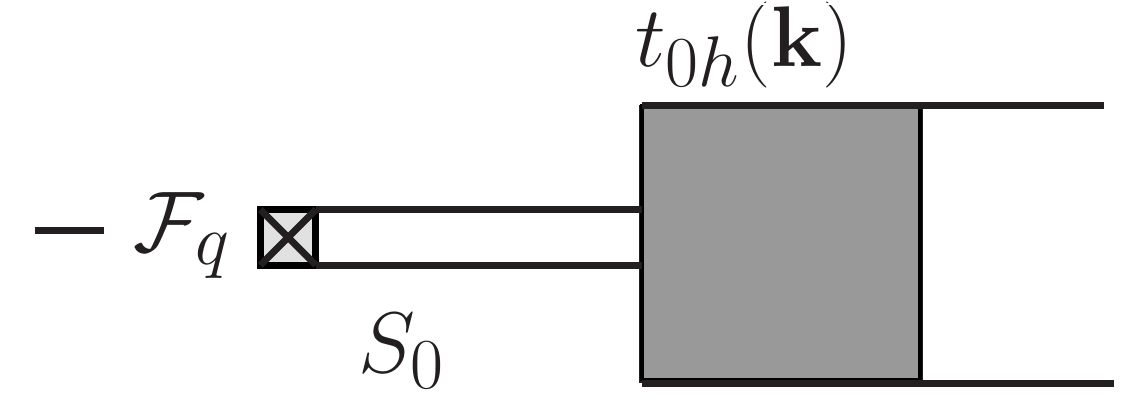, width=5.5cm}}
\hspace*{15mm}\epsfig{file=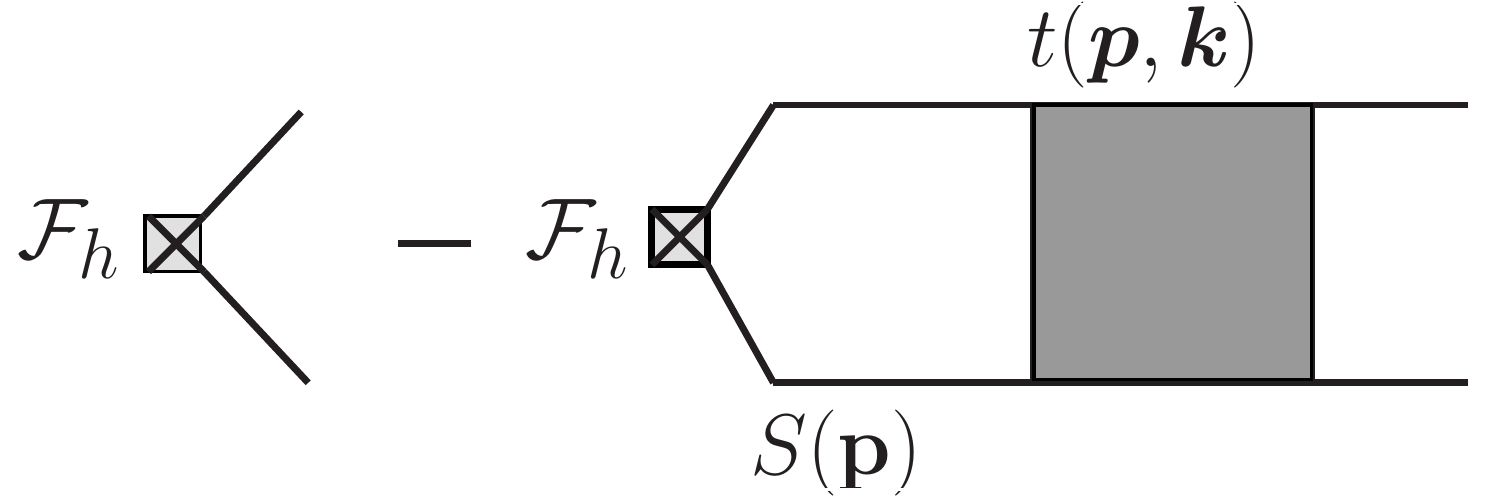,width=7.5cm}}
\caption{Diagrams for the two meson production via the quark component (left plot) and via the
hadronic component (right plot).}\label{diagfig}
\end{figure}

We conclude this subsection with the formulae for the production amplitude of the resonance under study. In fact, as the wave function of the system is 
multi-component, see Eq.~(\ref{state}), there are several production mechanisms possible,
namely a production of the resonance via its
quark component or via its hadronic components, as depicted in Fig.~(\ref{diagfig}). For our purposes it is sufficient to consider 
a point-like production source. Then, the production amplitude of the meson pair $(M_{i1}M_{i2})$ through
the hadronic component $j$ can be written as
\be
{\cal M}^i_j(E)={\cal
F}_{h_j}\delta_{ij}-{\cal F}_{h_j}\int S_j(\vep)t_{ji}(\vep,\vek_i,E)d^3p,
\label{Mh}
\ee
where ${\cal F}_{h_j}$ is the initial state production amplitude from a
point-like source, and $k=\sqrt{2\mu_i(E-\Delta_i)}$. The production amplitude from the quark component
is given by
\be 
{\cal M}^i_q(E)=-{\cal F}_qS_{0}(E)t_{0i}(\vek_i,E), 
\label{Mq} 
\ee
where ${\cal F}_q$ is the production amplitude for the bare quark state by a point-like source.

\subsection{Weinberg formulae, singularity structure and underlying dynamics}
\label{weinbergformula}

Consider a single mesonic channel (for this reason the channel index will be omitted) with the $S$ wave in the mesonic subsystem,
and take the low-energy limit defined by the inequality $k \ll \beta$, where $k$ is the momentum in the mesonic
subsystem and $\beta$ is the inverse range of force.

For convenience, let us define the loop functions
\begin{eqnarray}
g(E)=\int d^3q\frac{f^2(\veq)}{q^2/(2\mu)-E-i0}=f_0^2(R+4\pi^2\mu ik),
\nonumber\\[-2mm]
\label{ggpr}\\[-2mm]
g'(E)=\int d^3q\frac{f(\veq)}{q^2/(2\mu)-E-i0}=f_0(R'+4\pi^2\mu ik), \nonumber
\end{eqnarray}
where $\mu$ is the reduced mass, and for the real parts of the loops, $R$ and $R'$, only the leading contributions of the order $\mu\beta$ were retained while the corrections $O(k^2/\beta^2)$ were 
neglected;
we also defined $f_0\equiv f_1(0)$. Using the scattering
length approximation for the potential problem $t$-matrix,
\be
t_V(\veq,\vep,E)=-\frac{1}{4\pi^2\mu(-a_V^{-1}-ik)},
\ee
where the scattering length $a_V$ is treated as an input parameter, one arrives at the following form for the mesonic $t$-matrix element (\ref{tijsol}):
\be
t(E)=\frac{E-E_C}{[E-E_0]R_V+f_0^2[RR_V-R'^2]
+4\pi^2\mu ik[E-E_C]},
\label{tmatrix}
\ee
$$
E_C=E_0-f_0^2(R+R_V-2R'),
$$
where $R_V=(2\pi)^2\mu\gamma_V$, $\gamma_V=1/a_V$.
Notice that expression (\ref{tmatrix}) for the mesonic $t$-matrix has a zero at the energy $E=E_C$ --- to be discussed in detail in chapter \ref{sec:2chan}.

Expression (\ref{tmatrix}) allows one to perform a renormalisation procedure, that is, to redefine the ``bare'' parameters of the theory such way that they absorb the quantities $R$ and $R'$, which depend on the range of force (see
Ref.~\cite{Baru:2010ww} for further details). The new renormalised parameters $E_f$ and $g_f$ are introduced as
\be
[E_f-E_0]R_V+f^{(0)2}[RR_V-R'^2]=0
\ee
and
\be
g_f=8\pi^2\mu\frac{E_f-E_C}{R_V}=\frac{8\pi^2\mu
f^{(0)2}}{R_V^2}(R-R_V)^2.
\label{gf}
\ee

In terms of the renormalised quantities the $t$-matrix reads
\be
t(E)=\frac{1}{(2\pi)^2\mu}\;\frac{\frac12 g_f}{{\cal D}(E)},
\label{tf}
\ee
where
\be
{\cal D}(E)=E-E_f-\frac{(E-E_f)^2}{E-E_C}+\frac{i}{2}g_f k.
\label{Fden}
\ee

Equation (\ref{tf}) resembles a well-known Flatt{\'e} formula \cite{Flatte:1976xu} for the near-threshold scattering amplitude and,
if $|E_C| \gg |E_f|$, one indeed arrives at the standard Flatt{\'e} distribution,
\be
t(E)\mathop{\approx}_{|E_C| \gg |E_f|}\frac{1}{(2\pi)^2\mu}\frac{\frac12 g_f}{E-E_f+\frac{i}{2}g_f k}.
\label{tF1}
\ee

To establish a connection with the Weinberg's work \cite{Weinberg:1962hj,Weinberg:1963zza,Weinberg:1965zz} let us assume the existence of a near-threshold bound state with the binding energy $E_B$. 
Then, 
in the vicinity of the bound-state pole one has
\be
t(E)\simeq\frac{g_{\rm eff}^2}{E+E_B},
\label{tpole}
\ee
where (see Refs.~\cite{Weinberg:1962hj,Weinberg:1963zza,Weinberg:1965zz})
\be
g_{\rm eff}^2=\frac{\sqrt{2\mu E_B}}{4\pi^2\mu^2}(1-Z),
\label{boundstatevertex}
\ee
and the quantity $Z$ is defined as
\be
\frac{Z}{1-Z}=\frac{2E_B}{E_B +E_C}
\left(1-\frac{\gamma_V}{\sqrt{2\mu E_B}}\right).
\label{Z}
\ee
It should be noted that, for $Z=0$, expression (\ref{boundstatevertex}) turns to the standard formula which defines the coupling constant of a compound object through the binding energy of its 
constituent \cite{Landau}.

The zero of the elastic scattering amplitude is located at
\be
E_C=-E_B
\left(1-\frac{2(1-Z)}{Z}+\frac{2(1-Z)}{Z}\frac{\gamma_V}
{\sqrt{2\mu E_B}} \right).
\label{N27}
\ee

Finally, for the scattering phase in the mesonic channel, $\delta$, one can find
\be
k\cot\delta =
-\sqrt{2\mu E_B}+\frac{\sqrt{2\mu E_B}(E+E_B)
(E_B + E_C)} {2E_B(E-E_C)}\frac{Z}{1-Z}.
\label{kcot}
\ee

Generally, because of the aforementioned zero of the scattering amplitude, Eq.~(\ref{kcot}) does not admit an effective-range expansion. The
latter is recovered only if $|E_C|\gg E_B$, and then the most famous Weinberg formulae for the scattering length $a$ and the effective radius $r_e$ arise,
\be
a=\frac{2(1-Z)}{(2-Z)}\frac{1}{\sqrt{2\mu E_B}}+O\left(\frac{1}{\beta}\right),
\label{a}
\ee
\be
r_e=-\frac{Z}{(1-Z)}\frac{1}{\sqrt{2\mu E_B}}+O\left(\frac{1}{\beta}\right).
\label{re}
\ee

Thus, if the effective-range expansion exists, we have a powerful tool at our disposal which allows us to distinguish compact (elementary) bound states from molecular (composite) ones. Indeed, 
if 
$Z\approx 1$ and, therefore, the state is elementary, the effective range in Eq.~(\ref{re}) is large and negative, and the
scattering length is small --- see Eq.~(\ref{a}). If, on the contrary, the state is molecular ($Z\ll 1$), the situation is opposite: the effective range is small (it can be positive, if the
range-of-force corrections are taken into account), and the scattering length is large.

Complementary to the Weinberg formulae is the ``pole counting'' procedure \cite{Morgan:1992ge}. Rewriting the denominator of the Flatt{\'e}
formula (\ref{tF1}) in terms of the effective-range parameters with
\be
a=-\frac{g_f}{2E_f},\quad r_e=-\frac{2}{\mu g_f},
\ee
one finds for the $t$-matrix poles
\be
k_{1,2}=\frac{i}{r_e} \pm \sqrt{-\frac{1}{r_e^2}+\frac{2}{ar_e}}.
\ee
If there is a bound state, the pole positions are given by
\be
k_1=i\sqrt{2\mu E_B},\quad k_2=-i\frac{2-Z}{Z}\sqrt{2\mu E_B},
\ee
so that the first pole is on the physical sheet while the second one is on the unphysical sheet. If $Z \approx 1$ (elementary state) then the effective coupling $g_f$ is small while the effective 
range $r_e \to-\infty$, so that there are two almost symmetric near-threshold poles. If $Z \ll 1$ (molecular state), then the coupling $g_f$ is large, the effective range is small and, as a 
result, there is only one near-threshold pole.

If there is no bound state in the system (the scattering length is negative) the pole counting procedure works too. If the coupling $g_f$ is
small, $r_e \to-\infty$ and $k_{1,2}\rightarrow \pm \sqrt{2\mu E_f}$, so that there are two near-threshold poles, that corresponds to the
resonance case. For larger values of $g_f$ poles move deeper into complex plane and eventually collide at the imaginary axis. If the coupling is very large, the effective range is small and there is 
only one near-threshold pole, that corresponds to a virtual
state. One might argue that the probabilistic interpretation is
lost in the case of a resonance/virtual state. It is not the case however. In the low-energy approximation the spectral density (\ref{wE})
(which defines the probability to find a bare state in the continuum) is expressed in terms of Flatt{\'e} parameters as 
\be
w(E)=\frac{1}{2\pi}\frac{g_fk}{|E-E_f+\frac{i}{2}g_fk|^2}.
\label{wthreshold} 
\ee 
This, together with the normalisation condition (\ref{wnorm}) for the spectral density, allows one to draw conclusions on the relative weight of the molecular component in the wave function of the 
state: 
one simply integrates the expression (\ref{wthreshold}) over the near-threshold region --- see paper
\cite{Baru:2003qq} for further details.

Weinberg formulae look transparent when there is no near-threshold $t$-matrix zero. To understand what happens in the general case it is instructive to study the singularities of the amplitude in 
terms of the variables $E_f$, $g_f$, and $\gamma_V$. As
\be
E_C=E_f-\frac{1}{2}g_f\gamma_V,
\label{ecgf}
\ee
the expression for the $t$-matrix can be rewritten in the form
\be
t(E)=\frac{1}{4\pi^2\mu}\frac{E-E_f+\frac12 g_f\gamma_V}{(E-E_f)(\gamma_V+ik)+\frac{i}{2}g_f\gamma_Vk},
\label{tmat}
\ee
which shows that the $t$-matrix can have up to three near-threshold poles.
A detailed analysis \cite{Baru:2010ww} of these singularities shows that the appearance of the $t$-matrix zero corresponds to the case of all three poles reside near the threshold. This requires: 
i) the direct 
interaction in the mesonic channel to be strong enough to support a bound or virtual state and ii) a near-threshold bare quark state to exist, with a weak coupling to the mesonic channel. In other 
words, the pole generated by the direct interaction collides with the pair of poles corresponding to the bare state, and this collision results in a break-down of the effective range expansion.

If at least one of the conditions above is not met, there is no $t$-matrix zero in the near-threshold region, the effective-range
formulae are valid, and the conclusions of Refs.~\cite{Weinberg:1965zz,Baru:2003qq,Morgan:1992ge} hold
true: a large and negative effective range corresponds to a compact quark state while a small effective range implies that the state is composite. In the
former case there are two near-threshold poles in the $t$-matrix while in the latter case there is only one pole.

In the double-pole situation one can definitely state that the resonance is generated by an $s$-channel exchange, with a small coupling of this quark
state to the mesonic continuum. In the single-pole situation no model-independent insight on the underlying dynamics is possible and one can only conclude that the resonance is generated dynamically.

A final comment is in order here. Consider the case of a bound state and no direct interaction in the mesonic channel. It is straightforward to derive a formula which expresses the $Z$-factor through 
the binding energy and the ``bare'' coupling constant between the channels $f_0$ (see the definition in Eq.~(\ref{ggpr})),
\be
\frac{1}{Z}-1\approx
\frac{4\pi^2\mu^2f_0^2}{\sqrt{2\mu E_B}}.
\label{tgtheta}
\ee
From the formula just derived one can see that, for a decreasing binding energy, the $Z$-factor decreases too, thus approaching zero as $Z\propto\sqrt{E_B}$ in the limit $E_B\to 0$. In other words, 
the closer the resonance 
resides to the threshold, the larger is the contribution of the molecular component to its wave function, that is, the more ``compound'' it becomes. 
Furthermore, even a tiny ``bare'' coupling of the quark channel to the mesonic one $f_0$ is sufficient to reach any large, close to unity, value of the ``molecularness'' 
(that is, the value of $1-Z$) of the physical state due to its proximity to the threshold. Such a conclusion is quite natural given that proximity of the resonance to a threshold facilitates its 
transition to the corresponding final state, so that the probability to observe the resonance in the given hadronic channel is enhanced, that by definition increases the ``molecularness'' of this
near-threshold resonance.

It is also important to stress that the Weinberg approach is model-independent. To begin with, we notice that the $Z$-factor defines the residue of the $t$-matrix (and, therefore, that of the 
amplitude) 
at the bound state pole --- see Eqs.~(\ref{tpole}) and (\ref{boundstatevertex}). Since the probability evaluated from this amplitude is an observable quantity, then all the parameters entering it are 
model- and scheme-independent. This way, for a given binding energy $E_B$, the value of the $Z$-factor is defined uniquely and in a model-independent way. 

Furthermore, deeper reasons for the model-independence of the $Z$-factor can be identified. In the approach of effective field theories, parameters entering the Lagrangian are subject to 
renormalisation and as such can be model-dependent (in the sense of the dependence on the regularisation scheme). Such a renormalisation is done by adding to the Lagrangian counter terms which 
can only have an analytic dependence on the energy. This implies that the terms proportional to $\sqrt{E}$ cannot be affected by renormalisation and are fixed in a model-independent way. It is this 
square-root behaviour that is typical for the key quantities in the Weinberg approach, in particular, the $Z$-factor. Thus, as long as the binding energy is small enough for the binding momentum 
$\gamma=\sqrt{2\mu E_B}$ to be small compared to the inverse range of force $\beta$, the long-range part of the wave function of such a resonance
is model-independent and totally defined by the value of $\gamma$.

\subsection{Interplay of quark and meson degrees of freedom: two-channel case}
\label{sec:2chan}

Of particular relevance for the $\X$ is the case of two nearby $S$-wave thresholds, $D^0\bar{D}^{*0}$ and $D^+D^{*-}$ split by
$\Delta=M_{th_2}-M_{th_1}\approx 8$~MeV, so that the physical state $X$ is a mixture of a bare quark state and two mesonic components,
\be
|X\rangle=\left(
\begin{array}{c}
c|\psi_0\rangle\\
\chi_1(\vep)|M_{11}M_{12}\rangle\\
\chi_2(\vep)|M_{21}M_{22}\rangle\\
\end{array}
\right).
\label{2state}
\ee
Here the lower indices 1 and 2 denote the $D^0\bar{D}^{*0}$ and $D^+D^{*-}$ component, respectively.

We assume in what follows that the bare state is an isosinglet and set $f_1=f_2=f/\sqrt{2}$ in all general formulae, where the factor $1/\sqrt{2}$ is introduced for convenience in matching with the 
single-channel case. The low-energy reduction and renormalisation procedure are performed in a way quite similar to the one described in chapter~\ref{weinbergformula}. In particular, as the first 
step, we define the $t$-matrix for the direct potential $V$ and parameterise it
in the scattering length approximation,
\be 
t^V=\frac{1}{4\pi^2\mu}\frac{1}{\mbox{Det}} \left(
\begin{array}{cc}
\frac12(\gamma_s+\gamma_t)+ik_2&\frac12(\gamma_t-\gamma_s)\\
\frac12(\gamma_t-\gamma_s)&\frac12(\gamma_s+\gamma_t)+ik_1
\end{array}
\right), 
\label{tv} 
\ee 
where 
\be
\mbox{Det}=(\gamma_s\gamma_t-k_1k_2)+\frac{i}{2}(\gamma_s+\gamma_t)(k_1+k_2),
\label{detdef} 
\ee 
the quantities $\gamma_s$ and $\gamma_t$ are the inverse scattering lengths in the singlet and triplet (in isospin) channels, respectively,
\be 
k_1=\sqrt{2\mu E},\quad
k_2=\sqrt{2\mu(E-\Delta)}, 
\ee 
and $\mu$ is the reduced mass (since the splitting $\Delta$ is small compared to the masses of the mesons, the reduced masses in the charged and neutral channel are set equal to each other).

Then, setting
\bea
&&E_f=E_0-\frac{f_0^2}{R_s}(RR_s-R'^2),\label{Ef1}\\
&&E_C=E_0-f_0^2(R_s+R-2R'),\label{EC1}\\
&&g_f=\frac{8\pi^2\mu}{R_s^2}f_0^2(R_s-R')^2,\label{gf1} 
\eea 
where $R_s=4\pi^2\mu\gamma_s$ one readily arrives at the following expression for the full mesonic $t$-matrix:
\bea
t_{11}&=&\frac{1}{8\pi^2\mu}\frac{\gamma_s(E-E_f)+(E-E_C)\left(\gamma_t+2ik_2\right)}{D(E)},
\label{t11}\\
t_{12}&=&t_{21}=\frac{1}{8\pi^2\mu}\frac{\gamma_t(E-E_C)-\gamma_s(E-E_f)}{D(E)},\label{t12}\\
t_{22}&=&\frac{1}{8\pi^2\mu}\frac{\gamma_s(E-E_f)+(E-E_C)\left(\gamma_t+2ik_1\right)}{D(E)},
\label{t22} 
\eea 
where 
\be
D(E)=\gamma_s\left(\gamma_t+\frac{i}{2}(k_1+k_2)\right)(E-E_f)-\left(k_1k_2-\frac{i}{2}\gamma_t(k_1+k_2)\right)(E-E_C).
\label{Den}
\ee 

Expressions (\ref{t11})-(\ref{t22}) were derived in Ref.~\cite{Hanhart:2011jz}. An alternative derivation of these expressions can be found in  Ref.~\cite{Artoisenet:2010va}.

It is easy to demonstrate that the isosinglet element of the $t$-matrix has a zero at $E=E_C$. Similarly to the single-channel case
(see Eq.~(\ref{ecgf})), the coupling constant $g_f$ and the parameter $E_C$ are interrelated as
\be 
E_C=E_f-\frac{1}{2}g_f\gamma_s, 
\ee 
so that the zero at $E_C$ is generated in the near-threshold region if the condition $|\gamma_s|\lesssim \Delta/g_f$ holds true; otherwise it leaves
the region of interest.

Another phenomenon which affects the observables is the momenta entanglement: the momenta $k_1$ and $k_2$ enter the expressions for
the $t$-matrix in a complicated nonlinear way, including the product $k_1k_2$ (see also Ref.~\cite{Stapleton:2009ey} for the discussion of the related effect on the line shapes for the $X(3872)$). 
This entanglement of mesonic channels is governed by the triplet inverse scattering length $\gamma_t$ and becomes
strong for $|\gamma_t|\lesssim\sqrt{\mu\Delta}$.

In such a way one can identify the following limiting cases:
\begin{itemize}
\item Case (i): $|\gamma_s|\to\infty$ and $|\gamma_t|\to\infty$.
\item Case (ii): small $\gamma_s$ and $|\gamma_t|\to\infty$. 
\item Case (iii): $|\gamma_s|\to\infty$ and small $\gamma_t$. 
\item Case (iv): both $\gamma_s$ and $\gamma_t$ are small.
\end{itemize}

Case (i) corresponds to a weak direct interaction and is described by the two-channel Flatt{\'e} formula with the denominator containing a plain sum of the contributions from the two 
channels, $\frac12g_f(k_1+k_2)$. For a small coupling $g_f$ one has $Z \to 1$ and the state is mostly compact while a large $g_f$ corresponds to $Z \to 0$, and the state is predominantly molecular. 
Case (ii) is, effectively, a single-channel one:
with $|\gamma_t| \to \infty$ the interaction in the isotriplet channel is weak, channel entanglement is irrelevant, the interplay between the quark and mesonic degrees of freedom takes place in the 
isosinglet channel, and the dynamics is defined by the $t$-matrix zero. On the contrary, Case (iii) is free from the $t$-matrix zero, dynamics is
defined by the channel entanglement which, in turn, is governed by the 
isotriplet scattering length. Case (iv) is a generic one when all aforementioned effects play a role for the dynamics of the system.

Since there is no hope that any data on the $D^{(*)}$-meson scattering could become available, our knowledge comes from studies of resonance production 
in various reactions. Therefore, we consider 
production line shapes (production differential rates) of a given resonance under various assumptions concerning its production mechanism. In particular, we study the production of the neutral $D^0 
\bar D^{*0}$ mesons through the hadronic and quark component. The
corresponding amplitudes are obtained by a low-energy reduction of the general
formulae (\ref{Mh}) and (\ref{Mq}). We assume that the $t$-matrix has a near-threshold pole that allows us to neglect the Born terms
in the production through the hadronic components (the first term on the r.h.s. of Eq.~(\ref{Mh}));
a detailed discussion of this issue can be found in Ref.~\cite{Baru:2010ww}.

\begin{figure}
\centerline{\epsfig{file=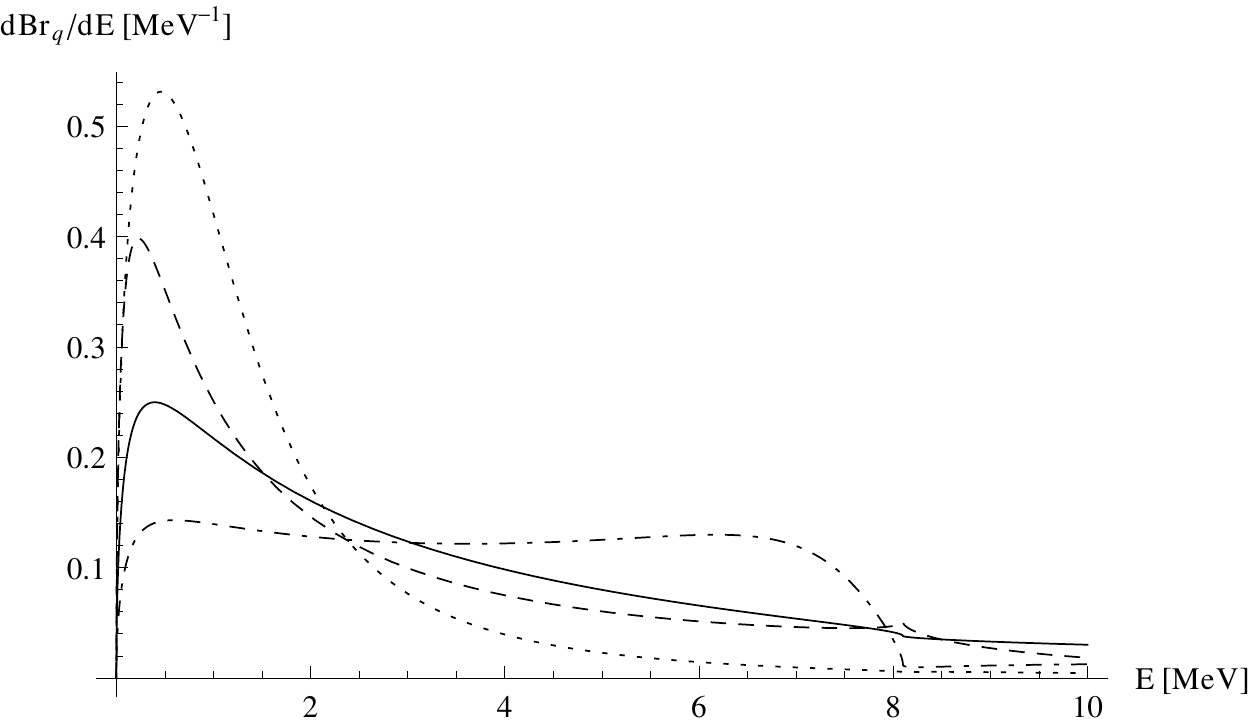, width=0.6\textwidth}}
\caption{Differential production rate through the quark component (\ref{BrqX}). Cases (i)-(iv) are given by the solid, dashed,
dashed-dotted, and dotted line, respectively. Adapted from Ref.~\cite{Hanhart:2011jz}.}\label{brqFig}
\medskip
\centerline{\epsfig{file=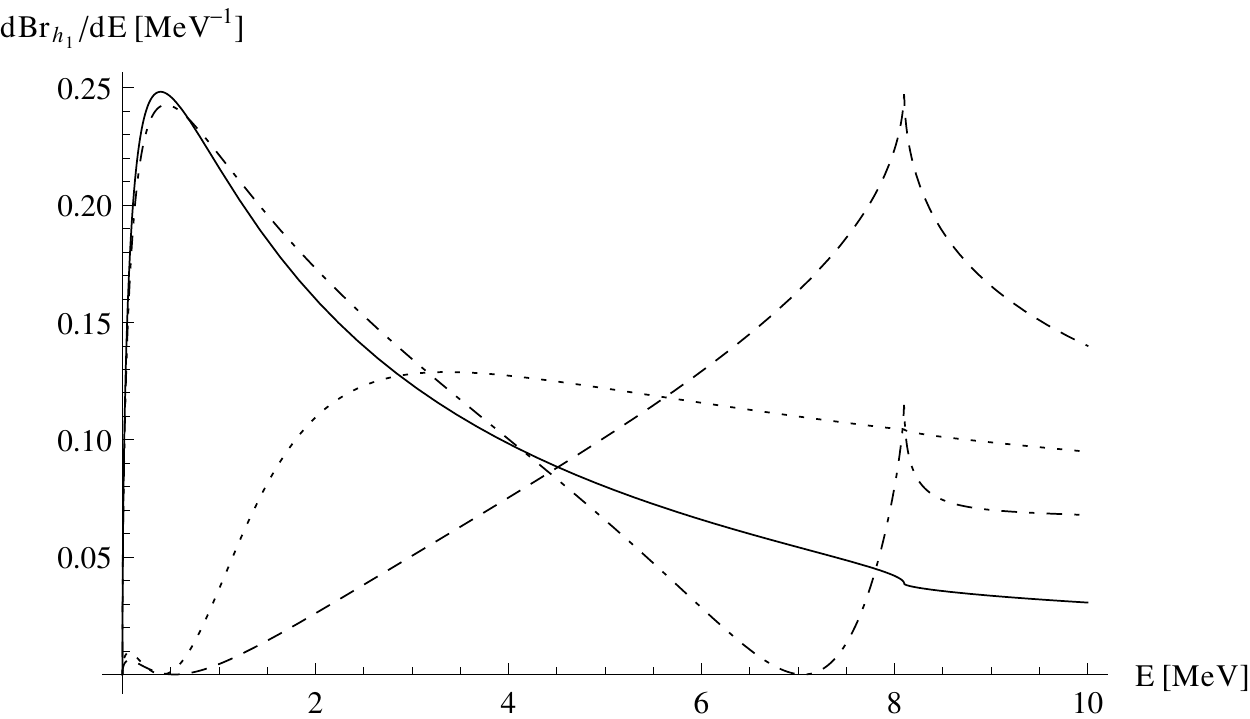, width=0.6\textwidth}}
\caption{Differential production rate through the first mesonic component(\ref{Brh1X}). Cases (i)-(iv) are given by the solid,
dashed, dashed-dotted, and dotted line, respectively. Adapted from Ref.~\cite{Hanhart:2011jz}.}\label{brh1Fig}
\medskip
\centerline{\epsfig{file=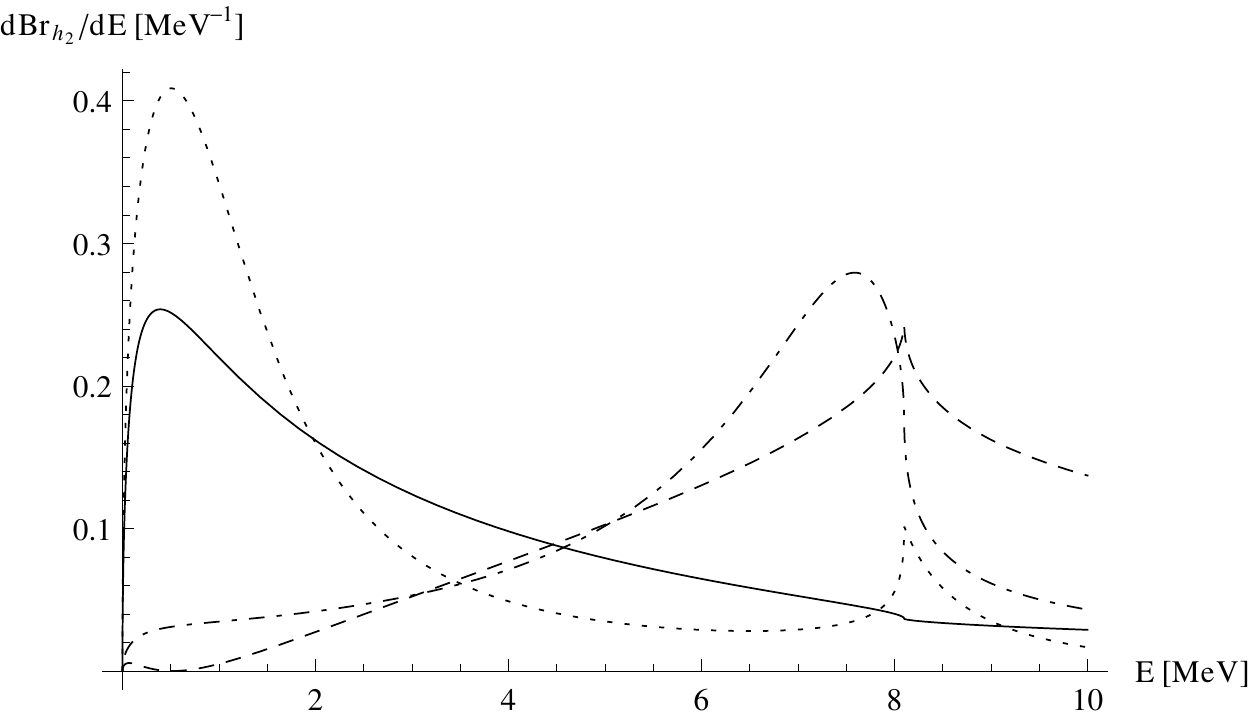, width=0.6\textwidth}}
\caption{Differential production rate through the second mesonic component (\ref{Brh2X}). Cases (i)-(iv) are given by the solid,
dashed, dashed-dotted, and dotted line, respectively. Adapted from Ref.~\cite{Hanhart:2011jz}.}\label{brh2Fig}
\end{figure}

If one neglects the interference between the three possible production mechanisms it is straightforward to find the corresponding differential
branchings,
\bea
&&\frac{d\Br_q}{dE}=\Theta(E)\frac{{\cal
B}_0\sqrt{E}}{|D(E)|^2}{\gamma_s}^2\left|\gamma_t+ik_2\right|^2,
\label{BrqX}\\
&&\frac{d\Br_{h_1}}{dE}=\Theta(E)\frac{{\cal
B}_1\sqrt{E}}{|D(E)|^2}\left|\vphantom{\int_0^1}\gamma_s
\left(E-E_f\right)
+(\gamma_t+2ik_2)\left(E-E_f+\frac12g_f\gamma_s\right)\right|^2,\label{Brh1X}\\
&&\frac{d\Br_{h_2}}{dE}= \Theta(E)\frac{{\cal
B}_2\sqrt{E}}{|D(E)|^2}\left|\vphantom{\int_0^1}\gamma_s\left(E-E_f\right)
-\gamma_t\left(E-E_f+\frac12g_f\gamma_s\right)\right|^2,\label{Brh2X}
\eea
where the denominator $D(E)$ is given by Eq.~(\ref{Den}). The coefficients ${\cal B}_0$, ${\cal B}_1$, and ${\cal B}_2$ only define the overall normalisation of the distributions.

In paper \cite{Hanhart:2011jz} examples of the line shapes are given corresponding to all four Cases (i)-(iv). For definiteness, the parameters of the problem were fixed as 
\be
\mu=966.5~\mbox{MeV},\quad\Delta=8.1~\mbox{MeV},\quad g_f=0.25,
\label{set}
\ee
and the values of $\gamma_s$, $\gamma_t$, and $E_f$ are listed in Table~\ref{t1}. For each case above, the Flatt{\'e} parameter $E_f$ was tuned to provide a bound state with $E_B=0.5$~MeV.

\begin{table}[t]
\begin{center}
\begin{tabular}{|c|c|c|c|c|c|}
\hline
&$\vphantom{\int_0^1}\gamma_s$, MeV&$\gamma_t$, MeV&$E_f$, MeV&Line&Z\\
\hline
(i)&$\pm\infty$&$\pm\infty$&-10.47&solid&0.30\\
\hline
(ii)&-30&$\pm\infty$&-3.22&dashed&0.85\\
\hline
(iii)&$\pm\infty$&-30&-7.77&dashed-dotted&0.19\\
\hline
(iv)&-30&-30&-2.97&dotted&0.67\\
\hline
\end{tabular}
\end{center}
\caption{The scattering length parameters, the Flatt{\'e} parameter $E_f$, and the $Z$-factor for the bound-state with
$E_B=0.5$~MeV for Cases (i)-(iv).} \label{t1}
\end{table}

In Figs.~\ref{brqFig}-\ref{brh2Fig} we plot the line shapes for the $X$ production through the quark component and both mesonic components for all four above Cases (i)-(iv). For each individual curve 
in Figs.~\ref{brqFig}-\ref{brh2Fig} the integral over the near-threshold region (chosen to be from 0 to 10~MeV) 
equals to unity (in the units of MeV$^{-1}$) which fixes the overall normalisation factors ${\cal B}_0$, ${\cal B}_1$, and ${\cal B}_2$.

As seen from the examples above, the interplay of various degrees of freedom gives rise to highly complicated resonance line shape. If
experimental data display such phenomena one concludes that the resonance is generated by complicated nontrivial dynamics. The 
reverse is, generally speaking, not correct: due to the interference between production mechanisms the resulting near-threshold line
shape could be smooth enough even in the presence of the competing quark and meson dynamics. This, however, requires a suitable fine-tuning of the production parameters that seems very implausible.

\section{A microscopic quark model for the $\X$}

As was already mentioned in the Introduction, the $\bar{c}c$ assignment for the $\X$ seems to be ruled out by its mass: the state is too low to be a $2^3P_1$ charmonium. However, a possibility exists 
that the strong coupling to the $D^{(*)}$ pairs could distort the bare $\bar{c}c$ spectrum. In this section we address the question of whether or not the existing quark models realise such 
a possibility. To this end, the general formalism presented in Subsection~\ref{general} is employed.

As a prerequisite, one is to calculate transition form factors which describe the coupling of the $D$-meson pairs to the bare $\bar{c}c$ charmonium. This
requires a model for the light-quark pair creation. The simplest model of such kind is the so-called $^3P_0$ model suggested many years ago in 
Ref.~\cite{Micu:1968mk}. It assumes that the light-quark pair is created uniformly in space with the vacuum quantum numbers $0^{++}$, that is, this is a $^3P_0$ pair, and this gives the name to
the model. Applications of this purely phenomenological model have a long history --- see, for example, Refs.~\cite{LeYaouanc:1972vsx,Busetto:1982qz,Kokoski:1985is} as well as more recent 
developments in 
Refs.~\cite{Barnes:1996ff,Barnes:2005pb}.

More sophisticated models for the pair-creation operator also exist which construct the current-current interaction due to the confining force and one-gluon exchange. Examples of such calculations 
can 
be found in the Cornell model \cite{Eichten:1978tg} which assumes that confinement has a Lorentz vector nature. The model used in Ref.~\cite{Ackleh:1996yt} assumes that the confining interaction is 
a scalar one. 
Possible mechanisms of strong decays were studied in the framework of the Field Correlator Method (FCM) (see, for example, the review
\cite{DiGiacomo:2000irz}) and an effective $^3P_0$ operator for the pair creation emerged from 
this study, with the coupling computed in terms of the FCM parameters \cite{Simonov:2002uf}. For the purposes of the present review it is sufficient to stick to the generic version of the $^3P_0$ 
model to calculate the 
transition form factors. Further details of the approach can be found in Ref.~\cite{Kalashnikova:2005ui}.

It is assumed in the model that the pair-creation Hamiltonian for a given quark flavour $q$ is a nonrelativistic reduction of the expression
\be 
H_q=g_q\int d^3x {\bar \psi_q} \psi_q,
\label{hcr} 
\ee 
so that the spatial part of the amplitude of the decay $A\to B+C$ in the
centre-of-mass frame of the initial meson $A$ is given by the operator
\be 
{\hat f}(\vep)=\int d^3k \phi_A(\vek-\vep){\hat O}(\vek)\phi_B^*(\vek-r_q\vep) \phi_C^*(\vek-r_q\vep), \quad
r_q=\frac{m_q}{m_q+m_c}, 
\label{amplitude} 
\ee
where $\phi_A$ is the wave function of the initial meson in momentum space, $\phi_B$ and $\phi_C$ are those
of the final mesons $B$ and $C$, $\vep=\vep_B=-\vep_C$ while $m_c$ and $m_q$ are the charmed and light quark mass, respectively.
The $^3P_0$ pair creation operator is taken in the form
\be 
{\hat O}(\vek)=-2\gamma({\bm \sigma}\cdot\vek),\quad \gamma=g_q/2m_q,
\ee
and the matrix ${\bm \sigma}$ acts on the spin variables of the light quark and antiquark (the heavy quarks are treated as spectators). 
To arrive at the vertex $f_i(\vep)$ for a particular mesonic channel $i$, which enters equations from Subsection~\ref{general}, one is to evaluate the matrix element of the operator ${\hat f}(\vep)$ 
between the spin wave functions of the initial and final state.  

The behaviour of the form factors (\ref{amplitude}) is defined by the scales of the wave functions involved which, in turn, are defined by the quark model. In the  standard nonrelativistic potential 
model the 
Hamiltonian reads
\be 
H_0=\frac{p^2}{m_c}+V(r)+C,
\quad V(r)=\sigma r-\frac{4}{3}\frac{\alpha_s}{r},
\label{potential}
\ee 
where $\sigma$ is the string tension, $\alpha_s$ is the strong coupling constant, and the constant $C$ defines an overall shift of the spectrum. The
Hamiltonian (\ref{potential}) should be supplied by the Fermi-Breit-type relativistic corrections including the spin-spin, spin-orbit and tensor force
which cause splittings in the $^{2S+1}L_J$ multiplets. In the leading approximation these splitting should be calculated as perturbations averaged over the eigenfunctions of the Hamiltonian 
(\ref{potential}). The same interaction $V(r)$ is used in the calculations of the spectra and wave functions of the $D$ mesons.

In the calculations of Ref.~\cite{Kalashnikova:2005ui} the coupled-channel scheme of Subsection~\ref{general} was employed, with the direct interaction in the mesonic channels neglected. The decay 
channels participating in the coupled-channel scheme were chosen to be $D \bar D$, $D \bar D^*$, $D^* \bar D^*$, $D_s \bar D_s$, $D_s\bar D_s^*$, and $D_s^* \bar D_s^*$, and mass difference between
the charged and neutral charmed mesons was not taken into account. The parameters of the underlying quark model and the pair-creation strength $\gamma$ were chosen to reproduce, with a reasonable 
accuracy, the masses of the $1S$, $1P$, and $2S$ charmonium as well as the mass and the width of the $^3D_1$ $\psi(3770)$
state.

The masses of the charmonia lying below the open-charm threshold are given by the poles of the $t$-matrix (see Eqs.~(\ref{t00sol})-(\ref{Gsol})),
that is, they come as solutions of the equation
\be
M-M_0+{\cal G}(M)=0,\quad {\cal G}(M)=\sum_i\int f_i^2(\veq)S_i(\veq)d^3q,
\ee
where $M$ is the physical mass and $M_0$ is the bare mass evaluated in the potential model (\ref{potential}) with the Fermi-Breit corrections taken into account. As to the above-threshold state
$\psi(3770)$, its ``visible'' mass $M_R$ was defined from the equation 
\be
M_R-M_0+{\rm Re}~{\cal G}(M_R)=0,
\label{visiblemass}
\ee
and its ``visible'' width was computed as
\be
\Gamma=2\Re~{\rm Im}~{\cal G}(M_R),\quad\Re=\left(\left.1+\frac{\partial ~{\rm Re}~ {\cal G}(M)}{\partial M}\right|_{M=M_R}\right)^{-1}.
\label{visiblewidth}
\ee

Hadronic shifts (the difference between the bare mass $M_0$ and the physical mass $M$) for lower-lying charmonia were calculated to be of
the order of 200~MeV, and $Z$-factors vary from $\approx 0.9$, for the $1S$ states, to $\approx 0.75$-0.8, for $2S$ ones. In other words, although the coupled-channel effects are substantial, the 
resulting spectra are not changed beyond recognition. 

The situation with the $2P$ levels promises more as $2P$ charmonia are expected to populate the mass range of 3.90-4.00 GeV where more charmed meson channels start to open, and some of these channels 
are $S$-wave ones.

\begin{figure*}[t]
\begin{center}
\begin{tabular}{ccc}
\epsfig{file=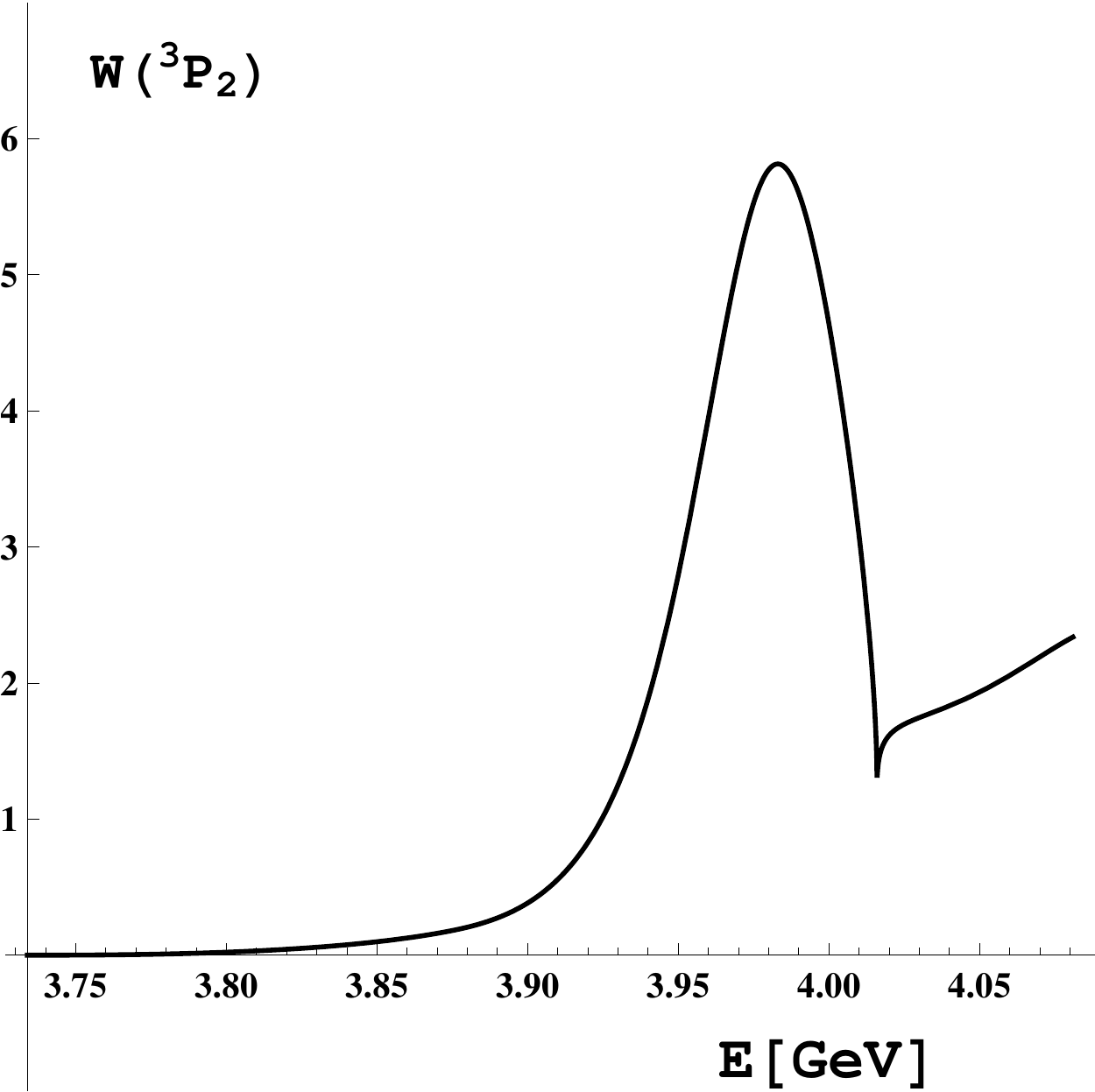, width=0.43\textwidth}&\hspace*{0.03\textwidth}&\epsfig{file=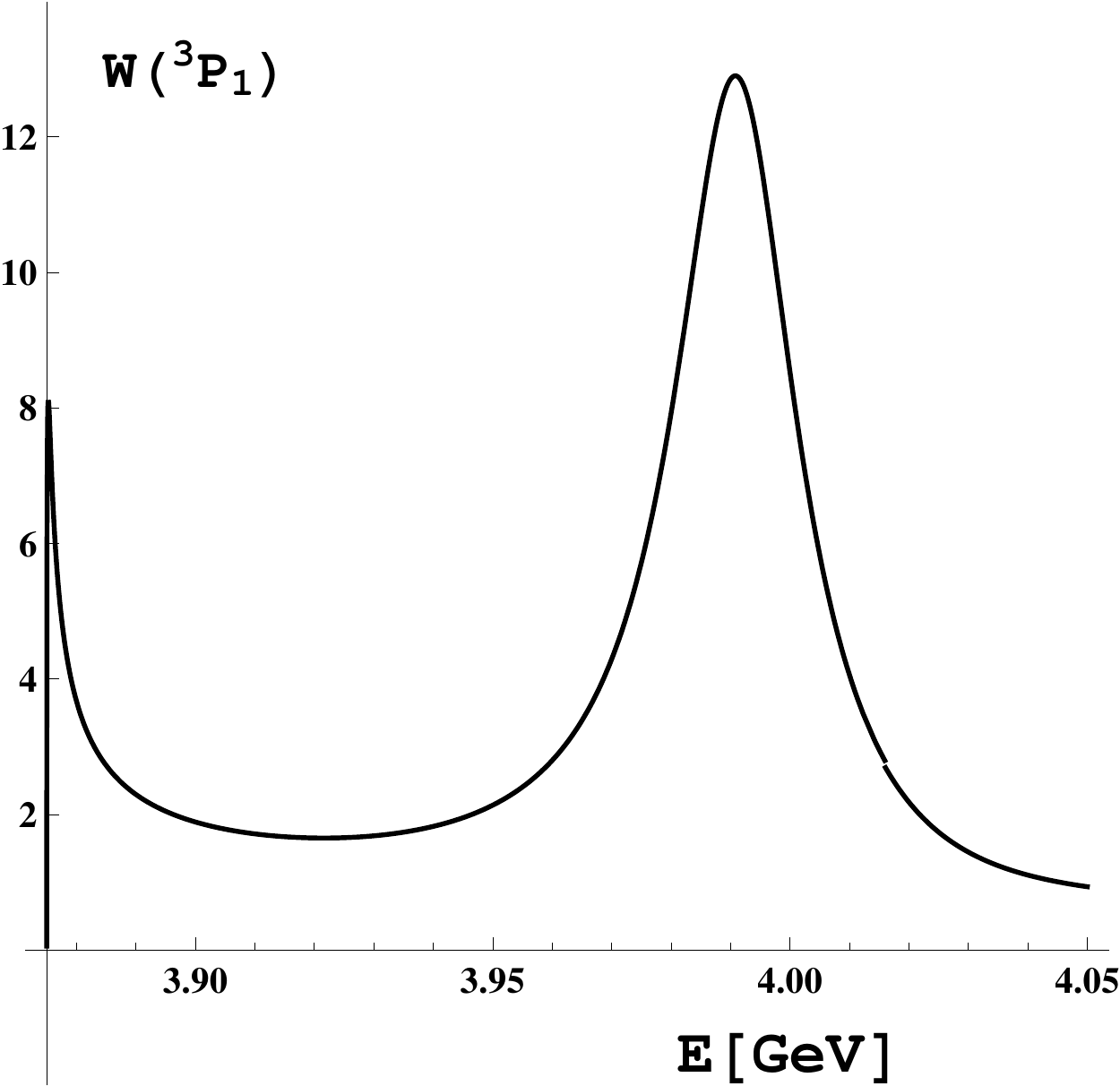, width=0.43\textwidth}\\
\epsfig{file=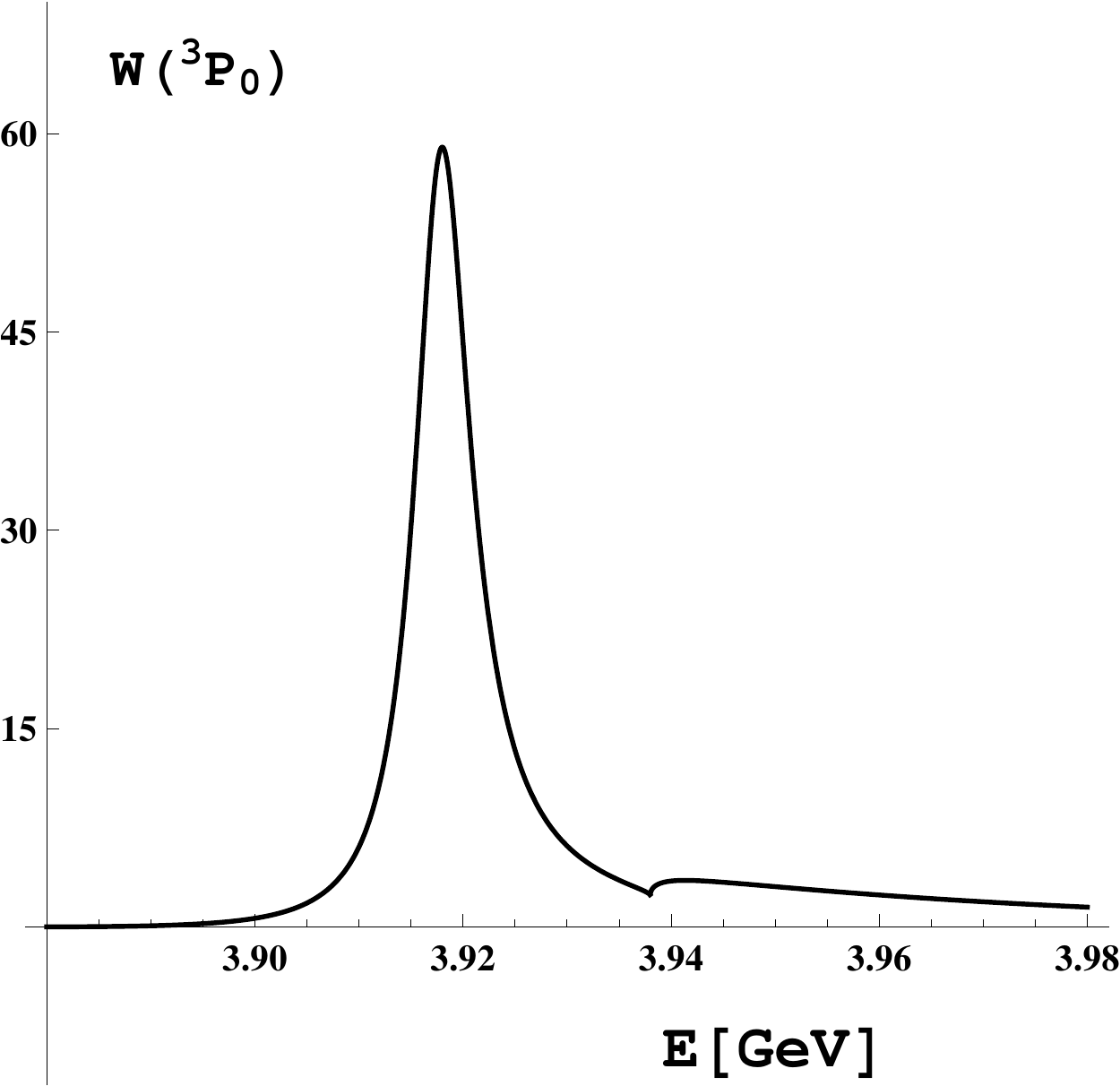, width=0.43\textwidth}&&\epsfig{file=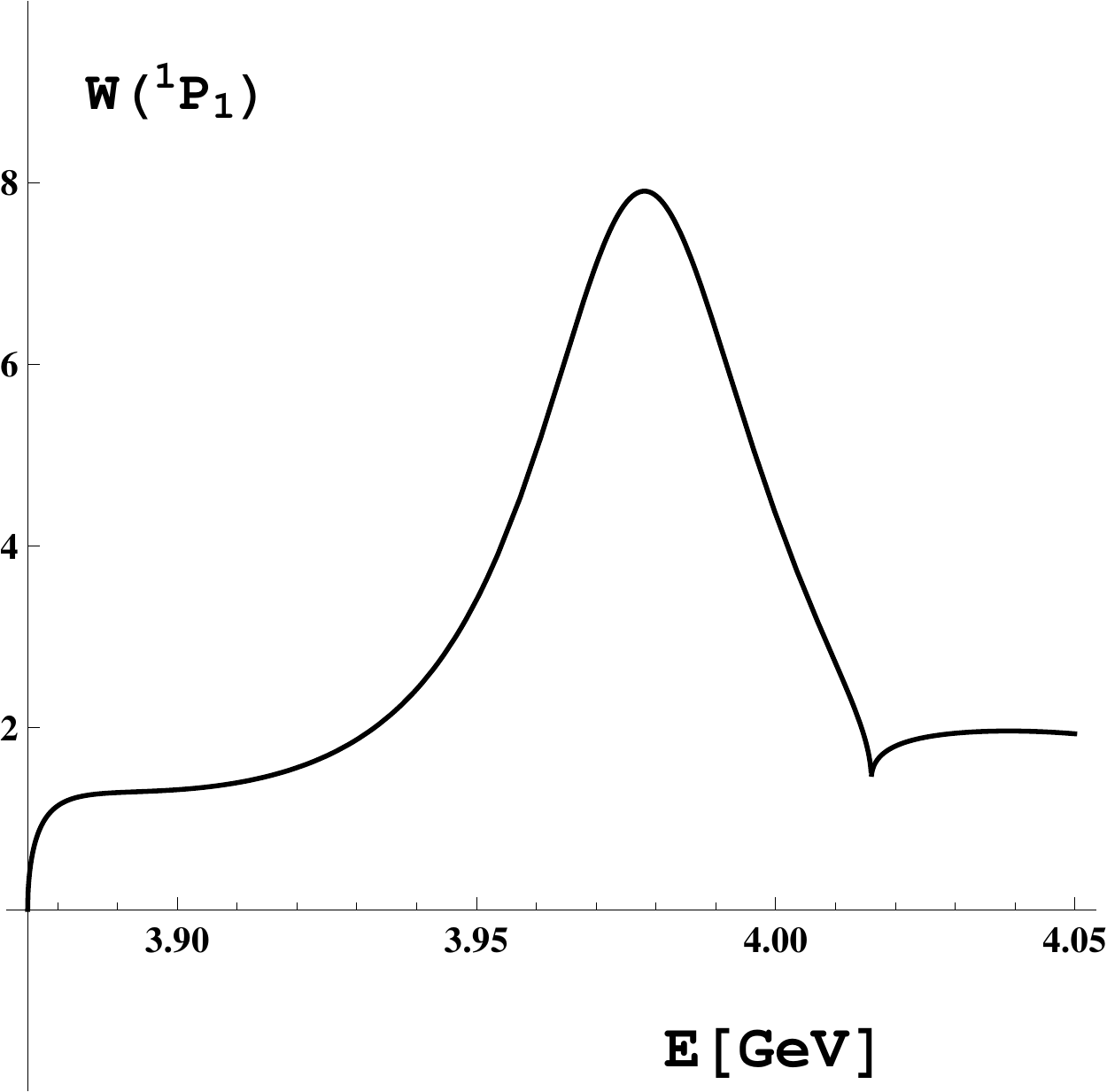, width=0.43\textwidth}\\
\end{tabular}
\caption{The spectral density of the bare $2P$ states in the units of GeV$^{-1}$. Adapted from Ref.~\cite{Kalashnikova:2005ui}.}\label{w2p}
\end{center}
\end{figure*}

The visible masses and widths of the $2P$ states were also calculated with the help of Eqs.~(\ref{visiblemass}) and (\ref{visiblewidth}) --- the results
are given in Table \ref{2P}. At first glance, nothing dramatic has happened due to $S$-wave thresholds. Indeed, similarly to the $S$-wave charmonia, the bare $P$-wave states suffer from hadronic 
shifts, but the latter are not too large, about 200 MeV, and the states acquire rather moderate finite widths.
To reveal the role of the $S$-wave thresholds one is to study the spectral density of the bare states which is depicted in Fig.~\ref{w2p}. Then one
observes, in addition to clean and rather narrow Breit-Wigner resonances in all considered channels, a near-threshold peak in the
$1^{++}$ channel (and only in this channel!) rising against a flat background.\footnote{There could be an interesting possibility, discussed in
Ref.~\cite{Kalashnikova:2005ui}, for the exotic line shape in the $0^{++}$ channel, if the bare $2^3P_0$ mass is shifted about 30
MeV upwards (which is not excluded by the quark model as the uncertainty in the spin-orbit splitting is large for a scalar). In
such a scenario, the physical $0^{++}$ state is landed at about 3.94 GeV, that is, just at the $D_s \bar D_s$ threshold. Because of this
threshold proximity the line shape is severely non-Breit-Wigner one, and a large admixture of the $D_s \bar D_s$ molecule is expected
in the wave function of the state.}

\begin{table}[t]
\begin{center}
\begin{tabular}{|c|c|c|c|}
\hline
$J^{PC}$&Bare mass&Physical mass&Width\\
\hline
$1^{+-}$&4200&3980&50\\
\hline
$2^{++}$&4230&3990&68\\
\hline
$1^{++}$&4180&3990&27\\
\hline
$0^{++}$&4108&3918&7\\
\hline
\end{tabular}
\end{center}
\caption{The masses and widths (in the units of MeV) of the $2P$ charmonium states evaluated in Ref.~\cite{Kalashnikova:2005ui} in the framework of a potential quark model.} \label{2P}
\end{table}

The elastic $D \bar D^*$ scattering length in the $1^{++}$ channel appears to be negative and large, 
\be 
a_{D \bar D^*}=-8~\mbox{fm},
\label{aoneplusplus} 
\ee 
signalling the presence of a virtual state very close to the $D \bar D^*$ threshold (with the energy
just 0.32 MeV relative to it). Slightly decreasing the value of the bare mass and/or increasing the value of the pair-creation strength one can
shift the virtual state closer to the threshold and, eventually, move the pole to the physical sheet thus forming a bound state.
Clearly, this near-threshold peak is to be identified with the $\X$. It should be noted that in a more sophisticated version of the
coupled-channel approach \cite{Danilkin:2009hr,Danilkin:2010cc}, where the mass difference between the neutral and charged $D \bar D^*$ thresholds was taken into account, the $\X$ peak persists, and 
resides at the neutral $D^0 \bar D^{*0}$ threshold.

Thus, the coupling to the mesonic channels generates not only a resonance but, in addition, a bound/virtual state very close to the threshold. In fact, in the presented dynamical scheme, the 
$X(3872)$ pole 
is an example of the so-called CC (coupled- channel) pole (see Ref.~\cite{Badalian:1981xj}): in the strong coupling limit an extra pole may come from infinity and move to the 
near-threshold region.

A natural question arises as to what extent the result just arrived at can be regarded as reliable? It was obtained in a naive nonrelativistic model whose relevance for light quarks is 
questionable. Applicability of the nonrelativistic coupled-channel formalism described in Subsection~\ref{general} and used here depends on whether or not the motion of the $D$-meson pairs can be 
treated as nonrelativistic in the considered energy range. Evaluation of the transition vertices $f_i(\vep)$ is an essentially relativistic problem, however the gross features of the vertices 
(\ref{amplitude}) are quite general. Firstly, at small $p$'s, the vertex function behaves as $p^{2l+1}$ where $l$ is the angular momentum in the $D$-meson system. Secondly, it decreases at large 
$p$'s, and its fall off is defined by the overlap of the initial- and final-state wave function. The first feature above defines the analytic properties 
of the self-energy $\cal G(M)$ while the large-momentum behaviour of the $f_i(\vep)$ is responsible, together with the effective coupling constant, for the hadronic shift of the bare state (we note 
here that 
the value of the effective coupling constant can be extracted from the experimental width of the state $\psi(3770)$). We expect, therefore, that 
the suggested coupled-channel scheme for the charmonium levels will not change qualitatively if more rigorous approaches are used, and resort to the $^3P_0$ model as to a simple illustration of a 
possible
scenario of the $\X$ binding.

Another question to be asked is why it is not possible to generate CC-states in the other $S$-wave $D^{(*)} \bar D^{(*)}$ channels? One could naively assume that the heavy-quark limit requires this. 
Indeed, 
the heavy quark limit implies that the initial states are degenerate in mass within a given $\{NL\}$-multiplet (here $N$ is the radial quantum number and $L$ is the quark-antiquark orbital angular 
momentum), 
and have the same wave functions. The final two-meson states exhibit the same degeneracy. The so-called loop theorems \cite{Barnes:2007xu} demonstrate that, in the heavy quark limit, 
the total strong open-flavour widths are the same within a given 
$\{NL\}$-multiplet. As a consequence, the hadronic
shift for a bare $\bar{Q}Q$ state is the same for all members of this multiplet.

Thus, in accordance with the loop theorems \cite{Barnes:2007xu}, the existence of the $1^{++}$ $D \bar D^*$ molecule would imply the existence of other molecules with the mesonic content given by 
the $S$-wave spin-recoupling coefficients of the $P$-wave levels of charmonium\footnote{These relations do not depend on the
pair-creation model and only assume that the spin of the heavy quark pair is conserved in the decay.}
\begin{eqnarray}
&&0^{++}\rightarrow \frac{\sqrt{3}}{2}D \bar D+\frac{1}{2}D^* \bar D^*,\nonumber\\
&&1^{++} \rightarrow \frac{1}{\sqrt{2}}(D \bar D^* +\bar D D^*),\nonumber\\[-3mm]
\label{spinrecoupling}\\[-3mm]
&&1^{+-} \rightarrow -\frac{1}{2}(D \bar D^* -\bar D D^*)+\frac{1}{\sqrt{2}}D^*\bar D^*,\nonumber\\
&&2^{++} \rightarrow D^* \bar D^*.\nonumber
\end{eqnarray}

Meanwhile, in actuality, the loop theorems are violated by the spin-dependent interaction both for the initial ($\bar{c}c$) and final
($D^{(*)}\bar{D}^{(*)}$) state. As the generation of a CC state is essentially a threshold phenomenon, this circumstance prevents generation of the $\X$
siblings in the coupled-channel dynamical scheme.  

Indeed, the effective coupling constants of a bare level of charmonium with different mesonic channels are proportional to the corresponding spin-recoupling coefficients which give the only difference 
between them in the heavy quark limit. As it is seen from the relations (\ref{spinrecoupling}), the full $S$-wave strength of the
$^3P_1$ decay is concentrated in the single $D \bar D^*$ channel while for the $^1P_1$ and $^3P_0$ decays it is diluted between two
channels with different thresholds. As a result, in the latter case, the coupling constant appears to be not strong enough to support a CC state: for
example, the scattering length in the $^1P_1$ channel, $|a|\approx 1$ fm, is much smaller than the value quoted in 
Eq.~(\ref{aoneplusplus}) for the
$^3P_1$ channel. The problem with the $^3P_2$ channel is of a different nature: the $D^* \bar D^*$ threshold with the mass of about
4.016 GeV lies too high that excludes a $2^{++}$ CC state. We conclude, therefore, that the $1^{++}$ molecular state is unique within the
coupled-channel model.\footnote{This statement applies only to the coupled-channel model of this chapter which lacks a direct interaction between the mesons. In most works on the molecular states 
in the spectrum of charmonium and bottomonium (see, for example, 
Refs.~\cite{AlFiky:2005jd,Voloshin:2011qa,Mehen:2011yh,Ohkoda:2011vj,Valderrama:2012jv,HidalgoDuque:2012pq,Nieves:2012tt,Guo:2013sya,Baru:2017gwo}), 
the problem of binding is solved by employing a short-range direct interaction between the heavy $D^{(*)}$ or $B^{(*)}$ mesons which is responsible for the formation of a near-threshold pole, 
typically residing on the first (physical) Riemann sheet, that is, a bound state. Description of the $\X$ in the framework of such an approach constitutes the subject of Subsection~\ref{Xthreedyn} of 
the present review; discussion of the spin partners of the $\X$ can be found in 
Refs.~\cite{Nieves:2012tt,Guo:2013sya,Hidalgo-Duque:2013pva,Albaladejo:2015dsa,Baru:2016iwj}.}

\section{Nature of the $\X$ from data}
\label{dataanalysis}

In this section, we discuss the nature of the charmonium-like state $\X$. Given the proximity of this resonance to a threshold, a considerable admixture of the $D\bar{D}^*$ component in its wave 
function is inevitable. Therefore, a realistic model for the description of the $\X$ developed in the previous chapter assumes that its wave function contains both a short-range (which can be 
associated with a genuine $\bar{c}c$ charmonium) and a long-range (defined by the molecular component $D\bar{D}^*$) part, and proximity to the threshold implies that the admixture of the latter 
component is not small.

Let us start from a qualitative phenomenological consideration which supports such a picture. To begin with, the isospin breaking which follows from the approximately equal probabilities of the $\X$
decays to 
the final states $\rho J/\psi$ and $\omega J/\psi$ finds a natural explanation in the framework of the molecular model as coming from the mass difference 
$\Delta\approx 8$~MeV between the charged and neutral thresholds $D\bar{D}^*$ --- see Refs.~\cite{Suzuki:2005ha,Gamermann:2007fi}.
Indeed, in the molecular picture, transition from the $\X$ to any final state can proceed through the intermediate $D\bar{D}^*$ loops and, for different isospin states, contributions from the loops 
with the
charged ($L_c$) and neutral ($L_0$) $D$ mesons come either in a sum or in a difference. 
In particular, for the ratio of the effective coupling constants of the $\X$ to the final states $\rho J/\psi$ and $\omega J/\psi$ one can find
\be
\frac{g_{X\to\rho J/\psi}}{g_{X\to\omega J/\psi}}\sim\left|\frac{L_0-L_c}{L_0+L_c}\right| \sim
\frac{\sqrt{m_D\Delta}}{\beta} \sim 0.1,
\label{Rmol}
\ee
where $m_D$ is the mass of the $D^{(*)}$ meson and $\beta\simeq 1$~GeV sets a typical scale of the real part of the loop defined by the range of force. If, in addition, utterly different phase 
spaces for the considered decays are taken into account then the relation (\ref{Rmol}) is sufficient to explain the experimental ratio (\ref{omega}).
For a pure $\bar{c}c$ state such a violation of the isospin symmetry would be very difficult to explain, if possible at all, since the ratio 
(\ref{Rmol}) for a genuine charmonium appears to be an order of magnitude smaller than for a molecule (see, for example, the discussion in 
Ref.~\cite{Hanhart:2011tn}).

On the other hand, as was already mentioned in the Introduction, the $\X$ is produced in $B$-meson decays with the branching fraction similar to those
for genuine charmonia (see, for example, relation (\ref{ccbarbr})), while this branching fraction was shown to be very small for a molecule \cite{Braaten:2004ai}. In a similar way, doubts were cast 
in 
Ref.~\cite{Bignamini:2009sk} on the molecular assignment based on the fact of a copious production of the $\X$ at high energies at hadron colliders. Recently this point has become a subject of vivid 
discussions --- see, for example, Refs.~\cite{Albaladejo:2017blx,Esposito:2017qef}. While the details of the $\X$ prompt production in the molecular picture are still to be clarified, 
the inclusion of a non-molecular seed could certainly be helpful in resolving this imbroglio.

We now turn to a quantitative description of the $\X$ based on the existing ex\-pe\-ri\-men\-tal data. In particular, in this section we will present a simple but realistic 
pa\-ra\-me\-te\-ri\-sa\-ti\-on for the line shape of the $\X$ \cite{Kalashnikova:2009gt,Kalashnikova:2010zz} and demonstrate that it allows one to make certain conclusions on the nature of the state 
$\X$.

The $\X$ production line shapes in the $B$-meson decays do not exhibit irregularities in the near-threshold region. Therefore, these line shapes can be described by simple Flatt{\'e} formulae 
\cite{Flatte:1976xu} which can be derived naturally in the coupled channel model which includes a molecular component and a bare charmonium.
More explicitly, the physical system contains a bare pole which will be associated with the genuine $\bar{c}c$ charmonium $\chi_{c1}'$ (a radially excited axial-vector $\bar{c}c$ state), two elastic 
channels (the charged and neutral $D\bar{D}^*$ channels split by $\Delta=M(D^+D^{*-})-M(D^0\bar{D}^{*0})\approx 8~\mbox{MeV}$) and a set of inelastic channels. It should be noted that the 
existing experimental data can be well described if the inelastic channels are taken into account through an effective width $\varGamma(E)$ \cite{Kalashnikova:2009gt},
\be
\varGamma(E)=\varGamma_{\pi^+\pi^-J/\psi}(E)+
\varGamma_{\pi^+\pi^-\pi^0J/\psi}(E)+\varGamma_0, \ee \be
\varGamma_{\pi^+\pi^-J/\psi}(E)=f_{\rho}\int^{M-m_{J/\psi}}_{2m_{\pi}}
\frac{dm}{2\pi}\frac{q(m)\varGamma_{\rho}}{(m-m_{\rho})^2+\varGamma_{\rho}^2/4},
\label{rhowidth}
\ee
\be
\varGamma_{\pi^+\pi^-\pi^0J/\psi}(E)=f_{\omega}\int^{M-m_{J/\psi}}_{3m_{\pi}}
\frac{dm}{2\pi}\frac{q(m)\varGamma_{\omega}}{(m-m_{\omega})^2+\varGamma_{\omega}^2/4 },
\label{omegawidth}
\ee
where
\be
q(m)=\sqrt{\frac{(M^2-(m+m_{J/\psi})^2)(M^2-(m-m_{J/\psi})^2)}{4M^2}},\quad M=M(D^0 \bar D^{*0})+E,
\ee
$f_{\rho}$ and $f_{\omega}$ are the coupling constants, and $m_\rho$, $m_\omega$ and $\varGamma_\rho$,
$\varGamma_\omega$ are the masses and the widths of the $\rho$ and $\omega$ meson, respectively \cite{Tanabashi:2018oca}.

The quantity $\varGamma_0$ is an additional inelasticity taken into account effectively. It is the short-range charmonium $\chi'_{c1}$
component of the $\X$ which is responsible for this contribution into the $\X$ decay modes. These are annihilation modes into light
hadrons, $\chi_{c1}(3515) \pi\pi$ mode and the radiative decay modes.\footnote{Assumption on the radiative modes being insensitive to the
long-range component of the $\X$ wave function is confirmed by calculations discussed below in Section~\ref{Xraddecs}.} In other
words, these are all the inelastic decay modes of the $\X$ but the $\rho J/\psi$ and $\omega J/\psi$ which are included explicitly because of
their substantial energy dependence --- see the expressions (\ref{rhowidth}) and (\ref{omegawidth}). Then, taking the value of the total width
for the ground-state axial-vector charmonium $\chi_{c1}(3515)$, equal to 0.84 $\pm$ 0.04~MeV \cite{Tanabashi:2018oca}, and an
estimate for the total width of the $\chi'_{c1}$ made in the framework of a quark model and yielding the value of 1.72~MeV
\cite{Barnes:2003vb}, it is quite natural to assume for the parameter $\varGamma_0$ to take a value of about 1-2 MeV. Correlating
this parameter with the upper limit (\ref{widthX}) of the  $\X$ 
width, we fix this parameter as
\be
\varGamma_0=1~\mbox{MeV}.
\label{constr3}
\ee

Therefore, for the denominator in the distribution for the $\X$ we arrive at
\be
D(E)=\left\{
\begin{array}{ll}
\ds
E-E_f-\frac12\Bigl(g_1\kappa_1+g_2\kappa_2\Bigr)+\frac{i}{2}\varGamma(E),&E<0\\
[3mm] \ds
E-E_f-\frac{1}{2}g_2\kappa_2+\frac{i}{2}\Bigl(g_1k_1+\varGamma(E)\Bigr),&0<E<\Delta\\[3mm]
\ds E-E_f+\frac{i}{2}\Bigl(g_1k_1+g_2k_2+\varGamma(E)\Bigr),
&E>\Delta,
\end{array}
\right.
\label{D}
\ee
where
$$
k_1=\sqrt{2\mu_1 E},\quad\kappa_1=\sqrt{-2\mu_1 E},\quad
k_2=\sqrt{2\mu_2(E-\Delta)},\quad\kappa_2=\sqrt{2\mu_2(\Delta-E)},
$$
and $\mu_1$ and $\mu_2$ are the reduced masses in the elastic channels. We set $g_1=g_2=g$, which is a good approximation in the isospin symmetry limit.

In accordance with the discussion in the beginning of this section, we assume that the $\X$ is produced in $B$-meson decays via the $\bar{c}c$ component. The short-range dynamics of the weak $B\to K$ 
transition is absorbed by the coefficient ${\cal B}$, and we estimate this quantity to be
\be
{\cal B}\equiv \Br(B \to K\chi'_{c1})=(3 \div 6) \cdot 10^{-4},
\label{constr1}
\ee
as it is reasonable to expect that the $\chi'_{c1}$ is produced in the $B \to K$ decays with the rate comparable to other similar charmonia. 
It should be noted that there exists a quark model prediction $Br(B \to K \chi'_{c1})=2 \times 10^{-4}$~\cite{Meng:2005er}, however, the model used is known to underestimate the production rate for 
the 
$\chi_{c1}$ by more than two times.

Then it is straightforward to find the following expressions for the differential production rates in the inelastic channels:
\be
\frac{d\Br(B \to K \pi^+\pi^- J/\psi)}{dE}={\cal B}\frac{1}{2\pi}\frac{\varGamma_{\pi^+\pi^-J/\psi}(E)}{|D(E)|^2},
\label{rhopsi}
\ee
\be
\frac{d\Br(B \to K \pi^+\pi^-\pi^0 J/\psi)}{dE}={\cal B}\frac{1}{2\pi}\frac{\varGamma_{\pi^+\pi^-\pi^0J/\psi}(E)}{|D(E)|^2}.
\label{omegapsi}
\ee

As to the elastic channels, we consider $D^{*0}$ meson to be unstable with the decay branching fractions to the final states 
$D^0\pi^0$ and $D^0\gamma$ with the following relative probabilities \cite{Tanabashi:2018oca}:
\be
\Br(D^{*0}\to D^0 \pi^0)=(61.9 \pm 2.9)\%,
\label{062}
\ee
\be
\Br(D^{*0} \to D^0 \gamma)=(38.1 \pm 2.9)\%,
\label{038}
\ee
so that for the final state $D^0 \bar D^0 \gamma$ one finds
\be
\frac{d\Br(B\to K D^0 \bar D^0 \gamma)}{dE}=0.38{\cal B}\frac{1}{2\pi}\frac{gk_1}{|D(E)|^2}.
\label{DDgamma}
\ee

In case of the $D^0 \bar D^0 \pi^0$ final state, we also take into account the signal-background interference which is defined by two parameters, the constant strength $c$ and the relative phase 
$\phi$, so that the differential production rate in this channel reads
\be
\frac{d\Br(B \to K D^0 \bar D^{0}\pi^0)}{dE}= 0.62\frac{k_1}{2\pi}\left[\left({\rm
Re}\frac{\sqrt{g \cal B}}{D(E)} +c\cos\phi\right)^2 +\left({\rm
Im}\frac{\sqrt{g \cal B}}{D(E)}+c\sin\phi\right)^2\right].
\label{DD}
\ee

The last ingredient needed for the data analysis is the spectral density, and it is easy to find the explicit expression for it,
\be
w(E)=\frac{1}{2\pi|D(E)|^2}(gk_1\Theta(E)+gk_2\Theta(E-\Delta)+\varGamma(E)).
\label{wE2}
\ee
In accordance with the approach outlined in Subsection~\ref{weinbergformula}, the integral from the quantity (\ref{wE2}), 
\be
W=\int_{E_{\rm min}}^{E_{\rm max}} w(E)dE,
\label{Wint}
\ee
taken over the near-threshold region defines the admixture of the genuine charmonium $\chi'_{c1}$ in the $\X$ wave function. For the problem under consideration, a natural definition of 
the near-threshold region is the energy interval $-10\;\mbox{MeV}\lesssim E\lesssim 10\;\mbox{MeV}$ which covers the neutral three-body threshold at $E_{D^0 \bar D^0 \pi^0}\approx -7$~MeV as well as 
the charged two-body one at $E_{D^+\bar{D}^{*-}}\equiv\Delta\approx 8$~MeV.

Now it is easy to estimate the total branching fraction for the $\X$ production as
\be
\Br(B \to K X)={\cal B}W<2.6 \cdot 10^{-4},
\label{constr2}
\ee
where the experimental bound (\ref{babartotal}) was used.\footnote{In papers \cite{Kalashnikova:2009gt,Kalashnikova:2010zz} a different, relevant at that time, upper bound $\Br(B \to K 
X)<3.2 \cdot 10^{-4}$ was used which had been established by the BABAR Collaboration \cite{Aubert:2006qua} in 2006. However, the results of the analysis performed are insensitive to a particular 
value of this bound.}

Finally, Table~\ref{widths} contains estimates for the widths of the radiative decays of the $\chi'_{c1}$ obtained in various quark models. They will be used in what follows. 

\begin{table}[t]
\begin{center}
\begin{tabular}{|l|c|c|c|}
\hline
&$\vphantom{\int_0^1}\varGamma(\chi_{c1}'\to \Jp)$ [keV] &$\varGamma(\chi_{c1}'\to \p)$ [keV] &$R$ (see definition (\ref{RLHCb}))\\
\hline
Ref.~\cite{Barnes:2003vb}&11&64&5.8\\
\hline
Ref.~\cite{Barnes:2005pb}&70&180&2.6\\
\hline
Ref.~\cite{Badalian:2012jz}&50-70&50-60&$0.8\pm 0.2$\\
\hline
Ref.~\cite{Badalian:2015dha}&30.8-42.7&70.5-73.2&1.65-2.38\\
\hline
\end{tabular}
\end{center}
\caption{Some typical estimates for the radiative decay width of the charmonium $\chi_{c1}'$ obtained in various quark models.}
\label{widths}
\end{table}

Therefore, the model is defined by the set of 8 parameters, 
\be
\varGamma_0,\quad g,\quad E_f,\quad f_\rho,\quad f_\omega,\quad {\cal B}, \quad\phi,\quad c, \label{params} 
\ee 
and entitled to describe the $\X$ line shape in the energy interval near the neutral $D\bar{D}^*$ threshold.

The data analysed in this section belong to the Belle Collaboration (see Ref.~\cite{Adachi:2008sua} for the $D^0\bar{D}^0\pi^0$ channel and 
Ref.~\cite{Adachi:2008te} for the $\pi^+\pi^-J/\psi$ channel) and the BABAR Collaboration (see Ref.~\cite{Aubert:2007rva} for the combined $D^0 \bar
D^{*0}$ mode which is a sum of the $D^0\bar{D}^0\pi^0$ and $D^0\bar{D}^0\gamma$ channels, and Ref.~\cite{Aubert:2008gu} for the
$\pi^+\pi^- J/\psi$ channel). Parameters of the experimental distributions are listed in Table~\ref{setuptab}.

\begin{table}
\begin{center}
\begin{tabular}{|c|c|c|c|c|c|}
\hline
Collaboration&Channel&$N_{\rm tot}$&$\Br_{\rm tot}$&$E_{\rm bin}$, MeV&$\sigma$, MeV\\
\hline
Belle&$\pi^+\pi^- J/\psi$&131&$8.3\cdot 10^{-6}$&2.5&3\\
Belle&$D^0\bar{D}^0\pi^0$&48.3&$0.73\cdot
10^{-4}$&2.0&$0.172\sqrt{m-M(D\bar{D}^*)}$\\
BABAR&$\pi^+\pi^- J/\psi$&93.4&$8.4\cdot 10^{-6}$&5&4.38\\
BABAR&$D\bar{D}^*$&33.1&$1.67\cdot 10^{-4}$&2.0&1\\
\hline
\end{tabular}
\end{center}
\caption{Parameters of the experimental distributions from Belle \cite{Adachi:2008sua,Adachi:2008te} and BABAR \cite{Aubert:2008gu,Aubert:2007rva}.}
\label{setuptab}
\end{table}

To be directly compared with the experimental distributions, the theoretical differential production rates are to be convolved with the detector resolution function which has the form of a Gaussian 
with the parameter $\sigma$ quoted in Table~\ref{setuptab} for all data sets.\footnote{Since the BABAR resolution function in 
the $D^0 \bar D^{*0}$ channel takes a very complicated form and it is not available in the public domain, in the present analysis we use in the calculations
a Gaussian function with the parameter $\sigma=1$ MeV.} Then, the resulting differential production rates are to be converted to the number-of-events distributions by means of the relation
\be
N(E)=E_{\rm bin}\left(\frac{N_{\rm tot}}{\Br_{\rm tot}}\right)\frac{d\Br}{dE},
\label{noe}
\ee
where $N_{\rm tot}$ is the total number of the events, $\Br_{\rm tot}$ is the total branching, and $E_{\rm bin}$ is the bin size. 
All these parameters are listed in Table~\ref{setuptab}.

The procedure comprises a simultaneous fit for the data from one and the same collaboration on the $\pi^+\pi^-J/\psi$ and $D^0 \bar D^{*0}$ channels. 
For each data set two possibilities are considered, a bound state and a virtual level. These two situations are distinguishable by the sign of real part of the $D^0 \bar D^{*0}$ scattering length, 
for 
which the following formula can be easily obtained:
\be
a=-\frac{1}{\sqrt{2\mu_2\Delta}+(2E_f-i\varGamma(0))/g}.
\label{scatlenX}
\ee

\begin{table}[t]
\centering
\begin{adjustbox}{width=\columnwidth}
\begin{tabular}{|c|c|c|c|c|c|c|c|c|c|c|}
\hline
{\small Set}&$g$&$E_f$&$f_\rho\cdot 10^3$&$f_\omega\cdot 10^3$&
${\cal B}\cdot 10^4$&$\phi$, $^0$&$W$&${\cal B}W\cdot 10^4$&Re($a$)&$\varGamma(\gamma\psi')$\\
\hline
{\small Belle}$_{\rm v}$&0.3&-12.8&7.7&40.7&2.7&$180$&0.19&0.5&-5.0&$\sim 10^3$\\
{\small Belle}$_{\rm b}$&0.137&-12.3&0.47&2.71&4.3&$153$&0.43&1.9&3.5&60\\
{\small BABAR}$_{\rm v}$&0.225&-9.7&6.5&36.0&3.9&$113$&0.24&1.8&-4.9&800\\
{\small BABAR}$_{\rm b}$&0.080&-8.4&0.2&1.0&5.7&$0$&0.58&3.3&2.2&25\\
\hline
\end{tabular}
\end{adjustbox}
\caption{The sets of parameters for the Belle \cite{Adachi:2008te,Adachi:2008sua} and
BABAR \cite{Aubert:2008gu,Aubert:2007rva} data. The lower index labels a bound state (b) or a virtual level (v). Quantities $E_f$ and $\varGamma(\gamma\psi')$ are given in MeV, the scattering length 
$a$ is given in fm.}\label{setstab}
\end{table}

\begin{figure}
\begin{center}
\begin{tabular}{|c|c|c|c|}
\hline
${\rm Belle}_{\rm v}$&${\rm Belle}_{\rm b}$&${\rm BABAR}_{\rm v}$&${\rm BABAR}_{\rm b}$\\
\hline
\epsfig{file=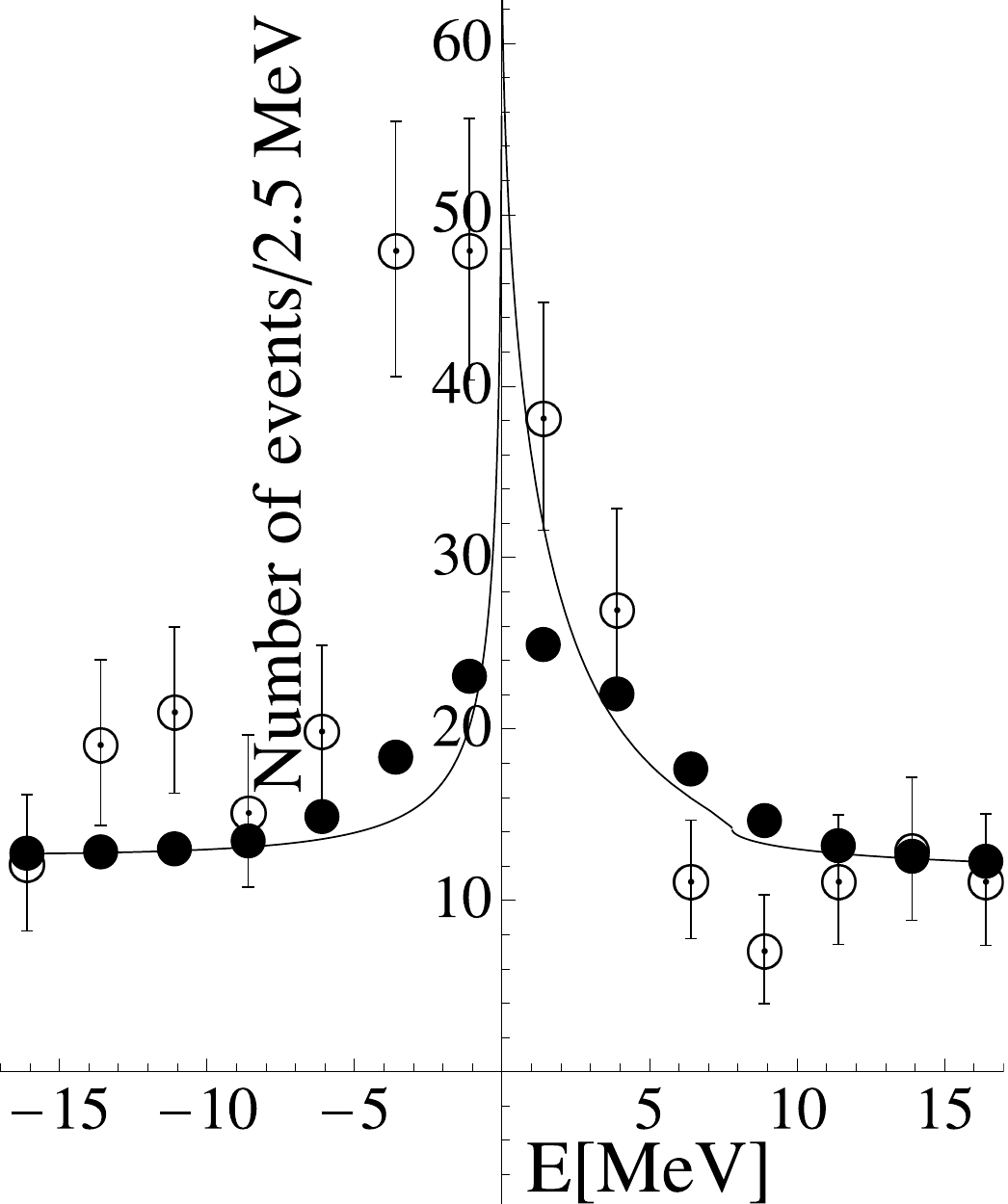,width=0.22\textwidth}&
\epsfig{file=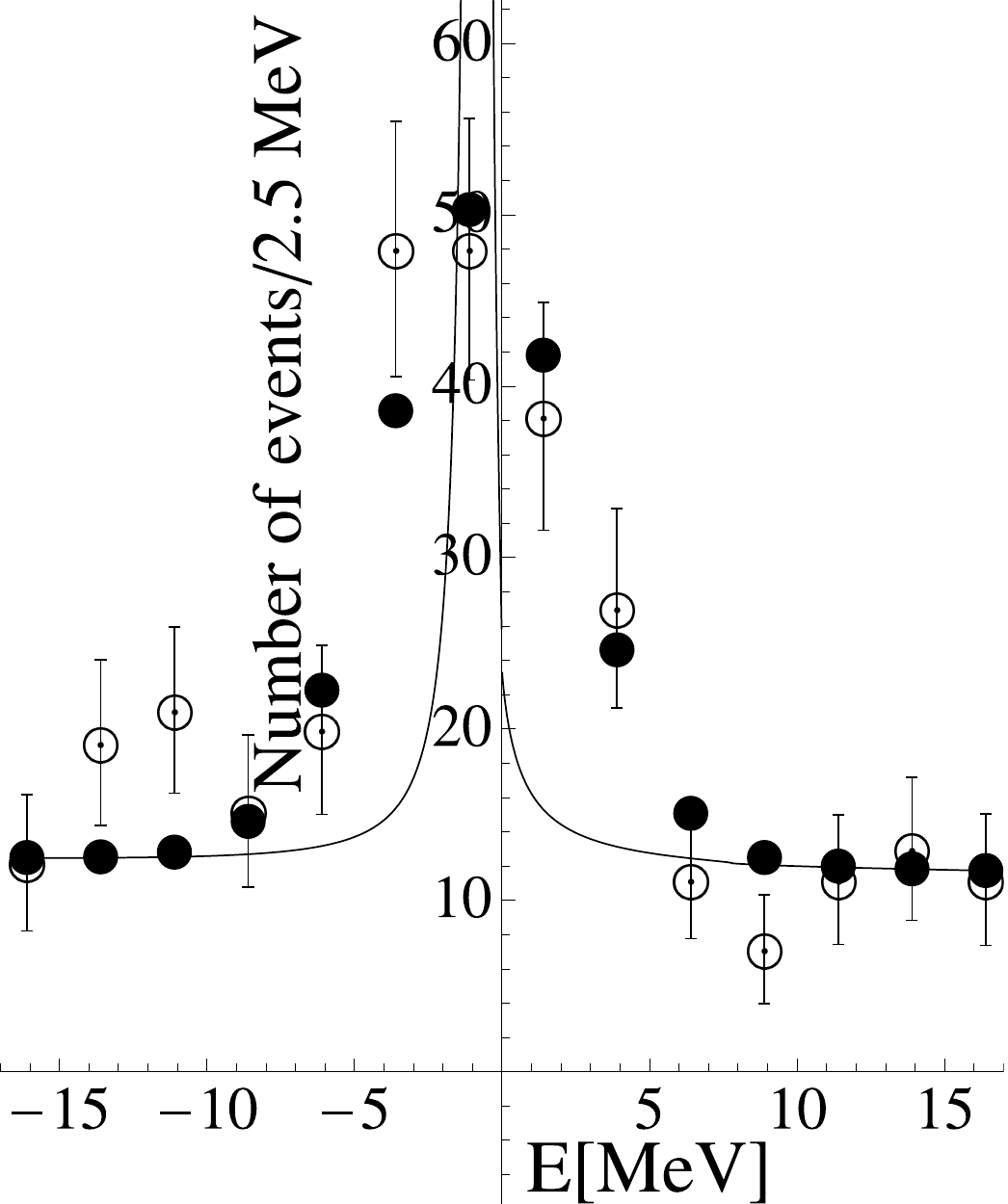,width=0.22\textwidth}&
\epsfig{file=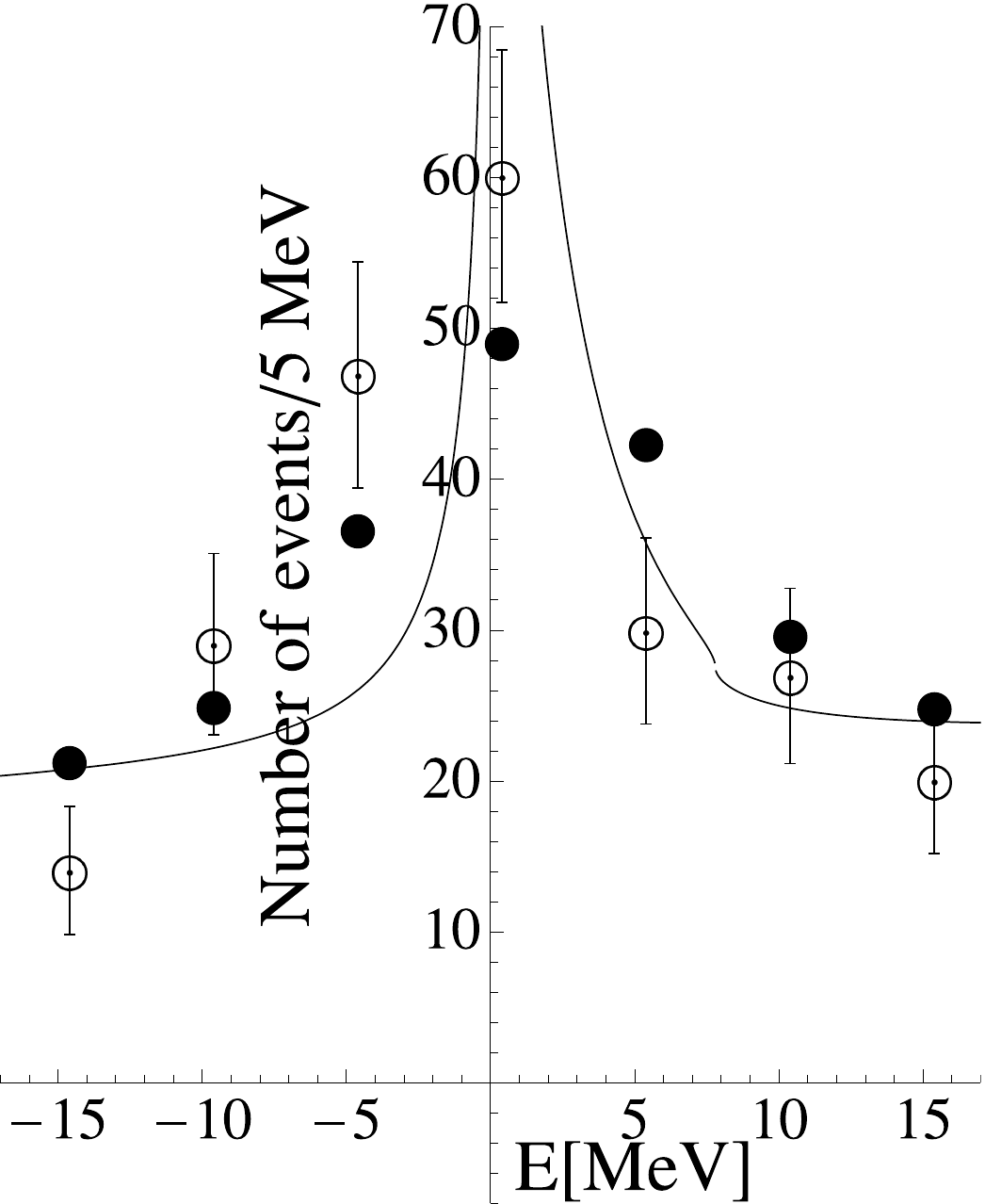,width=0.22\textwidth}&
\epsfig{file=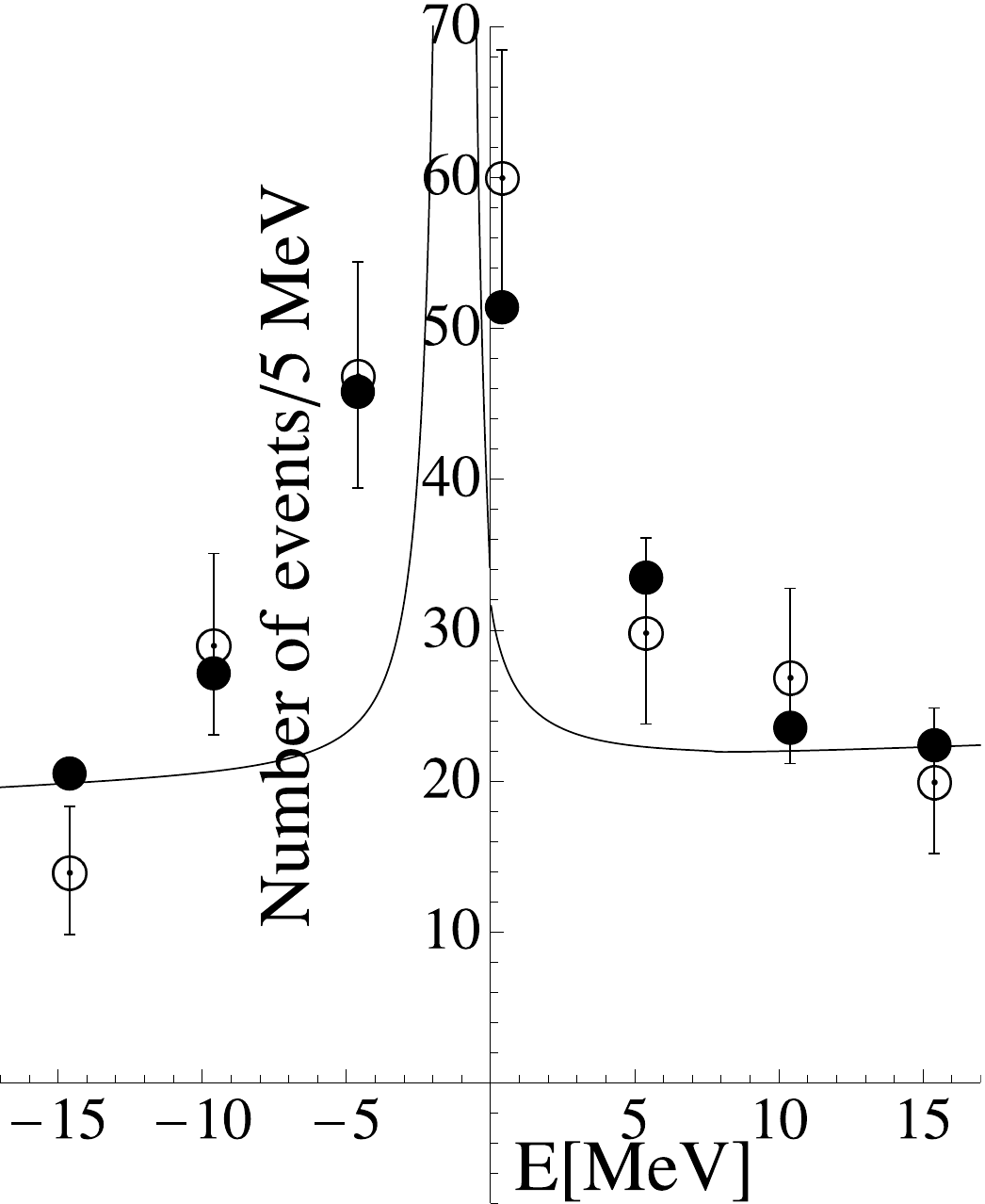,width=0.22\textwidth}\\
\hline
\epsfig{file=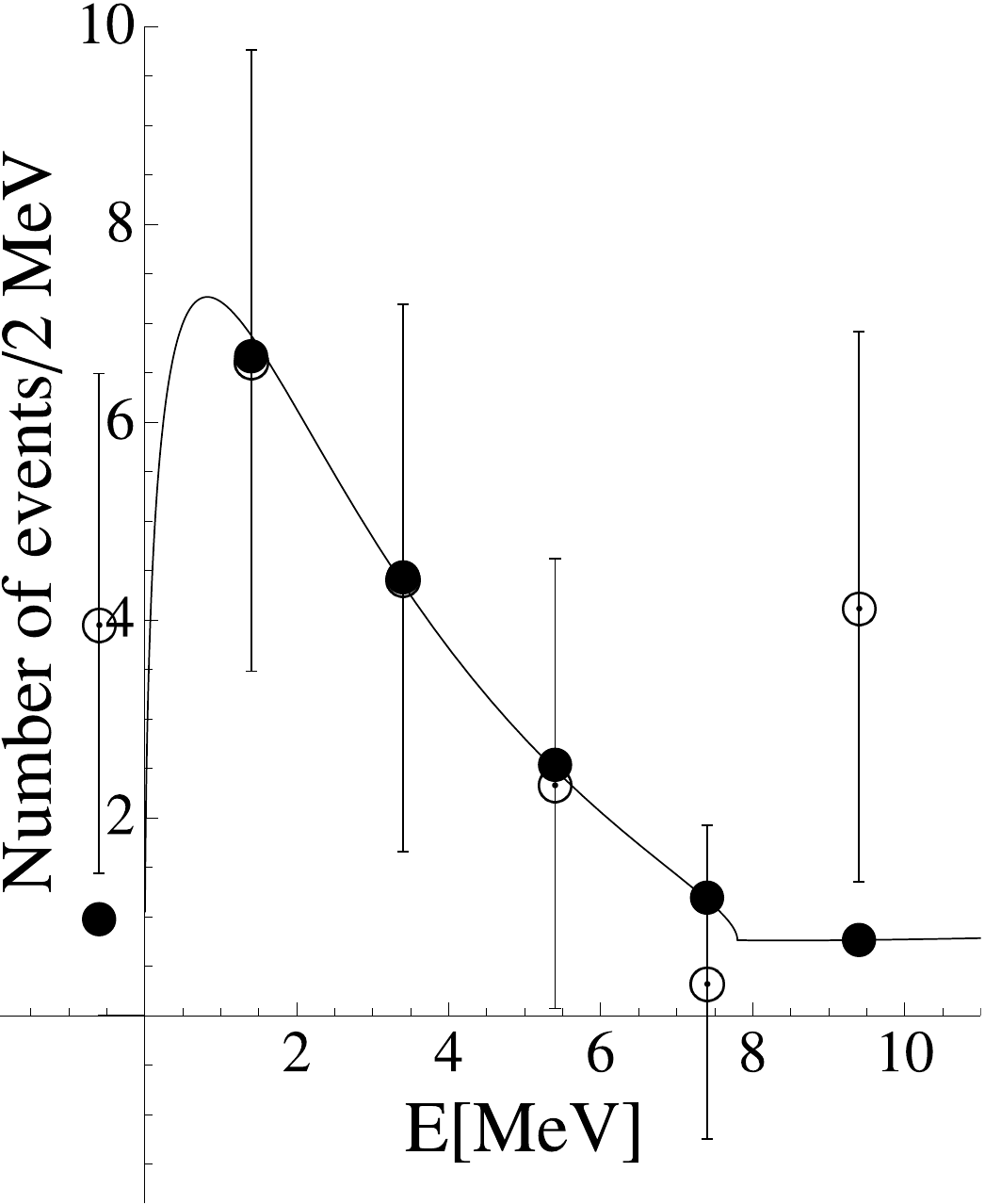,width=0.22\textwidth}&
\epsfig{file=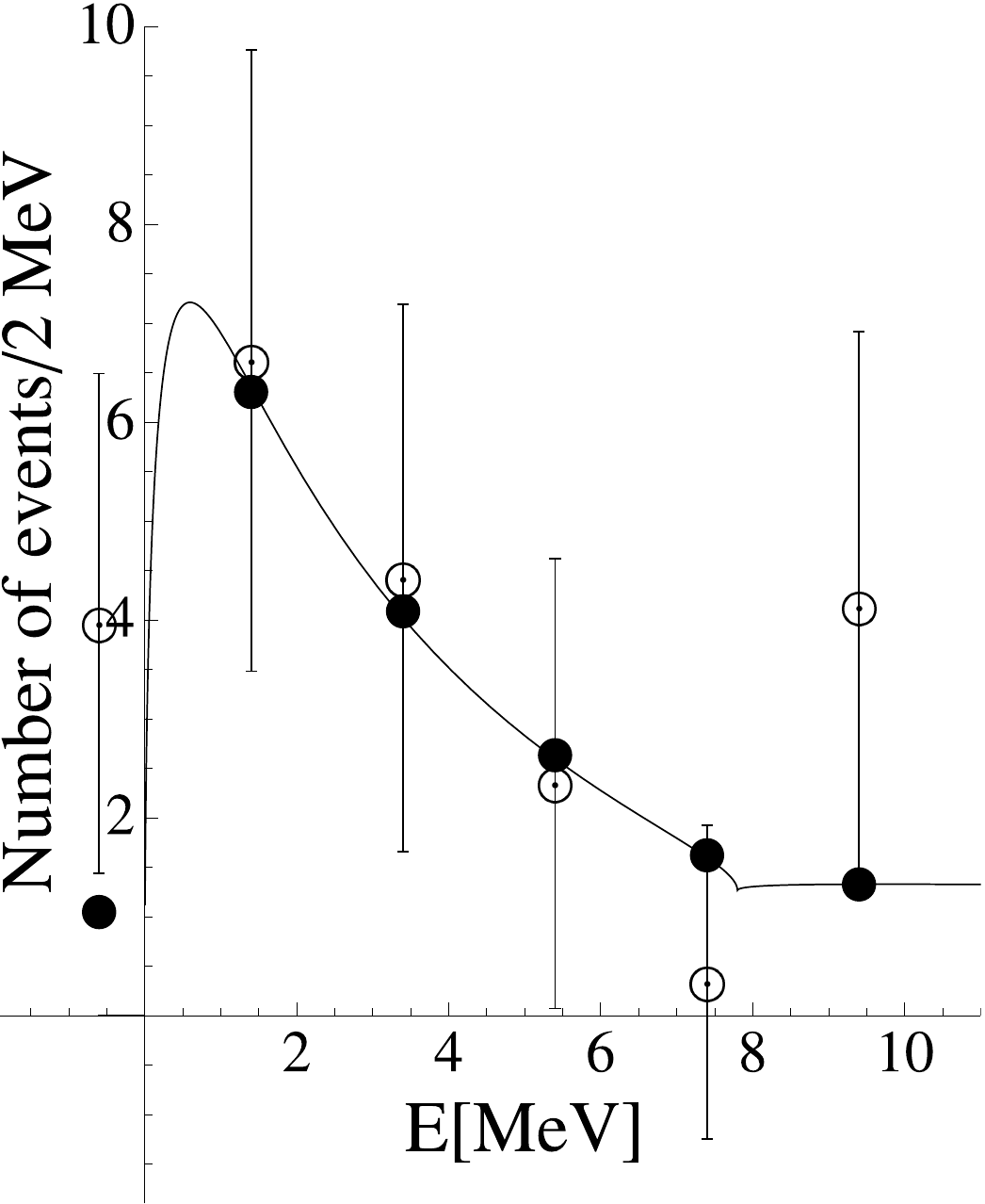,width=0.22\textwidth}&
\epsfig{file=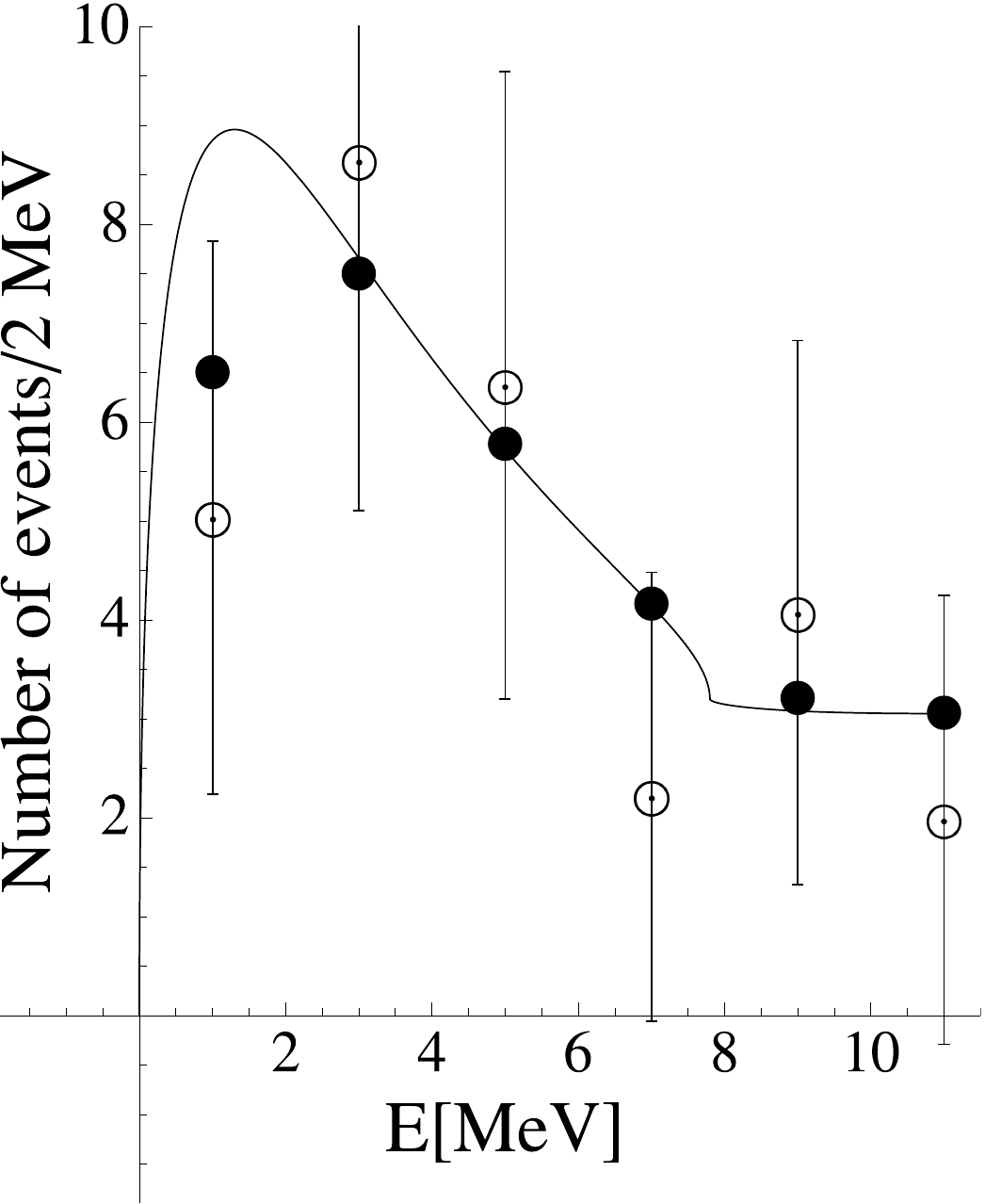,width=0.22\textwidth}&
\epsfig{file=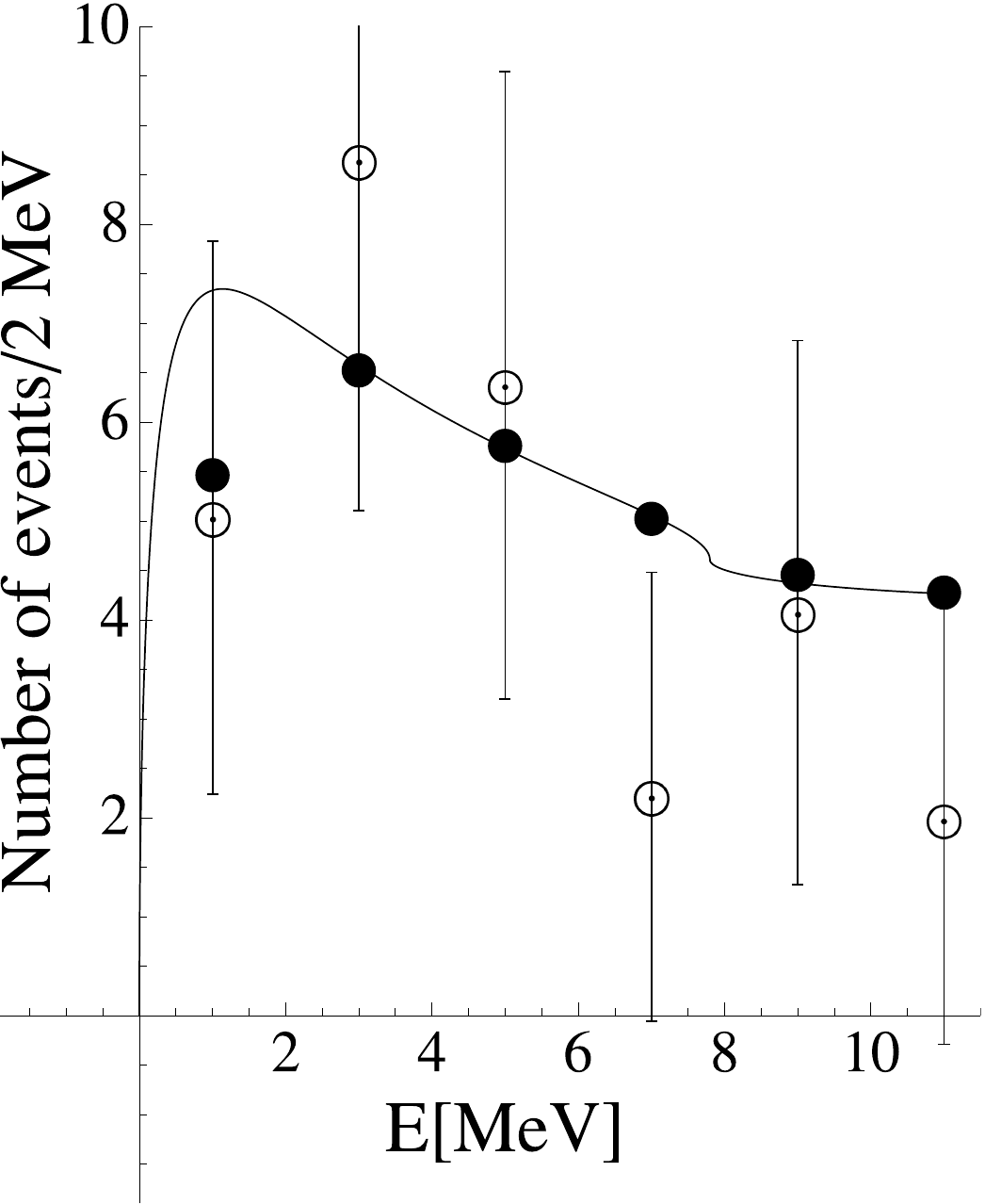,width=0.22\textwidth}\\
\hline
\end{tabular}
\end{center}
\caption{Description of the experimental data on the $\pi^+\pi^-J/\psi$ (upper panel) and $D^0 \bar D^{*0}$ (lower panel) with the parameters given in Table~\ref{setstab}. Adapted 
from Ref.~\cite{Kalashnikova:2009gt}.}\label{fig:Xfit}
\end{figure}

Several sets of parameters are listed in Table
~\ref{setstab}, which yield best description of the data. The name of each set indicates the collaboration whose data are processed and the type of the description: bound state (b) or virtual level 
(v).\footnote{In view of the inelasticity present in the system, the $\X$ pole in the complex energy plane does not lie on the real axis and it is somewhat shifted to the complex plane, so that, 
strictly speaking, it describes a resonance rather than a bound or virtual state. Nevertheless, as was explained in Section~\ref{sec:def}, it proves convenient to stick to the notions of the 
bound and virtual levels defined for the Riemann 
surface for the elastic channels only.}
The scattering length is calculated according to Eq.~(\ref{scatlenX}) and, in order to estimate the width $\varGamma(\gamma\psi')$, the experimental relation $\Br(\gamma\psi')\simeq \Br(\pipi J/\psi)$ 
is used (see Eq.~(\ref{raddecprime})).

The theoretical curves for the resonance line shapes for all four cases are shown in Fig.~\ref{fig:Xfit}, together with the corresponding integrals over bins (black dots), in comparison with the 
experimental data (open circles with error bars) \cite{Kalashnikova:2009gt,Kalashnikova:2010zz}.

Description obtained allows one to draw certain conclusions on the nature of the $\X$. First, the Belle data unambiguously point to the $\X$ being a
bound state. Indeed, the description of the data under a virtual level assumption is of a considerably worse quality than that under a bound
state assumption: the virtual level corresponds to a threshold cusp in the inelastic channel which is not compatible with the $\pipi
J/\psi$ data. The BABAR data prefer to some extent the virtual state description, though the bound state assumption is quite acceptable, too. Besides, the estimates of the $\varGamma(\gamma\psi')$ 
decay width for the virtual level look unnaturally large in comparison with the theoretical results collected in Table~\ref{widths} while the estimates of the $\varGamma(\gamma\psi')$ width in the 
case of a bound state is in a qualitative agreement with the results from Table~\ref{widths}.

The value $W$ of the integral from the spectral density over the near-threshold region given in Table~\ref{setstab} for the sets ${\rm Belle}_{\rm v}$ and ${\rm BABAR}_{\rm v}$ indicates a rather 
small admixture of the $\chi_{c1}'$ charmonium in the wave
function of the $X$ while, for the sets ${\rm Belle}_{\rm b}$ and
${\rm BABAR}_{\rm b}$ corresponding to a bound state, this admixture amounts to about 50\%. 

In the case of the bound state it is
instructive to consider the quantity 
\be 
{\cal Z}=\int_{E_{\rm min}}^{E_{\rm max}} w_{\rm inel}(E)dE, 
\ee
with
\be
w_{\rm inel}(E)=\frac{1}{2\pi|D(E)|^2}\varGamma(E). 
\ee 
In the limit of a vanishing inelasticity, the spectral density below the $D^0 \bar D^{*0}$ threshold becomes proportional to a $\delta$-function,
\be 
w(E)\to Z\delta(E-E_{\rm bound}),\quad E<0, 
\ee 
with the coefficient $Z$ being nothing else but the $Z$-factor for a true bound state. Then the factor ${\cal Z}$ can be viewed as the $Z$-factor
of the $\X$, as a bound state, smeared due to the presence of the inelasticity, and it takes the value 
\be 
{\cal Z}=0.37
\ee 
for the Belle data set.

Thus we conclude that the $X$ is not a {\it bona fide} charmonium accidentally residing at the $D^0 \bar D^{*0}$ threshold. Had it been the case, the integral of the spectral density over the 
resonance region would have been unity while, for the found bound-state solutions, it does not exceed 50\%, and is very small for the virtual-state solutions. The $\X$ is rather a state generated 
dynamically by a strong coupling of the bare $\chi'_{c1}$ charmonium to the $D \bar D^*$ hadronic channel, with a large admixture of the
$D \bar D^*$ molecular component.

Now let us comment on the role of the parameter $\varGamma_0$. As was already discussed above, this parameter accounts for the contribution of the numerous decay modes of the charmonium 
$\chi_{c1}'$ 
and, for this reason, it was fixed based on the estimates of the width of this genuine charmonium. Consider the ratio of the elastic branching to the inelastic one 
\cite{Stapleton:2009ey,Hanhart:2007yq}. In the gedanken limit $\varGamma_0=0$, the inelastic modes are exhausted by the $\pipi J/\psi$ and $\pipipi J/\psi$ ones. As the measured branchings for these 
modes are approximately equal to each other (see Eq.~(\ref{omega})), it is enough to consider the ratio  
\be 
\frac{\Br(X \to D^0 \bar D^0 \pi^0)}{\Br(X \to \pi^+\pi^- J/\psi)}, 
\label{Sdpsi} 
\ee 
which can be estimated with the help of the relation 
\be 
\Br(B^+\to K^+ X)\,\Br(X\to \pi^+\pi^-J/\psi)=(7\div 10)\times 10^{-6}
\label{BrA} 
\ee 
and the value (\ref{BrB}) for the elastic branching \cite{Adachi:2008te}. One finds in such a way that the experimental ratio (\ref{Sdpsi}) is large and equals approximately 10-15. Let us 
reproduce this value for the case of the bound state. The differential rate for the inelastic channels is given by 
\be 
{\cal B}\frac{1}{2\pi}\frac{\varGamma(E)}{|D(E)|^2},
\label{inelrate} 
\ee 
and for $\varGamma_0=0$ the total width $\varGamma(E)$ is defined entirely by the modes $\rho J/\psi$ and $\omega J/\psi$ and as such is small; thus it is reasonable to take the limit $\varGamma(E)\to 
0$. In this limit and in the case of the bound state, the distribution (\ref{inelrate}) becomes proportional to a $\delta$-function and the denominator in the ratio (\ref{Sdpsi}) becomes a constant. 
It is easy to verify then that the ratio (\ref{Sdpsi}) is numerically small (see, for example, Ref.~\cite{Swanson:2003tb}). If the condition $\varGamma_0=0$ is relaxed and the inelastic modes are not 
exhausted by the $\pipi J/\psi$
and $\pipipi J/\psi$ ones, the ratio (\ref{Sdpsi}) is rewritten as 
\be 
\frac{\Br(X \to D^0 \bar D^0 \pi^0)}{\Br(X \to {\rm non}D^0 \bar D^0 \pi^0)}\sim 1, 
\label{Sdpsi2} 
\ee 
which is quite attainable. Thus, the experimental data are compatible with the assumption of the $\X$ being a bound state only if an extra non-zero
contributions to the total width exist which come from the decay modes of the charmonium $\chi_{c1}'$. This conclusion is fully in line with the analysis presented above.

On the contrary, for a virtual level, the ratio (\ref{Sdpsi}) is not restrictive even in the limit $\varGamma_0=0$. Indeed, for the virtual level the denominator of the distribution (\ref{inelrate}) 
does not vanish, so that in the limit $\varGamma(E)\to 0$ the denominator in Eq.~(\ref{Sdpsi}) could be arbitrary close to zero and, consequently, the ratio (\ref{Sdpsi}) could be large (for example, 
in Ref.~\cite{Hanhart:2007yq} this ratio was found to be $\approx$ 9.9 for a virtual level). The idea that including an extra inelastic width one can fit the data on the $\X$ both with the virtual 
and bound state was first put forward in Ref.~\cite{Bugg:2008wu}.

To conclude this section, let us comment on the width of the $D^*$ meson. In the formulae above the $D^*$ meson was assumed to be stable while, if a finite width of the $D^*$  is taken into account, 
the $D^0 \bar D^{*0} \to D^0 \bar D^0 \pi^0$ decay chain feeds the mass region below the nominal $D^0 \bar D^{*0}$ threshold, distorting in such a way the $D^0 \bar D^{*0}$ line shape. A refined 
treatment of such a finite width of a molecule constituent is presented in Ref.~\cite{Hanhart:2010wh} while here we estimate these effects using a simple ansatz suggested in 
Ref.~\cite{Nauenberg:1964zz} and
re-invented in Ref.~\cite{Braaten:2007dw,Stapleton:2009ey}. The recipe is to make a replacement in the expressions for
the $D^0 \bar D^{*0}$ momentum entering the formulae for the differential rates, 
\be 
\Theta (E)k_1(E) \to \sqrt{\mu_1}\sqrt{\sqrt{E^2+\varGamma_*^2/4}+E}, 
\label{keff} 
\ee
and 
\be 
\Theta (-E)\kappa_1(E) \to \sqrt{\mu_1}\sqrt{\sqrt{E^2+\varGamma_*^2/4}-E}, 
\label{kappaeff}
\ee 
where $\varGamma_*$ is the width of the $D^{*0}$ meson. It can be shown \cite{Hanhart:2010wh} that these formulae are valid if the resonance is well-separated from the three-body threshold (the $D^0 
\bar D^{*0} \pi^0$ threshold in this case).

\begin{figure}
\begin{center}
\begin{tabular}{ccc}
\epsfig{file=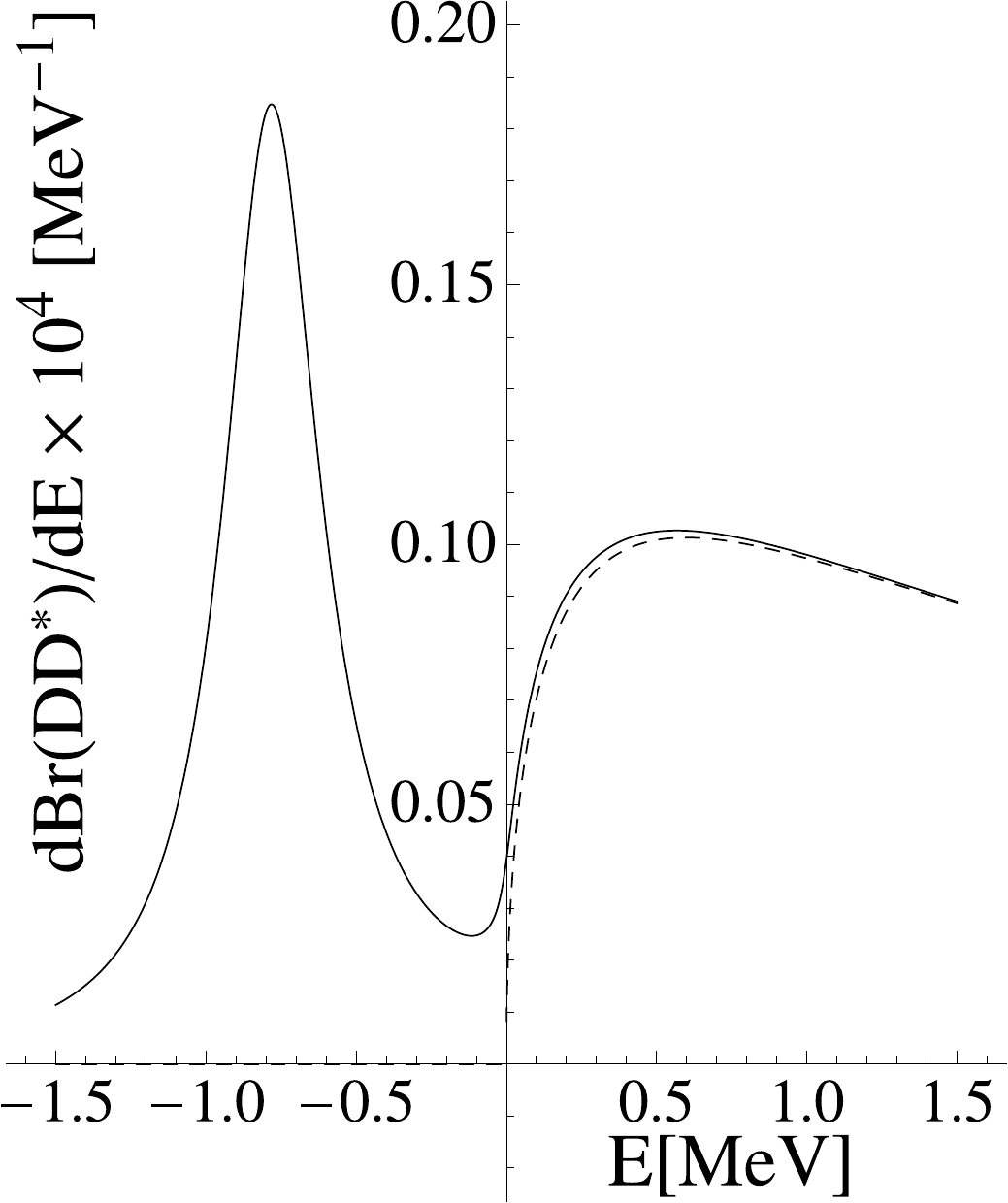,width=0.4\textwidth}&\hspace*{2cm}&
\epsfig{file=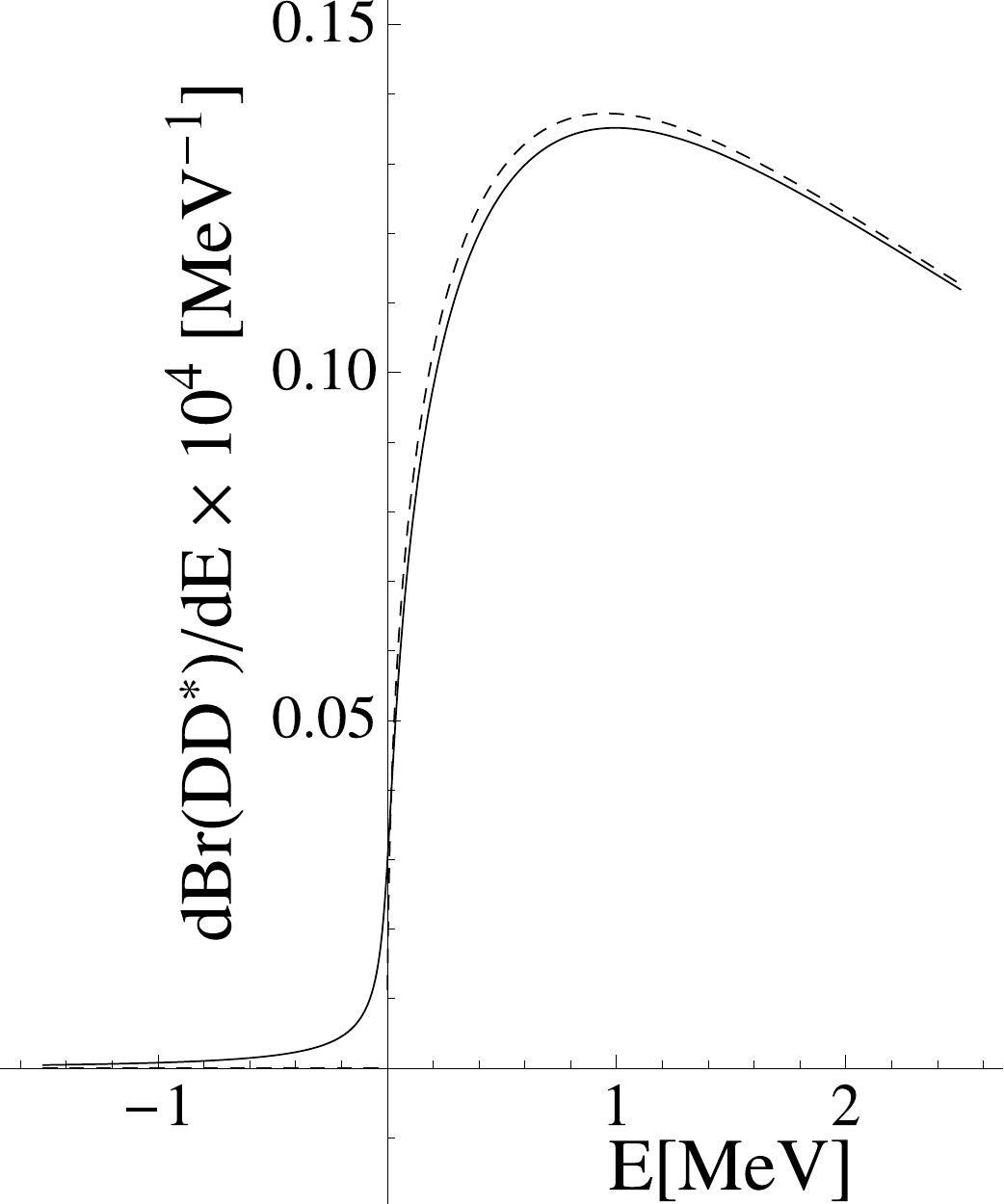,width=0.4\textwidth}\\
\end{tabular}
\end{center}
\caption{The differential rates for the $D^0\bar{D}^{*0}$ channel given by the formula with a finite width $\varGamma_*$ included (solid lines) and with a vanishing width (dashed lines) for 
the parameter sets ${\rm Belle}_{\rm b}$ (left) and ${\rm BABAR}_{\rm v}$ (right).}
\label{fw}
\end{figure}

To assess the role of the finite $D^{*0}$ width we evaluate the $D^0 \bar D^{*0}$ differential rates, with $\varGamma_*=63$ keV, for the set ${\rm Belle}_{\rm b}$ (the bound state scenario for the 
Belle data) and for the set ${\rm BABAR}_{\rm v}$ (the virtual state scenario for the BABAR data) and plot them in Fig.~\ref{fw} together with the
zero-width rates. As seen from Fig.~\ref{fw}, account for a small finite width of the $D^{*0}$ meson does not change the line shape in the case of the virtual state while, for the bound state, a 
below-threshold $D^0 \bar D^0 \pi^0$ peak is developed around the bound-state position.

\begin{figure}
\begin{center}
\epsfig{file=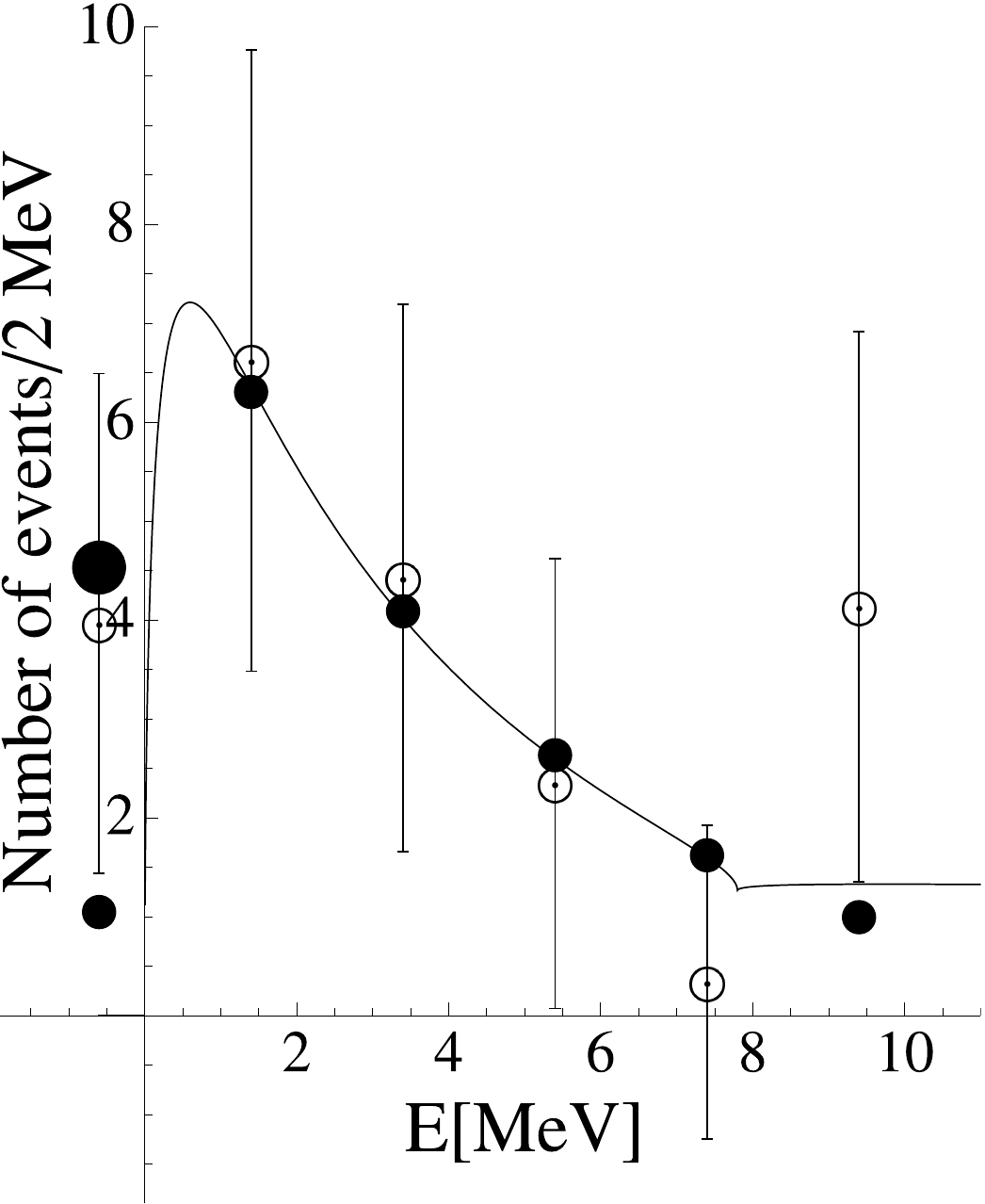,width=0.4\textwidth}
\end{center}
\caption{Theoretical description of the Belle data in the $D\bar{D}^*$ channel (parameter set  ${\rm Belle}_{\rm b}$). 
Open circles with error bars corrspond to the experimental data, 
filled circles are for the theoretical results; the big filled circle gives the result of the finite width of the $D^*$ meson taken into account in the formulae. Adapted 
from Ref.~\cite{Kalashnikova:2009gt}.}\label{finwidfig}
\end{figure}

One could suggest that, in order to distinguish between the bound- and virtual-state solution, one should look for a below-threshold peak in the $D^0\bar{D}^0\pi^0$ distribution. Indeed, it is 
clear from Fig.~\ref{fw} that the only effect expected is an increase of the number of events in the first near-threshold bin. For the Belle bound-state parameter set ${\rm Belle}_{\rm b}$, we have 
calculated the ratio $\tilde{N}_i/N_i$ of the number of events in the first ($i=1$) and second ($i=2$) bin, with ($\tilde{N}_i$) and
without ($N_i$) inclusion of the finite width,
\be
\tilde{N}_1/N_1=4.31,\quad \tilde{N}_2/N_2=1.01.
\ee
This is illustrated in Fig.~\ref{finwidfig}. The agreement with the experimental data is apparently improved, as the Belle bound-state solutions underestimate the number of events in the lowest bin 
only, and the number of events in higher bins is not affected by the finite-width effects.

Unfortunately, the present experimental situation does not allow one to identify the bound-state peak. First, the peak is very narrow, and the existing experimental resolution is too coarse to observe 
it. Besides, as was noted in Ref.~\cite{Stapleton:2009ey}, both the BABAR and Belle Collaboration assume that all the $D^0 \bar D^0 \pi^0$ events come from the $D^0 \bar D^{*0}$ distorting in such a 
way the kinematics of the below-threshold events and feeding artificially the above-threshold region at the expense of the below-threshold one.

Related to the question of the finite $D^{*0}$ width is the problem of the interference in the decay chains 
\be
\X
\genfrac{}{}{0pt}{}{\nearrow}{\searrow}
\raisebox{0mm}{$
\begin{array}{l}
\bar{D}^0D^{0*}\to\bar{D}^0[D^0\pi^0]\\[5mm]
D^0\bar{D}^{0*}\to D^0[\bar{D}^0\pi^0]
\end{array}
$}
\genfrac{}{}{0pt}{}{\searrow}{\nearrow}
D^0\bar{D}^0\pi^0.
\label{2ways}
\ee

According to the estimates made in Ref.~\cite{Voloshin:2003nt}, the interference effects could enhance the below-threshold $D^0 \bar D^0 \pi^0$ rate up to two times, however the effect is much more 
moderate above threshold
\cite{Hanhart:2010wh}. A proper account for the interference cannot be done in the over-simplified framework presented in this section but, as was demonstrated above, this effect lies far beyond the 
accuracy of the existing experimental data.

\section{$\X$ and pionic degrees of freedom}

\begin{figure}[t!]
\centerline{\epsfig{file=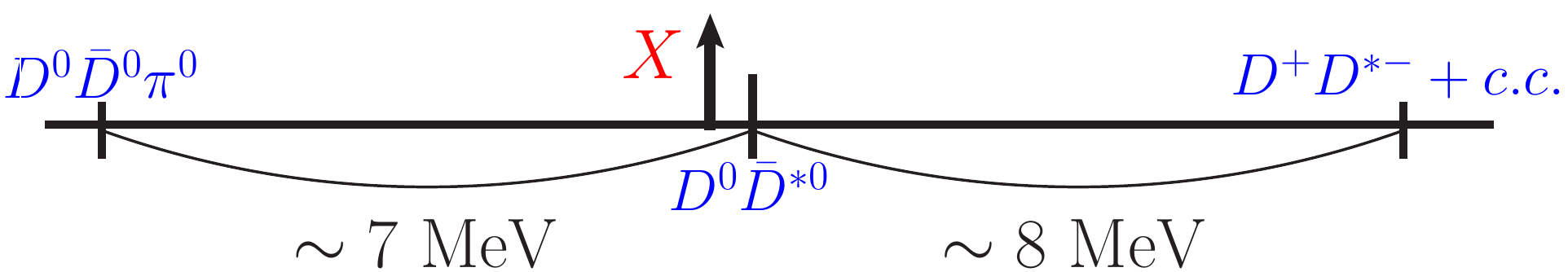, width=0.7\textwidth}}
\caption{The $\X$ mass relative to the relevant two- and three-body thresholds.} 
\label{fig:Xmass}
\end{figure}

From the discussion at the end of the previous section we conclude that the three-body $D \bar D \pi$ threshold is yet another one relevant to the $\X$ physics. Indeed, the mass difference 
between the $D^*$ and $D$ meson is very close to the pion mass, so that the two-body $D \bar D^*$ cut is close to the three-body $D \bar D \pi$ one, and both are close to the $\X$ mass (see Fig. 
\ref{fig:Xmass}). Because of this threshold proximity it is necessary to generalise the coupled-channel scheme used above to include the three-body channel explicitly \cite{Baru:2011rs}.

While reading Subsection~\ref{threebodyX}, it is important to pay attention to the form of the coupled-channel system which incorporates the three-body dynamics (formula (\ref{aa})), and to the 
partial wave decomposition of the amplitude described in Eq.~(\ref{angular}) and in the text below it. A detailed derivation of these equations can be skipped at the first reading. 
Subsection~\ref{XOPE} and, especially, its technical part can also be skipped for sketchy reading of the review. 
It is, however, important to pay attention to the main conclusion made that the one-pion exchange (OPE) is not binding enough to form the $\X$ even as a shallow bound state. Another important aspect 
to notice is the difference between the OPE in a charmonium system at hand and in the deuteron, the latter being quite often (erroneously) considered as a universal standard for such an exchange --- 
see Eq.~(\ref{vstatic}) and the text below it. Moreover, it follows from the discussions of this chapter that the quantum mechanical problem of the OPE without supplementary short-range interactions 
is not intrinsically self-consistent; an adequate approach based on the effective field theory technique is given in Subsection~\ref{Xthreedyn}.

\subsection{$D\bar{D}^*\leftrightarrow D\bar{D}\pi \leftrightarrow \bar{D}D^*$ coupled-channel system}
\label{threebodyX}

The basis of the model which incorporates the three-body dynamics in the $\X$ comprises three channels,
\be
|2\rangle=D\bar{D}^*,\quad|\bar{2}\rangle=\bar{D}D^*,\quad|3\rangle=D\bar{D}\pi,
\label{chandef}
\ee
coupled by the pion exchange.

In the centre-of-mass frame the momenta in the two-body $D \bar D^*$ and $\bar D D^*$ system are defined as
\bea
\vep_D&=&\vep,\quad\vep_{\bar D^*}=-\vep,\\
\vep_{\bar D}&=&\bar {\vep},\quad\vep_{D^*}=-\bar {\vep},
\eea
while in the three-body $D\bar{D}\pi$ system it is convenient to define two sets of Jacobi variables, $\{\vep,\veq\}$ and $\{\bar{\vep},\bar{\veq}\}$,
\be
\vep_D=\vep,\quad\vep_{\bar D}=-\veq-\frac{m}{m+m_{\pi}}\vep,\quad\vep_{\pi}=\veq-\frac{m_{\pi}}{m+m_{\pi}}\vep 
\ee 
or 
\be
\vep_D=-\bar{\veq}-\frac{m}{m+m_{\pi}}\bar{\vep},\quad \vep_{\bar D}=\bar{\vep},\quad\vep_{\pi}=\bar{\veq}-\frac{m_{\pi}}{m+m_{\pi}}\bar{\vep},
\ee
where $m$ is $D$-meson mass and $m_{\pi}$ is the pion mass. The Jacobi variables from different sets are related to each other as
\be
\bar{\veq}=\alpha \veq+\beta \vep,\quad\bar {\vep}=-\veq-\alpha \vep,\quad
\veq=\alpha \bar {\veq}+\beta \bar {\vep},\quad\vep=-\bar
{\veq}-\alpha\bar{\vep},
\ee
where
\be
\alpha=\frac{m}{m+m_{\pi}},\quad\beta=\alpha^2-1=-\frac{(2m+m_{\pi})m_{\pi}}{(m+m_{\pi})^2}.
\label{alphabeta}
\ee

The vertex $D^*D\pi$ is defined as 
\be
v_{D^*D\pi}(\veq)=g\; ({\bm \epsilon}\cdot \veq),
\label{g1}
\ee
where ${\bm \epsilon}$ is the polarisation vector of the $D^*$ meson, $\veq$ is the relative momentum in the $D\pi$ system, and $g$ is the coupling constant which can be fixed from the 
experimentally measured $D^*\to D\pi$ decay width.

The two-body and three-body channels communicate via the interaction potentials,
\be 
V_{32}^{m}(\vep,\veq;\vep')=g q_m\delta(\vep-\vep'),\quad V_{3 \bar 2}^{m}(\bar {\vep},\bar{\veq};
\bar {\vep}')=g\bar{q}_m\delta(\bar {\vep}-\bar{\vep}'),
\label{V32} 
\ee 
and similar potentials $V_{23}^{m}$ and $V_{\bar 2 3}^{m}$.

The equation for the $t$-matrix reads (schematically)
\be
t=V-Vg_0t,
\label{fulleq}
\ee
where $g_0$ is diagonal matrix of free propagators,
\be
g_{03}(\vep,\veq,M)=\frac{1}{D_3(\vep,\veq,M)},\quad
g_{02}(\vep,M)=g_{0 \bar 2}(\vep,M)=\frac{1}{D_2(\vep,M)}.
\ee
The inverse two- and three-body propagator are defined as
\be
D_2(\vep)=m+m_*+\frac{p^2}{2\mu_*}-M,\quad D_3(\vep,\veq)=
2m+m_{\pi}+\frac{p^2}{2\mu_p}+\frac{q^2}{2\mu_q}-M,
\label{g0}
\ee
where $M$ is the total mass of the system, $m_*$ is the mass of the $D^*$ meson, the reduced masses are
\be
\mu_*=\frac{mm_*}{m+m_*},\quad\mu_p=\frac{m(m+m_{\pi})}{2m+m_{\pi}},
\quad\mu_q \equiv \mu_q(D\pi)=\frac{mm_{\pi}}{m+m_{\pi}},
\ee
and in what follows we will set $\mu_*=\mu_p$.

Finally, the self-energy part $\Sigma(p)$ due to the virtual $D\pi$ loop can be written as
\be
\Sigma(p)=\frac{g^2}{3}\int \frac{q^2d^3q}{D_3(\vep,\veq)}.
\label{Sigma} 
\ee

Using the definition (\ref{Sigma}), after some algebraical manipulations, one arrives at the system of equations
\be
\left\{
\begin{array}{rcl}
t_{22}&=&\ds-\Sigma+\Sigma D_2^{-1}t_{22}+V_{23}D_3^{-1}V_{32}D_2^{-1}t_{\bar{2}2}\\[3mm]
t_{\bar{2}2}&=&\ds-V_{\bar{2}3}D_3^{-1}V_{32}+\Sigma
D_2^{-1}t_{\bar{2}2}+ V_{\bar{2}3}D_3^{-1}V_{32}D_2^{-1} t_{22}
\end{array}
\right.
\label{s102}
\ee
and a similar pair of equations for the $t_{2\bar{2}}$ and $t_{\bar{2}\bar{2}}$ components. A detailed derivation can be found in Ref.~\cite{Baru:2011rs}.

As the interaction respects $C$-parity, it is convenient to define proper combinations of the amplitudes which possess a given $C$-parity,
\be
t_{\pm}=t_{22}\pm t_{\bar{2}2},
\ee
and satisfy the equations
\be
\Delta D_2^{-1}t_\pm=-\Sigma\mp V_{23}D_3^{-1}V_{32}\pm V_{23}D_3^{-1}V_{32}D_2^{-1}t_\pm,
\ee
where the inverse dressed $D^*$-meson propagator $\Delta(p)$ is introduced as
\be
\Delta(p)=m_*+m+\frac{p^2}{2\mu_*}-M-\Sigma(p).
\label{Deltadef}
\ee

Then, substituting
\be
t_{\pm}=-\frac{\Sigma D_2}{\Delta}+\frac{D_2}{\Delta}a_{\pm}\frac{D_2}{\Delta}
\label{intra0}
\ee
one arrives at the expression for the new function $a_{\pm}^{mn}(\vep,\vep')$,
\be
a_{\pm}=V_{\pm}-V_{\pm}\Delta^{-1}a_{\pm}.
\label{a0}
\ee
In what follows only $C$-even states are considered, that corresponds to the ``$+$'' sign above. In addition, to simplify notations, we set $V_+ \equiv V$.

Now we define the full form of the one-pion exchange potential $V$ which enters Eq.~(\ref{a0}). To this end we introduce explicitly the neutral 
($D^0 \bar D^{*0}+\bar{D}^0 D^{*0}$) and charged ($D^-D^{*+}+D^+D^{*-}$) two-body channels and take into account the mass splittings between the charged and neutral particles both for the $D$ mesons 
and 
pions. This yields
\be
V^{mn}_{ik}(\vep,\vep')=(\vep'+\alpha_{ik}
\vep)^m(\vep+\alpha'_{ik}\vep')^nF_{ik}(\vep,\vep'), \quad
F_{ik}(\vep,\vep')=-\frac{g^2}{D_{3ik}(\vep,\vep')},
\label{Vmn2}
\ee
where the indexes $i,k$ stand for the neutral ($0$) and charged ($c$) component, coefficients $\alpha$ are equal to
\bea
\alpha_{00}=\alpha'_{00}=\frac{m_0}{m_{\pi^0}+m_0},\quad
\alpha_{cc}=\alpha'_{cc}=\frac{m_c}{m_{\pi^0}+m_c},\nonumber\\[-2mm]
\label{alphass}\\[-2mm]
\alpha_{0c}=\alpha'_{c0}=\frac{m_c}{m_{\pi^c}+m_c},\quad\alpha_{c0}=\alpha'_{0c}=\frac{m_0}{m_{\pi^c}+m_0},\nonumber
\eea
and the inverse three-body propagators are
\begin{eqnarray}
D_{300}(\vep,\vep')&=&2m_0+m_{\pi^0}+\frac{p^2}{2m_0}+\frac{p'^2}{2m_0}+\frac{(\vep+\vep')^2}
{2m_{\pi^0}}-M-i0,\nonumber\\
D_{3cc}(\vep,\vep')&=&2m_c+m_{\pi^0}+\frac{p^2}{2m_c}+\frac{p'^2}{2m_c}+\frac{(\vep+\vep')^2}
{2m_{\pi^0}}-M-i0,\nonumber\\[-3mm]
\label{D3s}\\[-3mm]
D_{30c}(\vep,\vep')&=&m_c+m_0+m_{\pi^c}+\frac{p^2}{2m_0}+\frac{p'^2}{2m_c}+\frac{(\vep+\vep')^2}
{2m_{\pi^c}}-M-i0,\nonumber\\
D_{3c0}(\vep,\vep')&=&m_c+m_0+m_{\pi^c}+\frac{p^2}{2m_c}+\frac{p'^2}{2m_0}+\frac{(\vep+\vep')^2}
{2m_{\pi^c}}-M-i0.\nonumber
\end{eqnarray}

We will be interested in the processes with the neutral mesons in the final state. The relevant components of the matrix $a$ satisfy the
set of equations
\be 
\left\{
\begin{array}{l}
a_{00}^{mn}(\vep,\vep')=\lambda_0V_{00}^{mn}(\vep,\vep')-\ds
\sum_{i=0,c}\lambda_i\int\frac{d^3s}{\Delta_i(s)}V_{0i}^{mp}(\vep,\ves)a_{i0}^{pn}(\ves,\vep')\\[5mm]
a_{c0}^{mn}(\vep,\vep')=\lambda_cV_{c0}^{mn}(\vep,\vep') -\ds
\sum_{i=0,c}\lambda_i\int\frac{d^3s}{\Delta_i(s)}V_{ci}^{mp}(\vep,\ves)a_{i0}^{pn}(\ves,\vep'),
\end{array}
\right.
\label{aa}
\ee
where $\lambda_0=1$ and $\lambda_c=2$ are the coefficients which take into account the isospin content of the OPE.

The inverse propagators $\Delta_0$ and $\Delta_c$ entering the system (\ref{aa}) take the form
\be
\Delta_0(p)=m_{*0}+m_0+\frac{p^2}{2\mu_{0*}}-M-\frac{i}{2}\varGamma_0(p),
\quad
\Delta_c(p)=m_{*c}+m_c+\frac{p^2}{2\mu_{c*}}-M-\frac{i}{2}\varGamma_c(p),
\label{Deltas0c}
\ee
where $\mu_{0*}$ and $\mu_{c*}$ are the reduced masses in the $D^0 \bar D^{*0}+\bar{D}^0 D^{*0}$ and $D^-D^{*+}+D^+D^{*-}$ systems, respectively, and the loop operator (see 
Eq.~(\ref{Deltadef})) is 
replaced by the running width $\varGamma(p)$ (analytically continued below the threshold) which includes both the self-energy $\Sigma(p)$ and the  contributions from the other decay channels of the 
$D^*$ meson --- see Ref.~\cite{Baru:2011rs}.

The system of equations (\ref{aa}) (together with the system of equations for the quantities $a_{0c}^{mn}$ and $a_{cc}^{mn}$) is the central 
result of this chapter: the equations arrived at describe the interaction in the system $D \bar D^*$ which stems from the OPE and comply with the constraints from the three-body unitarity due to 
the channel $D \bar D \pi$.

It is convenient to perform a partial wave decomposition of the amplitude $a$ in terms of the spherical vectors $\veY_{JLM}(\ven)$,
\be
a_{ik}^{mn}(\vep,\vep')=\sum_J\sum_{L_1L_2}a_{ik}^{J,L_1,L_2}(p,p')\sum_M(\veY_{JL_1M}(\ven))^m(\veY^*_{JL_2M} (\ven'))^n,
\label{angular}
\ee
where $\ven$ and $\ven'$ are the unit vectors for the momenta $\vep$ and $\vep'$, respectively.
As the $\X$ quantum numbers are $1^{++}$, two partial waves, $S$ and $D$, contribute to the amplitude, so that one has $J=1$ and $L_1,L_2=0,2$. If, in addition, we are interested in the 
$S$ wave in the final state we need only the $a_{ik}^{SS}$ and $a_{ik}^{DS}$ matrix elements.
The explicit form of the potentials entering the equations for these quantities can be found in Ref.~\cite{Baru:2011rs}.

\begin{figure}[t]
\centerline{\epsfig{file=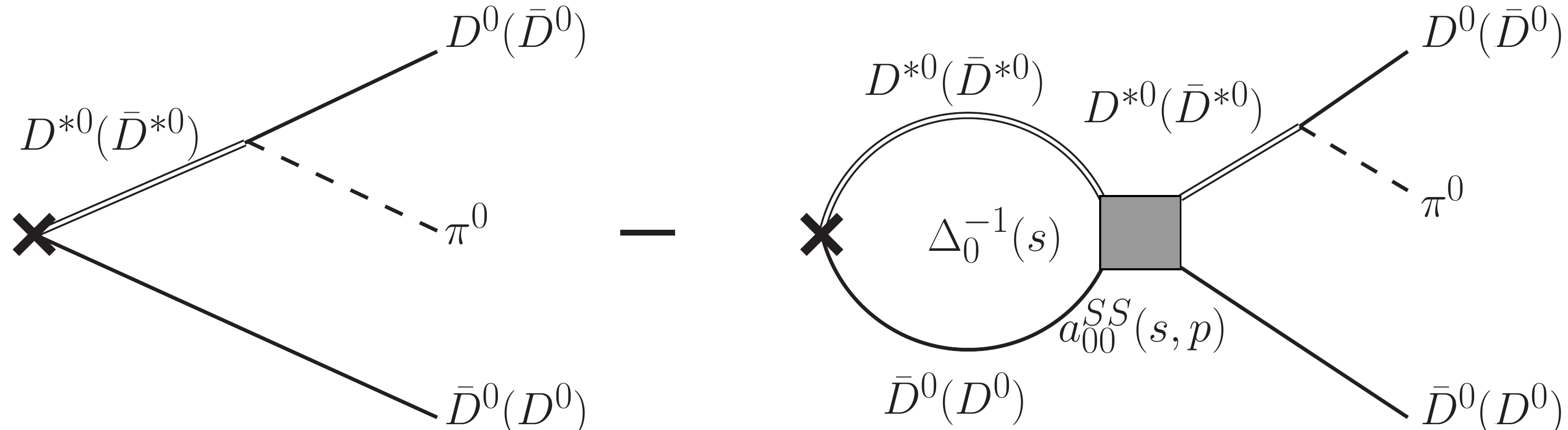, width=0.7\textwidth}}
\caption{The production amplitude for the $D^0 \bar D^0 \pi^0$ final state. The cross stands for a point-like source.}\label{diagram2}
\end{figure}

Under the assumption of a point-like source which produces a $C$-even pair $D^0\bar{D}^{*0}$ in the $S$ wave the production amplitude for the final state 
$D^0\bar D^0 \pi^0$ with the invariant mass $M$ (see Fig.~\ref{diagram2}) is totally defined through the component $a_{00}^{SS}$; the corresponding explicit expressions can also be found in 
Ref.~\cite{Baru:2011rs}.

\subsection{One-pion exchange in the $\X$}\label{XOPE}

One-boson exchange ($\rho$, $\omega$ and so on) was suggested long ago \cite{Voloshin:1976ap} as a mechanism able to form a bound state of charmed particles. A possibility to bind an isosinglet 
$C$-even $D \bar D^*$ system with OPE was studied in Ref.~\cite{Tornqvist:1991ks}. Shortly after the discovery of the $\X$,  the model was revisited in Refs.~\cite{Tornqvist:2004qy,Swanson:2003tb}  
where the $D \bar D^*$ system was treated in a deuteron-like fashion: OPE enters the problem in the form of a static potential. Almost immediately the
warning was issued in Ref.~\cite{Suzuki:2005ha} that, because of the three-body $D \bar D \pi$ threshold proximity, the pion may go on shell and fail to bind the $\X$. As was shown 
in Ref.~\cite{Filin:2010se}, the three-body cut effects could play a fatal role in the $D_{\alpha}\bar D_{\beta}$ system if the $D_\beta$
width is dominated by the $S$-wave $D_\beta\to D_\alpha\pi$ decay. In particular, the bound states in the $D_{\alpha}\bar
D_{\beta}$ system predicted in Refs.~\cite{Close:2009ag,Close:2010wq} in the static approximation disappear in the continuum if the full three-body treatment is applied. In case of the $\X$, the 
generic decay 
$D^*\to D \pi$ is a $P$-wave one, and this extra power of the momentum could, in principle, attenuate the influence of the cut. The problem, however, deserves a detailed investigation.

With the $P$-wave vertex envolved the $D$-meson loop integrals entering the equations derived above diverge and require regularisation. The latter can be performed introducing suitable form factors 
with a cut-off 
parameter $\Lambda$. Then the conclusion whether or not the OPE is able to bind the $\X$ can be made based on the values of the cut-off $\Lambda$ needed to produce a shallow bound state. Namely, the 
existence of a bound state for a relatively small value $\Lambda\lesssim 1$~GeV can be interpreted as a proof that the OPE provides enough attraction to produce a bound state as such values of 
$\Lambda$ can be justified in the quark model. In the meantime, larger values of $\Lambda$ should be disregarded as unphysical. The corresponding
calculations were performed in the static limit and results were presented in Refs.~\cite{Liu:2008fh,Thomas:2008ja}. It is important to note that, in the former paper, a bound state in the $D^0 \bar 
D^{*0}$ system was found for the values of $\Lambda$ about 6-8 GeV. In the latter paper, the charged channel was also taken into account which allowed the authors to find a bound state for a much 
smaller cut-off, $\Lambda\simeq 0.6$-0.8~GeV. Thus, the conclusion of Ref.~\cite{Thomas:2008ja} was that the $\X$ could
be bound by the OPE. In paper \cite{Kalashnikova:2012qf}, the OPE in the $\X$ system was considered beyond the static limit and it
was shown that no bound state existed in the $\X$ system for reasonable values of $\Lambda$. In view of the importance of this result for the future discussions we dwell on it here. 

To have a better contact with the previous works we define a covariant $D^* \to D \pi$ vertex, 
\be 
v_\mu=g_f\bar{u}^*_{\alpha}(\tau^a)^{\alpha}_{\beta}u^{\beta}\pi^a p_{\pi\mu},
\label{fvertex} 
\ee 
where $p_{\pi\mu}$ is the pion 4-momentum and $u^*$, $u$ and $\pi$ are the isospin wave functions of the $D^*$, $D$ and pion, respectively. In 
paper~\cite{Thomas:2008ja}, an effective coupling parameter $V_0$ was introduced as 
\be 
\frac{g_f^2}{4m_*^2}=\frac{6\pi V_0}{m_{\pi}^3},
\ee 
with $V_0 \approx 1.3$ MeV defined from the experimental value \cite{Tanabashi:2018oca} of the width $\varGamma(D^{*+}\to D^0\pi^+)$,
\be
\varGamma(D^{*+}\to D^0\pi^+)=\frac{g_f^2 q_{0c}^3}{12\pi m_{*c}^2}=2V_0\frac{q_{0c}^3}{m_{\pi}^3}, 
\ee
where $q_{0c}$ is the relative momentum in the $D^0\pi^+$ system.

Let us re-define the $D^*D\pi$ vertex (\ref{g1}) and introduce a form factor which regulates its behaviour at large momenta, 
\be
\veg(\veq)=g\veq\frac{\Lambda^2} {\Lambda^2+\veq^2},\quad g=\frac{\sqrt{6\pi V_0\vphantom{1^2}}}{m^{3/2}_{\pi}}, 
\label{ggg}
\ee 
where the cut-off $\Lambda$ is introduced, as discussed above. If one neglects the isospin symmetry breaking, the OPE potential in the static limit expressed in terms of this vertex takes the form  
\be
V_{\rm stat}^{mn}(\veq)=-\frac{3}{(2\pi)^3}\frac{g_m(\veq)g_n(\veq)}{\veq^2+[m_{\pi}^2-(m-m_*)^2]},
\label{vstatic} 
\ee 
where the factor 3 corresponds to the isosinglet state. We notice here that the potential (\ref{vstatic}) is two times weaker than the one used in Ref.~\cite{Thomas:2008ja}. This means that, 
effectively, 
the coupling $V_0$ used in Ref.~\cite{Thomas:2008ja} is two times larger than the value $V_0=1.3$ MeV found from the data on the $D^*$ pionic decays.

It is relevant to comment on a critical difference between the OPE potential in the nucleon physics (for example, in the deuteron) and in the physics of charmonium. If the masses of the particles 
exchanging the pion are close to one another ($m_*\approx m$) and their mass difference can be neglected compared to the pion mass (this condition is obviously met for the proton and neutron mass) 
then the potential (\ref{vstatic}) reduces to the Yukawa one --- the standard pion exchange potential in the physics of nucleons. To some extent this condition is true in the $b$-quark sector, too, 
since the mass difference for the $B^*$ and $B$ meson equals 45 MeV that is 3 times smaller than the pion mass. However, the situation is utterly different from the one described above in the physics 
of charmonium since the mass difference between the $D^*$ and $D$ meson nearly (with the accuracy about 7 MeV) coincides with the pion mass. Besides that, $m_*-m>m_\pi$, so that the parameter 
$\mu=\sqrt{m_{\pi}^2-(m-m_*)^2}$,
which enters the denominator of the potential (\ref{vstatic}) and defines its behaviour is not only small in the absolute value but, in addition, it is purely imaginary, so that instead of an 
exponential fall-off with the increase of the distance between the mesons the static potential oscillates. It is also important to notice that the numerical smallness of the parameter $\mu$ 
emphasises a low accuracy of the static approximation for the $\X$ in general since the neglect of the $D$-meson recoil terms in the denominator of the potential does not look like a 
well-justified 
procedure. In a more adequate approach such terms are retained --- see, for example, the expressions for the inverse three-body propagators (\ref{D3s}) with the three-body dynamics included. 
Below, this problem and its implications for the OPE potential are discussed in more detail. 

\begin{figure}[t]
\centerline{\epsfig{file=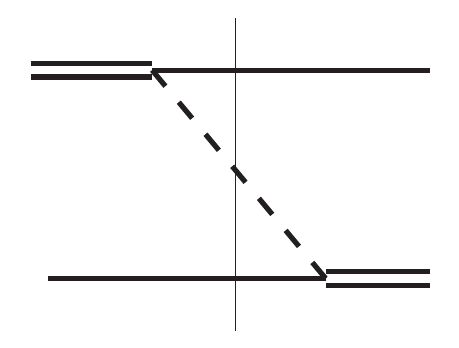,
width=0.3\textwidth}\hspace*{0.1\textwidth}\epsfig{file=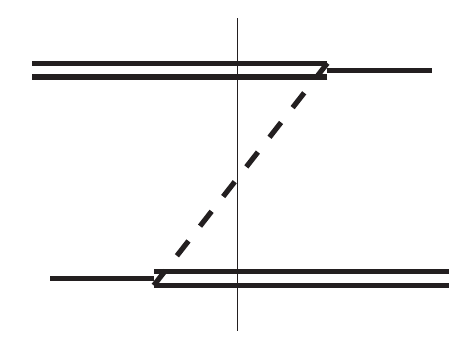,
width=0.3\textwidth}} 
\caption{The diagrams in the Time-Ordered Perturbation Theory corresponding to the potentials $V_1$ (left)
and $V_2$ (right). The double and single solid line are for the $D^*$ and $D$ meson, respectively, while the dashed line is for the pion.
The thin vertical line pin-points the intermediate state.}
\label{fig:D1D2}
\end{figure}

The potential (\ref{vstatic}) can be written as
$$
V_{\rm stat}^{mn}(\veq)=-\frac{3}{(2\pi)^3}g_m(\veq)g_n(\veq)\Bigl(V_1^{\rm stat}(\veq)+V_2^{\rm stat}(\veq)\Bigr),
$$
\bea
V_1^{\rm stat}(\veq)&=&\frac{1}{2E_\pi(E_\pi+m-m_*)},
\label{firstV}\\
V_2^{\rm stat}(\veq)&=&\frac{1}{2E_\pi(E_\pi+m_*-m)},
\label{secondV} 
\eea 
where $E_\pi=\sqrt{\veq^2+m_{\pi}^2}$. This corresponds to two different orderings (contributions of the intermediate states $D\bar{D}\pi$ and $D^*\bar{D}^*\pi$) in the framework of the Time-Ordered 
Perturbation Theory (see the graphs in
Fig.~\ref{fig:D1D2}).

As $m_* \approx m+m_{\pi}$, one has $V_1 \gg V_2$ for the momenta $q$ of the order of several hundred MeV. However, as demonstrated above, for such momenta the static approximation is not valid. 
On the other hand, for higher momenta one should account for the contribution from the second ordering. This opens a
Pandora box of the intermediate states available via the transition chain 
\be
\bar {D}^*D^* \pi \leftrightarrow \bar {D} D^* \pi \pi \leftrightarrow \bar{D} D \pi\pi\pi\ldots
\ee
invalidating, {\it inter alia}, the very notion of a single-pion exchange.

\begin{figure}[t]
\centerline{\epsfig{file=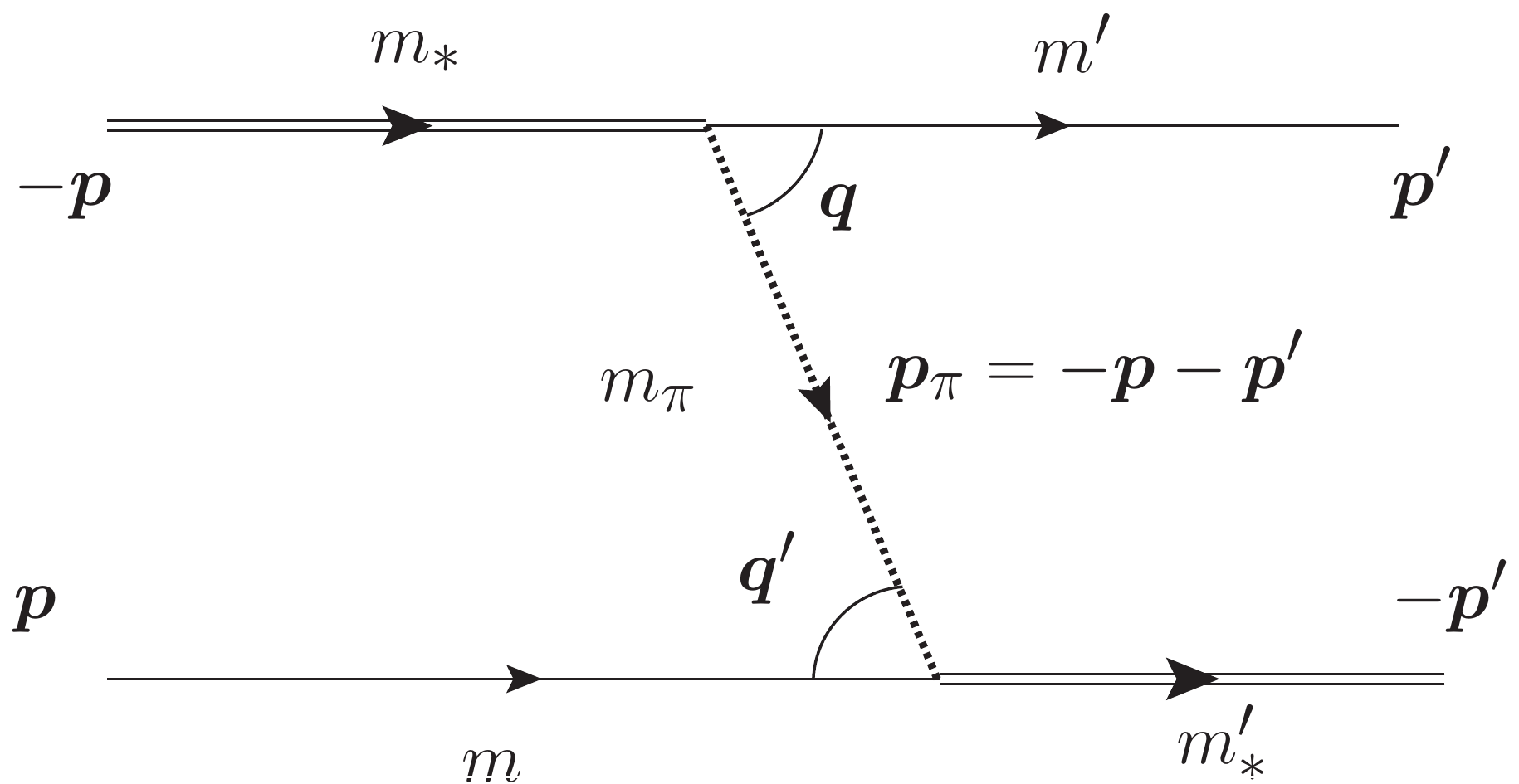, width=0.5\textwidth}}
\caption{Kinematics of the $D \bar D^*$ scattering due to the OPE.}\label{graphv1}
\end{figure}

Consider now the OPE potential beyond the static limit. Then, to begin with, we notice that, for a given formulation of the problem, all momenta 
$q\lesssim\Lambda$ are allowed. Therefore, once the expected values of the cut-off are large, $\Lambda\lesssim 1$~GeV, one has to resort to the relativistic kinematics and modify the potential 
accordingly, that is, to write it as
\be
V^{mn}_{ik}(\vep,\vep')=-\frac{1}{(2\pi)^3}\frac{g_m(\vep'+\alpha_{ik}\vep)g_n(\vep+\alpha_{ik}\vep')}{D_{3ik} (\vep,\vep')},
\label{VVmn}
\ee
where the vertex $\veg(\veq)$ is given in Eq.~(\ref{ggg}) and the relativised three-body propagator is 
\be 
D_3(\vep,\vep')=2E_\pi(E_\pi-\mu-i0),
\label{D3} 
\ee 
where 
\be
\mu=m_{*0}+m_0+E-\sqrt{m^2+p^2}-\sqrt{m'^2+p'^2} 
\label{muXDD} 
\ee
and 
\be 
E_\pi=\sqrt{(\vep+\vep')^2+m_\pi^2}. 
\label{Epi} 
\ee 
Here, as before, the energy $E$ is defined relative to the neutral two-body $D^0\bar{D}^{*0}$ threshold. the coefficients $\alpha$ and
$\alpha'$ can be found from the standard relativistic textbook formula (see, for example,
textbook \cite{Novozhilov:1975yt} or paper \cite{Kalashnikova:1996pu}),
\bea
&\ds\alpha=\frac{1}{\sqrt{\varepsilon'^2-p^2}}\left[\sqrt{m'^2+p'^2}+\frac{\vep\vep'}
{\varepsilon'+\sqrt{\varepsilon'^2-p^2}}\right],&\nonumber\\[-2mm]
\label{alphass2}\\[-2mm]
&\ds\alpha'=\frac{1}{\sqrt{\varepsilon^2-p'^2}}\left[\sqrt{m^2+p^2}+\frac{\vep\vep'}{\varepsilon+\sqrt{
\varepsilon^2-p'^2}}\right],&\nonumber
\eea 
where
$$
\varepsilon=\sqrt{m^2+p^2}+E_\pi,\quad \varepsilon'=\sqrt{m'^2+p'^2}+E_\pi.
$$
The kinematics corresponding to this potential is detailed in Fig.~\ref{graphv1}. It is clear that the potential (\ref{VVmn}) is nothing else but the $V_1$ part of the OPE beyond the static limit. 
The static 
approximation is obtained from the expression (\ref{VVmn}) by setting $\alpha=\alpha'=1$ and $\mu=m_{*0}+m_0-m-m'$. Both replacements are not insignificant, especially the former one: $\alpha$ 
decreases with the increase of the momentum that leads to an effective suppression of the potential as compared to the naive static approximation with $\alpha=\alpha'=1$.

Besides that, the two-body propagators $\Delta_0(p)$ and $\Delta_c(p)$ in Eq.~(\ref{aa}) as well as the three-body propagator $D_{3ik}(\vep,\vep')$
generate contributions to the imaginary part of the interaction which are to be kept in order to preserve unitarity but which are obviously
skipped in Refs.~\cite{Liu:2008fh,Thomas:2008ja}. These contributions also effectively weaken the interaction. 

The strategy employed in Ref.~\cite{Kalashnikova:2012qf} to resolve the problem of the OPE as a binding force is as
follows. The system of equations (\ref{aa}) is solved for various versions for the potential and the differential production rate is calculated then. 
A bound-state singularity, if exists, reveals itself as a below-threshold peak in the production rate. 
Once, for a fixed value of the coupling ($V_0=1.3$ MeV), the potential becomes more attractive with the cut-off growth there exists a minimal 
value of the cut-off $\Lambda$ for which a bound state appears right at the two-body threshold. The results for such boundary cut-offs are listed in Table~\ref{table1VV} from which one can see that, 
on 
the one hand, even inclusion of the second ordering and using the static limit do not lead to phenomenologically adequate values of $\Lambda$ and, on the other hand, the effects related to the 
dynamical pions increase the minimal $\Lambda$.

One is forced to conclude, therefore, that the OPE is not strong enough to bind the $\X$, and some other short-range dynamics is
responsible for its binding.

\begin{table}
\begin{center}
\begin{tabular}{|c|c|c|c|c|}
\hline
&$V_1^{\rm stat}$&$V_1^{\rm stat}+V_2^{\rm stat}$&$V_1$&$V_1+V_2^{\rm stat}$\\
\hline
$V_0$=1.3~MeV&2750&1650&3800&2100\\
\hline
\end{tabular}
\caption{The minimal cut-off parameter $\Lambda_{\rm min}$ (in MeV) consistent with a bound state in the $D \bar D^*$ system.}
\label{table1VV}
\end{center}
\end{table}

\subsection{Elastic line shape of the $X(3872)$ with the three-body dynamics included}
\label{Xthreedyn}

Results obtained in chapter \ref{XOPE} allow one to conclude that a more adequate approach to the description of the $\X$ is a field-theoretical approach in which the OPE interaction is 
well-defined only in a combination with the contact term which describes the short-range part of the
interaction (including the short-range part of the OPE) \cite{Baru:2011rs,Baru:2015nea}. In such a formulation, large momenta (of the order of the cut-off) do not enter, as they are absorbed by the renormalised 
contact term and the dynamics of the system takes place at the momenta of the order of the binding momentum scale $k_B\simeq\sqrt{E_B
m}$, that is, the momenta of the order of several tens MeV. In this case, it is quite legitimate to use nonrelativistic kinematics for all
particles (including the pions).

Following this idea, let us parameterise the short-range part of the potential by a constant $C_0(\Lambda)$ where $\Lambda$ is the regulator (for simplicity, a sharp cut-off in the three-dimensional 
momentum by the step-like function $\theta(\Lambda-|\vep|)$ is used in the integrals). For a given value of $\Lambda$ this contact term is fixed by the value of the $\X$ binding energy which, for the 
sake of definiteness, we set to be 0.5~MeV. In such a way, $C_0(\Lambda)$ absorbs the leading dependence of the observables on the cut-off while the
residual dependence is expected to be rather weak. Should it not be the case, this might indicate the need for additional, momentum-dependent counter terms.

Before we proceed towards the solution of the full problem, let us consider a simple analytically solvable model in which the potential is exhausted by the contact interaction. Then we have the 
following system of equations for the amplitudes $a_{00}\equiv a_{00}^{SS}$ and $a_{c0}\equiv a_{c0}^{SS}$:
\be
\left\{
\begin{array}{rcl}
a_{00}&=&C_0-C_0 a_{00}I_0-2C_0 a_{c0} I_c\\[2mm]
a_{c0}&=&2C_0-2C_0 a_{00} I_0 - C_0 a_{c0} I_c,
\end{array}
\right.
\ee
with the solution for the $a_{00}$ given by
\be
a_{00}=\frac{C_0(1-3C_0I_c)}{(1+C_0 I_0)(1+C_0 I_c)-4 C_0^2 I_0I_c}.
\label{a00couple}
\ee

For the loop integrals one can use the expansion
\bea
I_0&=&\int_0^{\Lambda} dq\frac{q^2}{q^2/(2\mu_{0*})-E-i0}\approx 2\mu_{0*} \left(\Lambda +\frac{i}{2}\pi
k_0-\frac{k_0^2}{\Lambda}\right) +O\left(\frac{k_0^4}{\Lambda^3}\right),\\
I_c&=&\int_0^{\Lambda} dq \frac{q^2}{q^2/(2\mu_{c*})+\Delta-E-i0}\approx 2\mu_{c*} \left(\Lambda + \frac{i}{2}\pi
k_c-\frac{k_c^2}{\Lambda}\right)
+O\left(\frac{k_c^4}{\Lambda^3}\right),
\label{Ic}
\eea
where $k_0^2=2\mu_{0*} E$, $k_c^2=2\mu_{c*} (E-\Delta)$ and $\Delta=(m_{*c}+m_c)-(m_{*0}+m_0)\approx 8$~MeV. 
The requirement for the pole of the scattering amplitude (\ref{a00couple}) to exist at $E=-E_B$ leads to a quadratic equation, 
\be
\Bigl(1+C_0I_0(-E_B)\Bigr)\Bigl(1+C_0I_c(-E_B)\Bigr)-4C_0^2I_0(-E_B)I_c(-E_B)=0,
\label{pole}
\ee
to fix the counter term $C_0$ which has a couple of opposite-sign solutions. These solutions correspond to different values of the isospin, so for the case of the isoscalar state $\X$ only one of 
them 
should be retained. 

\begin{figure}[t]
\begin{center}
\epsfig{file=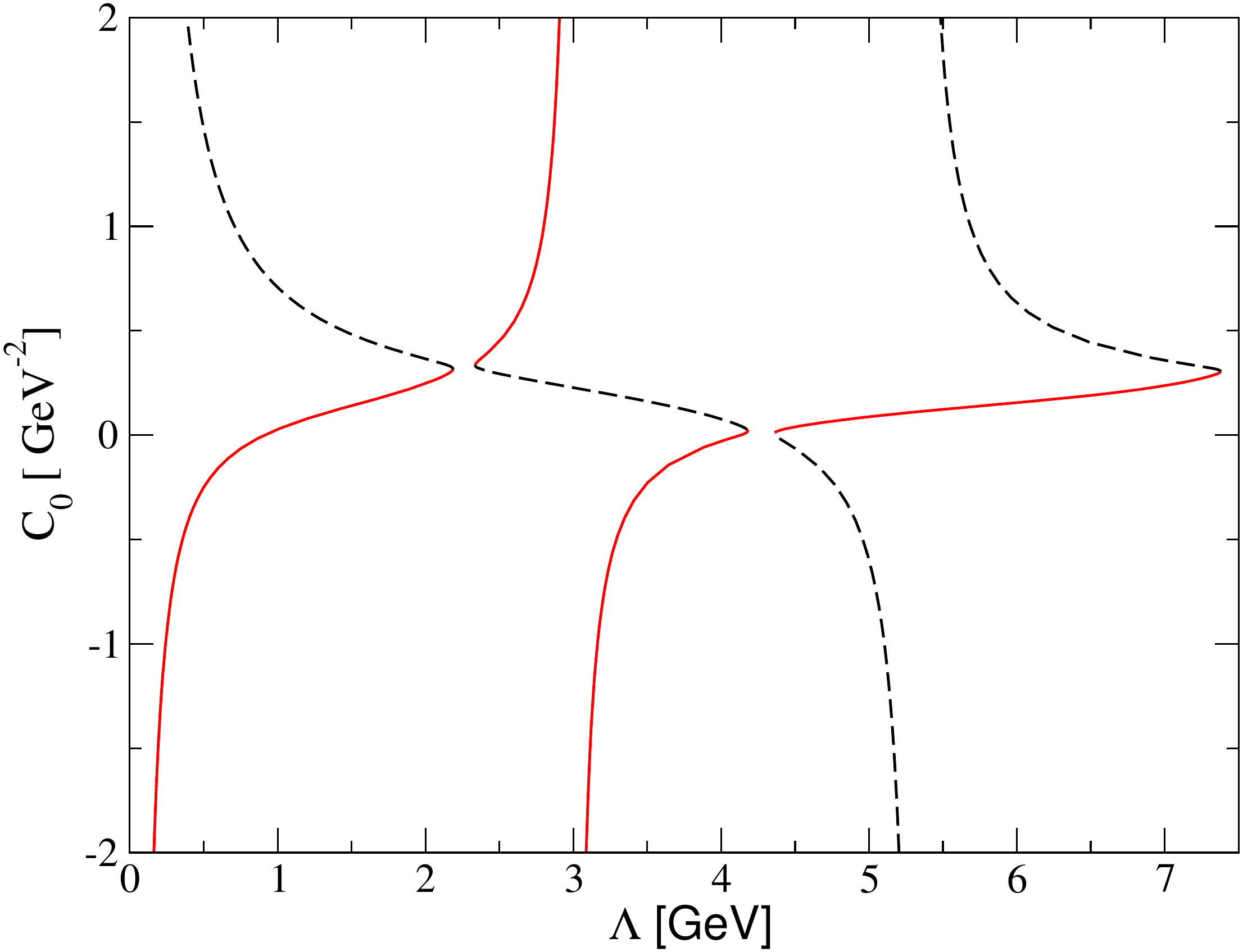, width=7.7cm}
\end{center}
\caption{The behaviour of the contact term $C_0$ as a function of the cut-off $\Lambda$ in the full problem with pions. Adapted from Ref.~\cite{Baru:2011rs}.}\label{c0fig}
\end{figure}

Now we consider the full problem which includes, together with the contact interaction, the OPE potential derived in Subsection~\ref{threebodyX}. As discussed above, the value of the contact term 
$C_0$ 
is 
chosen to ensure the existence of a bound state with the given energy. The dependence $C_0(\Lambda)$ shown in Fig.~\ref{c0fig} is rather nontrivial. This type of behaviour (the so-called limit cycle) 
is typical for such kind of calculations --- see, for example, 
Refs.~\cite{Nogga:2005hy,Kudryavtsev:1978af,Kudryavtsev:1979gq,Beane:2001bc,Braaten:2004pg},
where a similar behaviour was observed in nucleon systems. Indeed, with the increase of the cut-off $\Lambda$ the long-range part of the interaction varies, so that to compensate for this variation 
and 
to 
keep the bound state at place a new value of the contact term is needed. Near the discontinuity point of the curve $C_0(\Lambda)$ small variations of the cut-off lead to large variations of the 
$C_0$. 
Then, at some critical value of the cut-off it becomes impossible to adjust the latter, so the level under consideration sinks and a next level gets fixed at the given binding energy, thus performing 
a 
hop 
to the next branch of the function $C_0(\Lambda)$. Then this cycle repeats itself. As was clarified
before, the two solutions for the $C_0$ at each $\Lambda$ correspond to the two values of the isospin of the system. For the $\X$ case one should
consider $I=0$ which is described by the red curve in Fig.~\ref{c0fig}. It should be noted that $C_0$ vanishing at some value of the cut-off cannot be interpreted as the absence of the short-range 
dynamics in the system since the choice of the cut-off and the corresponding re-definition of the short-range and long-range contribution to the dynamics does not affect observable quantities that is 
guaranteed by the renormalisation group equations --- by the requirement $\partial E_B/\partial\Lambda=0$ in this case. 
The residual $\Lambda$-dependence can be removed by additional contact terms of higher order that is not required in this particular case since such a residual $\Lambda$-dependence is very weak 
(about 
a few per cent --- see the discussion below) and can be treated as a theoretical uncertainty of the calculation.

To increase the accuracy it is convenient to fix the value of the cut-off in the plateau region, so that a reasonable choice is $300~\mbox{MeV}\lesssim\Lambda\lesssim 1700$~MeV 
or $2500~\mbox{MeV}\lesssim\Lambda\lesssim 3800$~MeV. As was discussed in the previous chapter, using large cut-offs calls for the relativistic dynamics, so it makes sense to consider not very large 
cut-offs in order to have momenta in the renormalised theory small in comparison with the particle
masses, that is, for the nonrelativistic approximation to be applicable. Thus, it proves convenient to fix the value of $\Lambda=500$~MeV which falls into the first region above.

\begin{figure}[t]
\centerline{\epsfig{file=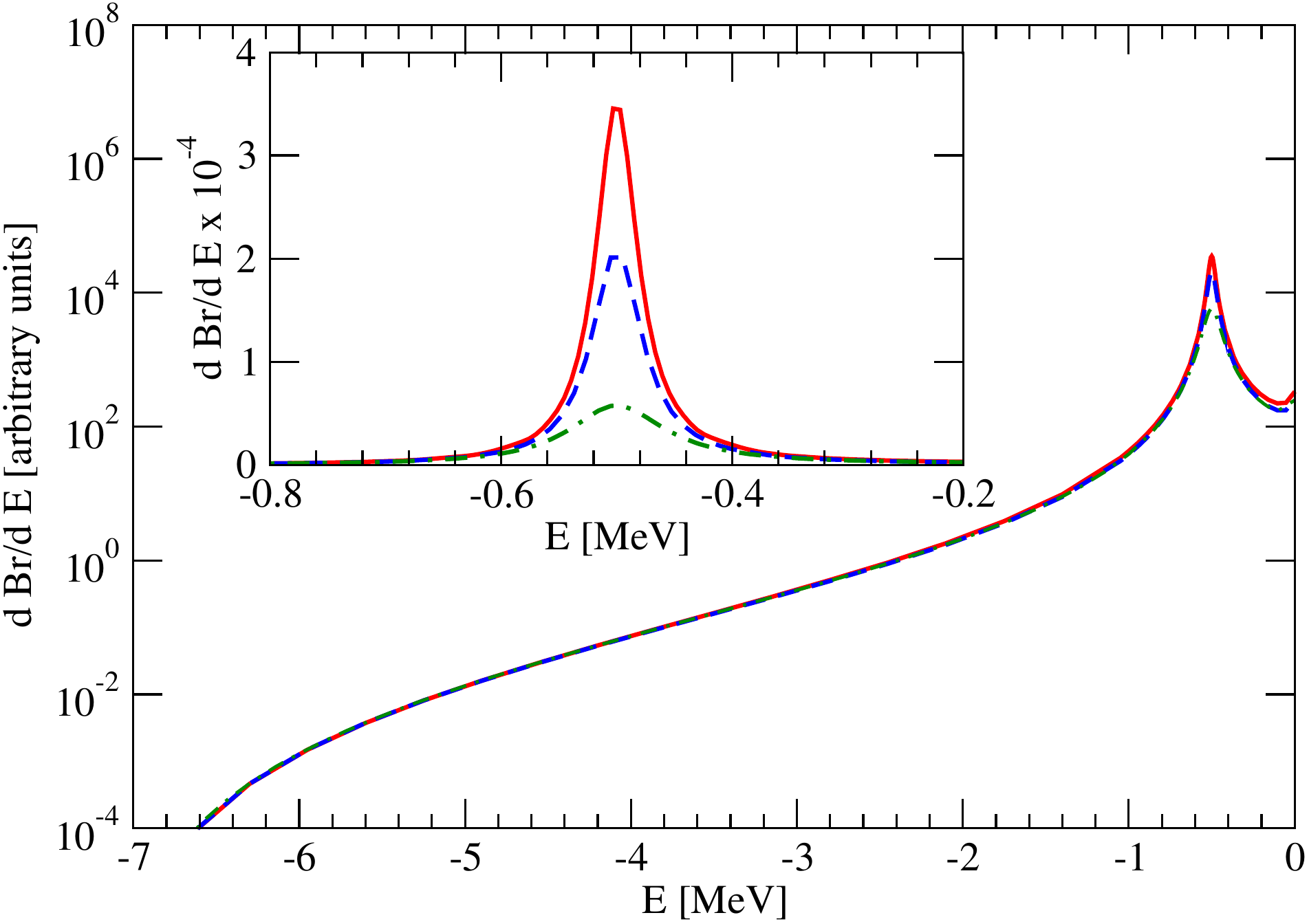, width=0.6\textwidth}}
\caption{The $\X$ production rate (in log scale) in the $D^0\bar D^0\pi^0$ channel for three cases: 1) single-channel calculation
in the static approximation (green dot-dashed line); 2) single-channel dynamical calculation (blue dashed line); 3) full two-channel
dynamical calculation (red solid line). The peak region is zoomed in the inlay in the linear scale. Adapted from Ref.~\cite{Baru:2011rs}.} \label{rate2}
\end{figure}

\begin{figure}
\begin{center}
\epsfig{file=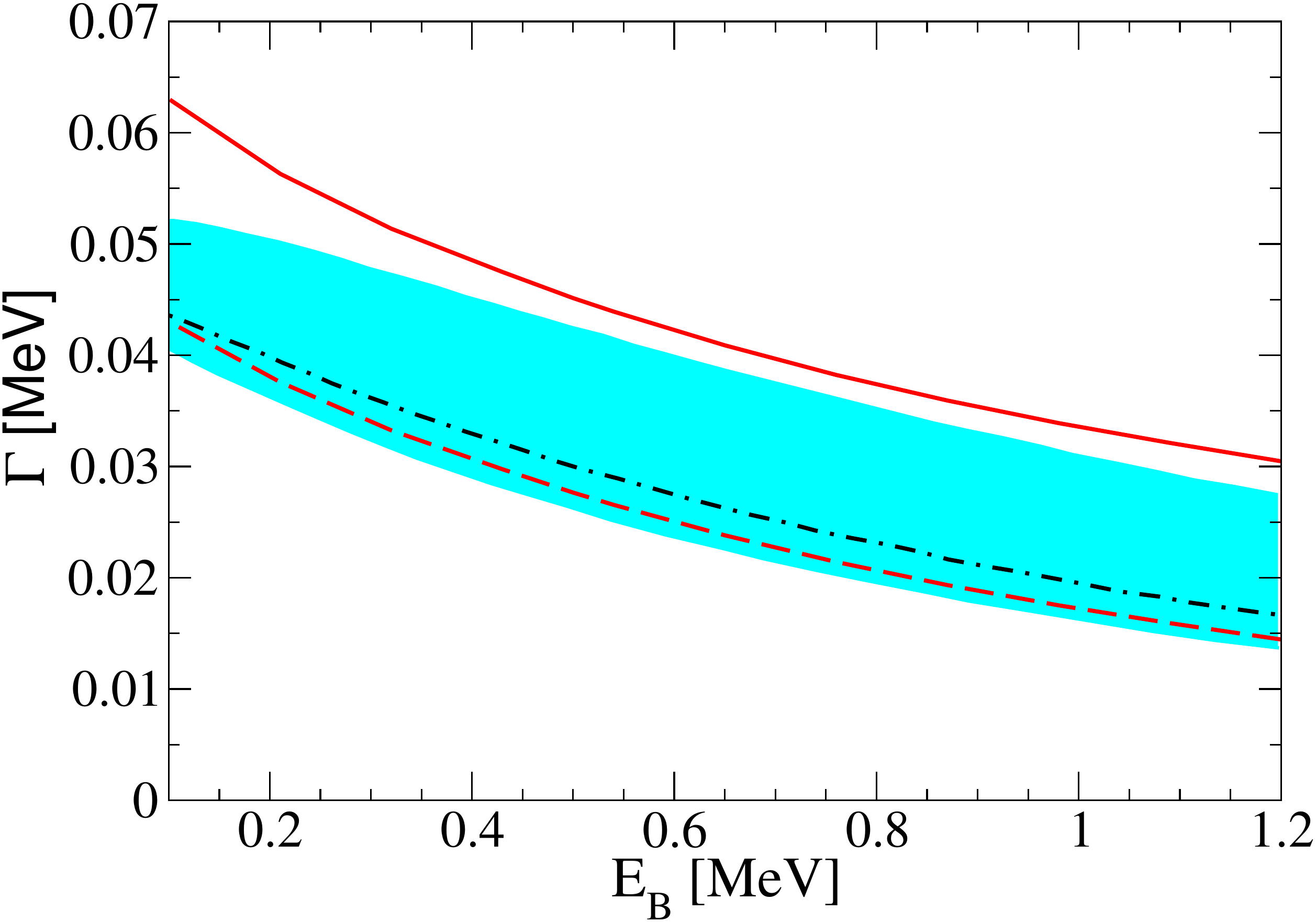, width=0.6\textwidth}
\end{center}
\caption{The width of the $X$ as a function of the binding energy $E_B$. The red dashed line shows the result of the full dynamical calculation,
the black dashed line and the blue band show the X-EFT results in the leading \cite{Voloshin:2003nt} and next-to-leading order \cite{Fleming:2007rp}, respectively. The solid 
red line shows the full width, including the contribution of the decay $D^*\to D\gamma$. Adapted from Ref.~\cite{Baru:2011rs}.} \label{width}
\end{figure}

The results for the calculated line shape of the $\X$ in the $D^0\bar D^0\pi^0$ channel are shown in Fig.~\ref{rate2}. The following three cases were investigated:
\begin{enumerate}
\item Single-channel problem (only the neutral channel $D^0 \bar D^{*0}$) in the static approximation with the imaginary part of the potential neglected and with a constant value used in the two-body 
propagator $\Delta_0$ instead of the running width.
\item Single-channel approximation for the full dynamical problem, including intermediate three-body $D\bar{D}\pi$ states and the dynamical width of the $D^*$ with the omitted contribution of 
the charged pions.
\item Full dynamical two-channel problem with the account for the three-body $D\bar{D}\pi$ dynamics and the dynamical (``running'') width of the $D^*$.
\end{enumerate}

All three curves in Fig.~\ref{rate2} are normalised near the three-body threshold $D^0\bar D^0\pi^0$ at $E=-7.15$ MeV. It is clearly seen that the difference between cases 2 and 3 above is not large 
that 
is readily explained by the position of charged three-body thresholds $D^+\bar D^0\pi^- $, $D^- D^0\pi^+$ and $D^+ D^-\pi^0$
which are located sufficiently far from the peak. In the meantime, the difference between cases 1 and 2 is substantial, and elucidates
the role of the three-body dynamics in the $\X$ --- inclusion of the three-body effects makes the resonance narrower by a factor of 2 (see the details of the calculation in Ref.~\cite{Baru:2011rs}).

We note that the dependence of the results on the value of cut-off is weak. The results shown were obtained for the cut-off $\Lambda=500$~MeV, however variations of the cut-off in a rather wide 
interval (test calculations were performed for $300~\mbox{MeV}\lesssim\Lambda\lesssim 1700$~MeV) result in the variations in the line shape at the level of several per cent \cite{Baru:2011rs}.

It is instructive to compare the results of the full two-body dynamical calculations with similar results obtained in the framework of a perturbative description of the pions --- the so-called 
effective field theory for the $\X$ (X-EFT). The width in the leading order was calculated in Ref.~\cite{Voloshin:2003nt} and
this result was improved in Ref.~\cite{Fleming:2007rp} by considering next-to-leading order corrections that comprises a natural
generalisation of the approach developed for nucleon-nucleon systems in Ref.~\cite{Kaplan:1998we}. The results of comparison
shown in Fig.~\ref{width} confirm a possibility to consider the pionic contribution in the framework of the perturbation theory, at least for the problem of the form of the below-threshold peak in 
the $\X$ system. It has to be noted, however, that the suggested approach has a much wider domain of applicability. In particular, it admits a natural generalisation to other 
near-threshold states, including those for which the threshold mass difference is not as small as for the $\X$. In this case the possibility to
treat the pions perturbatively is questionable. Examples of such states are the $D^{(*)}\bar D^{(*)}$ and $B^{(*)}\bar B^{(*)}$ systems,
and a universal approach to these states based on the heavy quark spin symmetry was suggested in Ref.~\cite{Nieves:2011vw} and further developed
in subsequent papers \cite{Nieves:2012tt,Albaladejo:2015dsa,Guo:2014hqa}). 
Examples of application of the given nonperturbative approach to the aforementioned systems can be found in Refs.~\cite{Baru:2016iwj,Baru:2017gwo}.

\section{Remark on the lattice calculations for the $\X$}

Lattice calculations performed from first principles of QCD are known to be an alternative source of information on hadronic states, in addition to the real experiment. Due to a fast growth of the 
computer power and the appearance of a large number of high productivity computer clusters contemporary lattice calculations are able to answer many questions related to the physics of strong 
interactions. 

Recently, the first results of lattice calculations for exotic states in the spectrum charmonium started to arrive, in particular, for the $\X$
\cite{Prelovsek:2013cra,Lee:2014uta,Padmanath:2015era}. Unfortunately, at the moment, all such calculations suffer from certain shortcomings. 
To begin with, the accuracy of such calculations is quite low. For example, the $\X$ binding energy found in Ref.~\cite{Prelovsek:2013cra} reads
\be
E_B=11\pm 7~\mbox{MeV},
\label{EBlat}
\ee
that is, the uncertainty in the binding energy is of the order of its central value. The results reported in Refs.~\cite{Lee:2014uta,Padmanath:2015era} are quite analogous. Besides that, the size 
of the lattice used, $L\simeq 2$~fm, is definitely not large enough to reliably describe near-threshold states with the wave function containing a long-range molecular component 
\cite{Jansen:2015lha}. Finally, all such calculations are performed for a relatively large current mass of the $u$ and $d$ quark (it proves convenient to deal with the pion mass instead of the 
quark mass, for they are uniquely interrelated through the Gell-Mann--Oakes--Renner relation \cite{GellMann:1968rz}).
Most of the above shortcomings can be improved only improving the numerical methods and increasing the computer power. However, the problem of the unphysical pion mass used on the lattice can be 
reduced to a large extent with the help of the approach used in Refs.~\cite{Baru:2013rta,Baru:2015tfa} and described below. Indeed, in all calculations performed above the pion mass enters as 
a parameter whose value can be varied freely (preserving the hierarchy of the scales which plays an important role for the formulation of the coupled-channel problem) and, in particular, can be 
increased up to the value of $m_\pi=266$~MeV used in the lattice calculations
\cite{Prelovsek:2013cra,Lee:2014uta,Padmanath:2015era}. An important prerequisite for such an adiabatic variation of the pion mass in the equations is a full (preferably nonperturbative) account 
for the three-body dynamics in the equations which is affected substantially by the pion mass change. The coupled-channel equations (\ref{aa}) derived above meet this condition that allows one to use 
them when solving the problem of the chiral extrapolation of the $\X$ binding energy from the physical pion mass to an unphysically large one used on the lattice \cite{Baru:2013rta,Baru:2015tfa}.

\begin{figure}[t]
\centerline{\epsfig{file=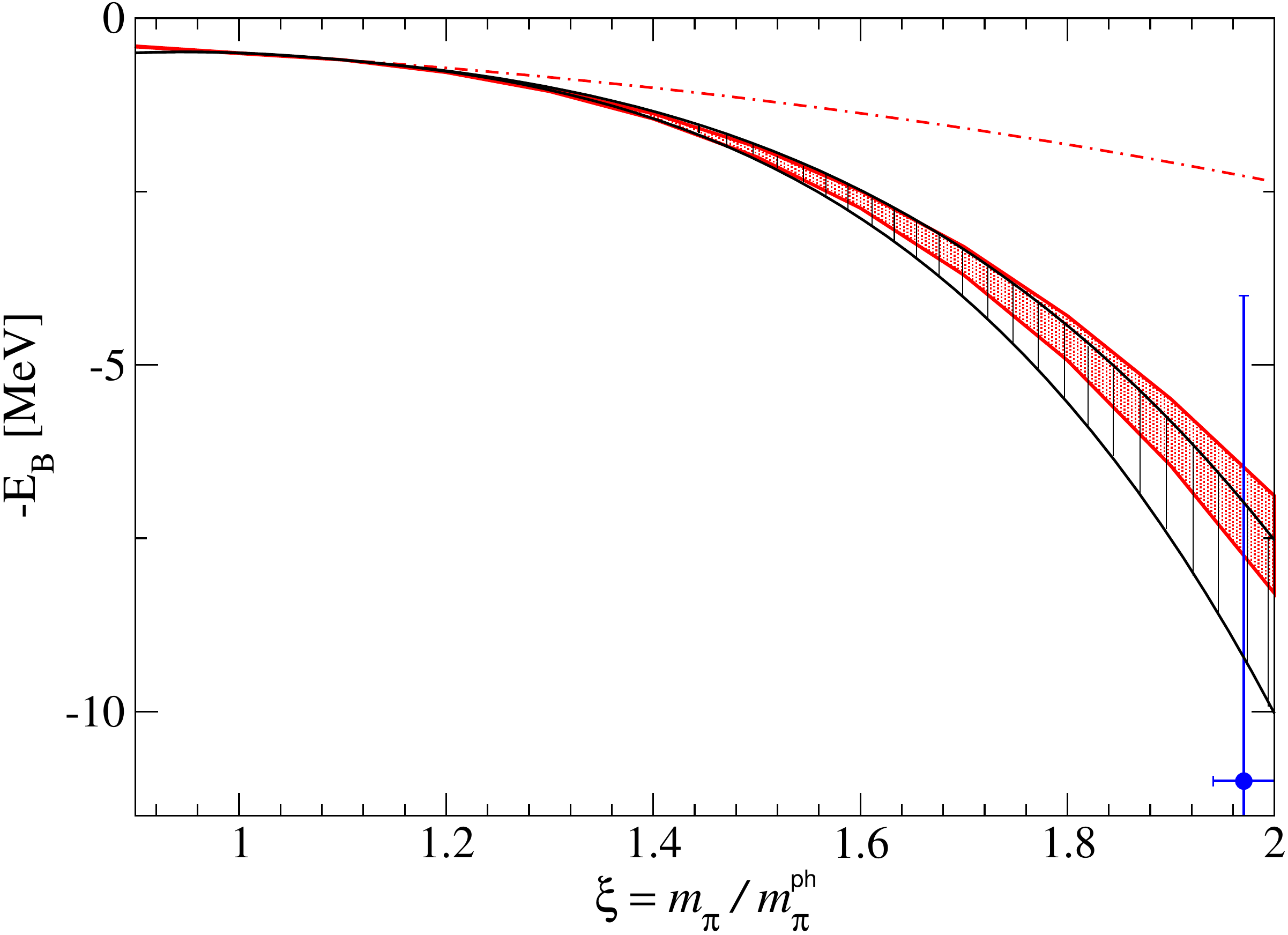,width=0.6\textwidth}}
\caption{Dependence of the $\X$ binding energy on the pion mass. The black and red band correspond to the calculations performed in the framework of a nonrelativistic \cite{Baru:2013rta} and 
relativised approach
\cite{Baru:2015tfa}. The dashed-dotted line shows the result obtained in a purely contact theory without pions. The blue dot with error bar gives the lattice result reported in 
Ref.~\cite{Padmanath:2015era}. Adapted from Ref.~\cite{Baru:2015tfa}.}\label{figNLO}
\end{figure}

The main problem one encounters when building the chiral extrapolation for the $\X$ binding energy is the unknown nature of the short-range forces responsible for its formation. As a result, the 
dependence of the contact term $C_0$ on $m_\pi$ can only be established resorting to the idea of its natural behaviour. 
If applied to the problem at hand, naturalness implies the absence of abnormally large or small mass scales and coefficients. This appears to be sufficient to explain the results obtained on the 
lattice --- see Fig.~\ref{figNLO} which demonstrates that, on the one hand, the increase of the binding energy with the pion mass growth predicted by the lattice can be explained naturally in the 
molecular model for the $\X$ and, on the other hand, the role of the pionic dynamics in this calculation is quite important --- it suffices to confront the predictions of the pionless theory (the 
dashed-dotted line in the figure) with the predictions of the full theory (the black and red bands). A detailed description of the chiral extrapolations for the $\X$ can be found in 
Refs.~\cite{Baru:2013rta,Baru:2015tfa}.

Further discussion of the lattice calculations for the $\X$ goes beyond the scope of the present review devoted to various phenomenological aspects of the description and interpretation of this 
exotic state. It has to be noticed, however, that regardless of the perspectives of the studies of exotic states on the lattice, building chiral extrapolations by itself can be viewed as an 
important theoretical problem for near-threshold resonances since it allows one to investigate the contribution and interplay of different dynamics defined and acting at different energy scales.

\section{Radiative decays of the $\X$}
\label{Xraddecs}

Radiative decay modes of the $\X$ into $\gamma J/\psi$ and $\gamma\psi'$ final states are among experimentally studied processes; in
particular, the ratio of the branching fractions of the said processes has been measured --- see Eq.~(\ref{RLHCb}). As radiative transitions offer a
rather powerful tool to study the meson wave function and to choose the most adequate model, then a theoretical study of these decays is quite a natural next step.

Qualitatively, such radiative modes were taken into account in the data analysis discussed in Section~\ref{dataanalysis}. Namely, the requirement was imposed on the fit for the $X\to\gamma\psi'$ 
decay width to be in accord with the model estimates for the $\chi_{c1}'$ charmonium. So, if one assumes that the $\X$ radiative decays proceed solely via the quark component of its wave function, 
one easily reproduces the experimental ratio (\ref{RLHCb}). Phenomenological estimates, presented in some papers (see, for example, \cite{Dong:2008gb,Dong:2009yp,Dong:2009uf}) demonstrate that even a 
small admixture of the $\bar{c}c$ quarkonium (around 5-12\%) in the $\X$ wave function is enough to explain the experimental data. On the other hand, there is an opinion expressed in the literature 
that it is impossible to obtain such kind of ratio (and, more generally, a ratio comparable with unity) in the molecular picture (see, for example, Ref.~\cite{Swanson:2004pp}). In this section we 
show 
that, with quite natural assumptions on the model parameters (in particular, on the ratio of the coupling constants of the charmonia $J/\psi$ and
$\psi'$ with $D$ mesons), the experimental ratio (\ref{RLHCb}) can be reproduced also in the molecular picture. Here we follow Ref.~\cite{Guo:2014taa}.
At the first or sketchy reading of this review one can skip this section, in particular, the technical details of the derivation of the radiative decay amplitude for the $\X$. It is, however, 
important to pay attention to the conclusions contained in the end of the section which state that the data on the radiative decays cannot be decisive in verification of the molecular nature of the 
$\X$ 
since they are much more sensitive to the short-range part of its wave function. 

\begin{figure*}[t]
\begin{center}
\begin{tabular}{ccc}
\raisebox{13mm}{(a)}~\epsfig{file=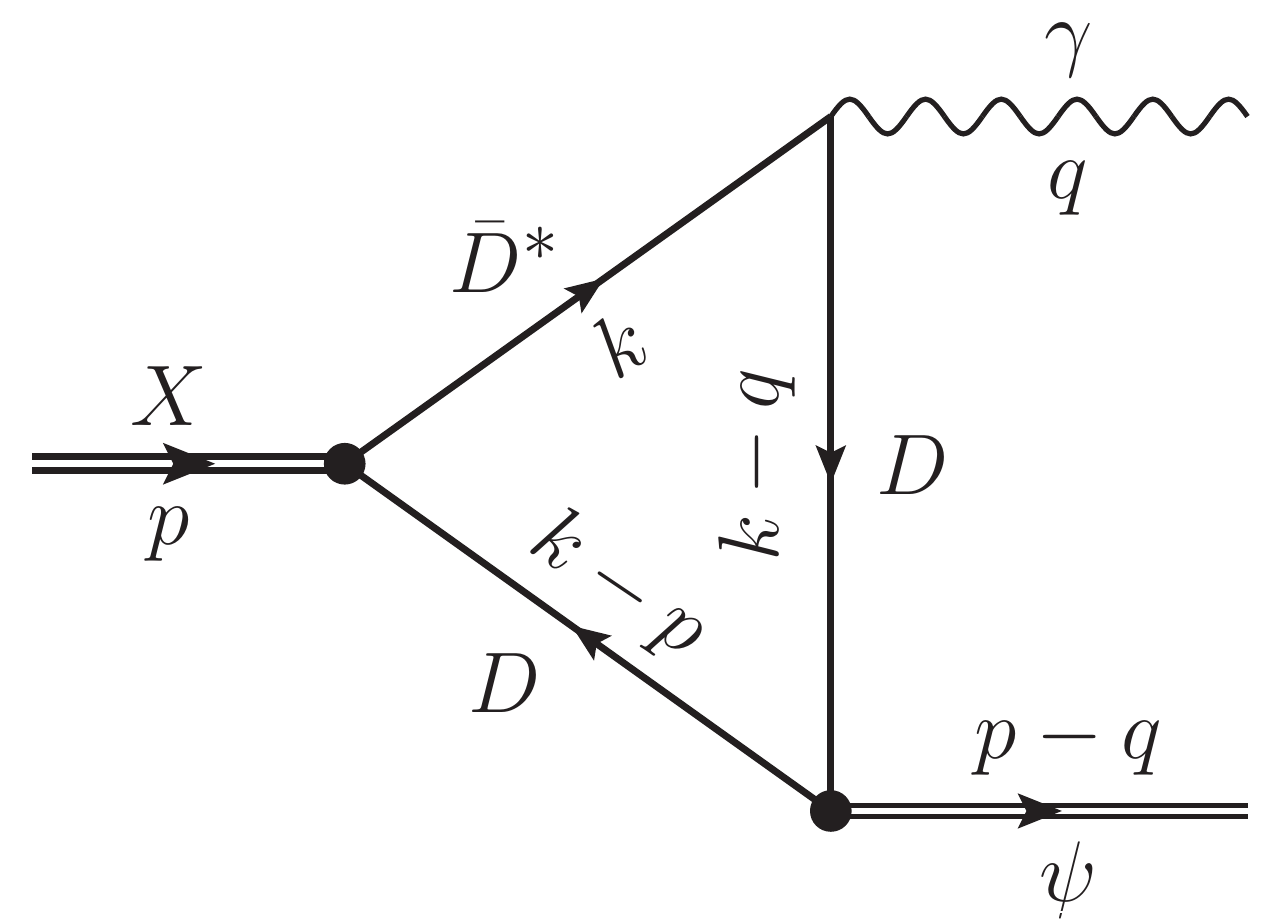,width=0.25\textwidth}&
\raisebox{13mm}{(b)} ~\epsfig{file=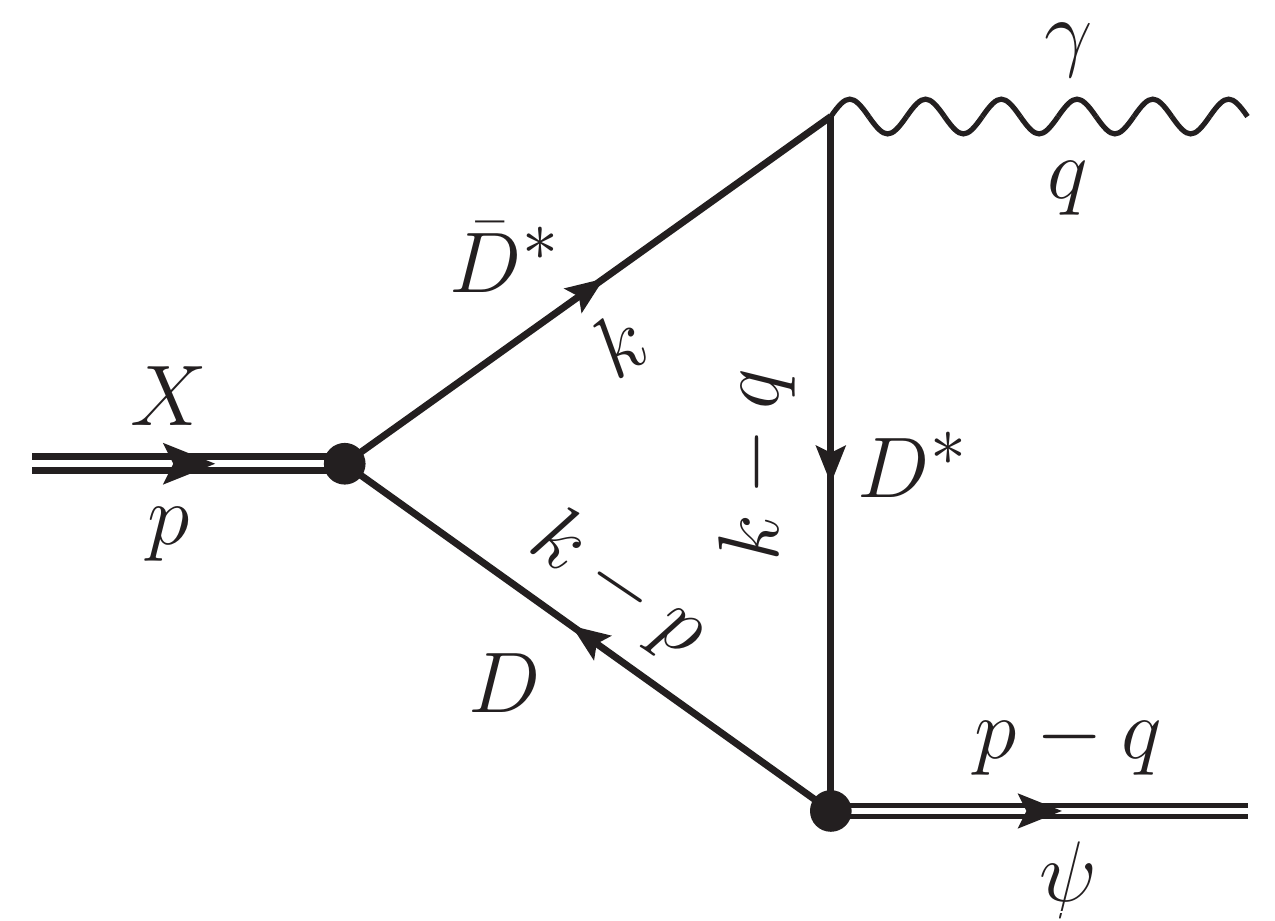,width=0.25\textwidth}&
\raisebox{13mm}{(c)} ~\epsfig{file=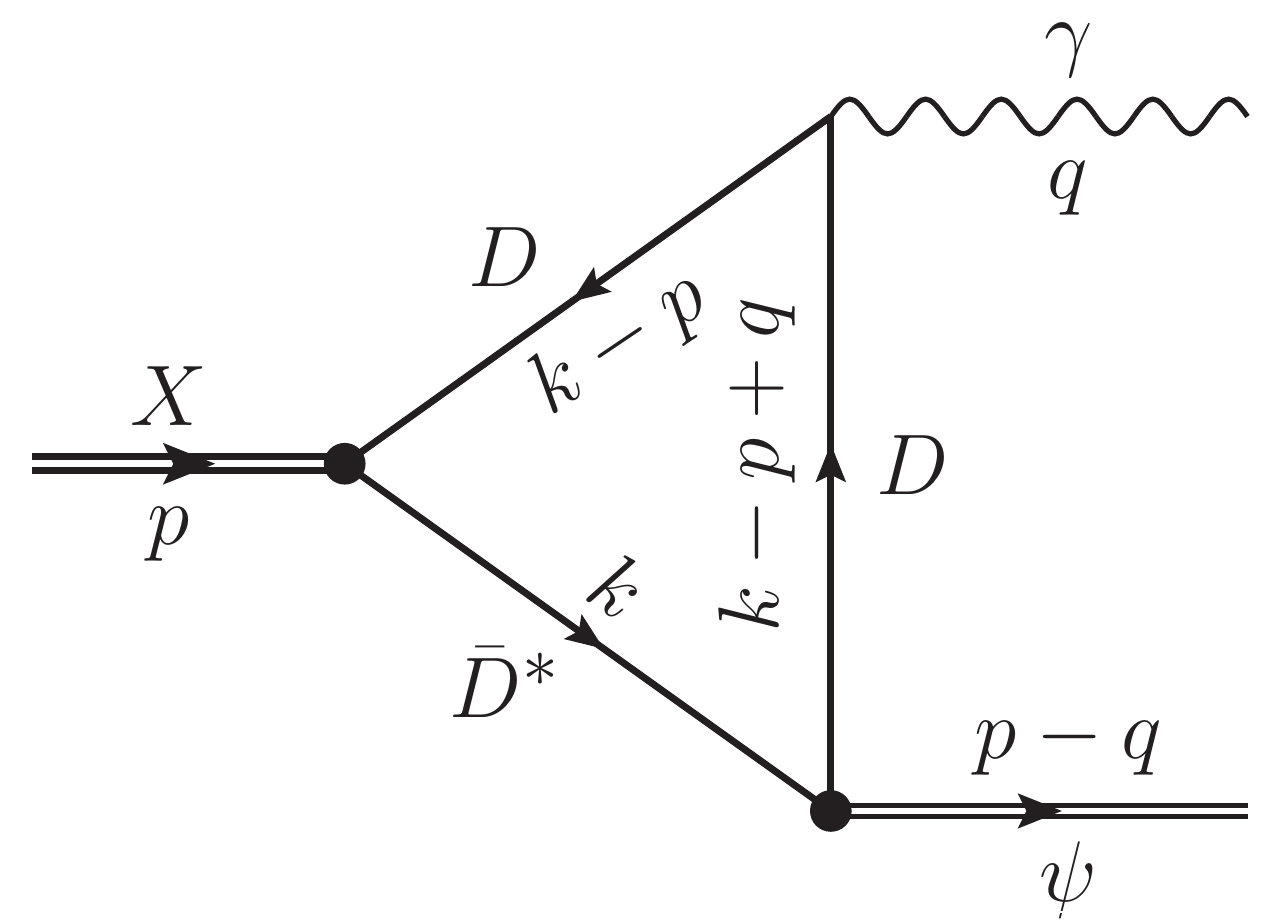,width=0.25\textwidth}\\
\raisebox{13mm}{(d)} ~\epsfig{file=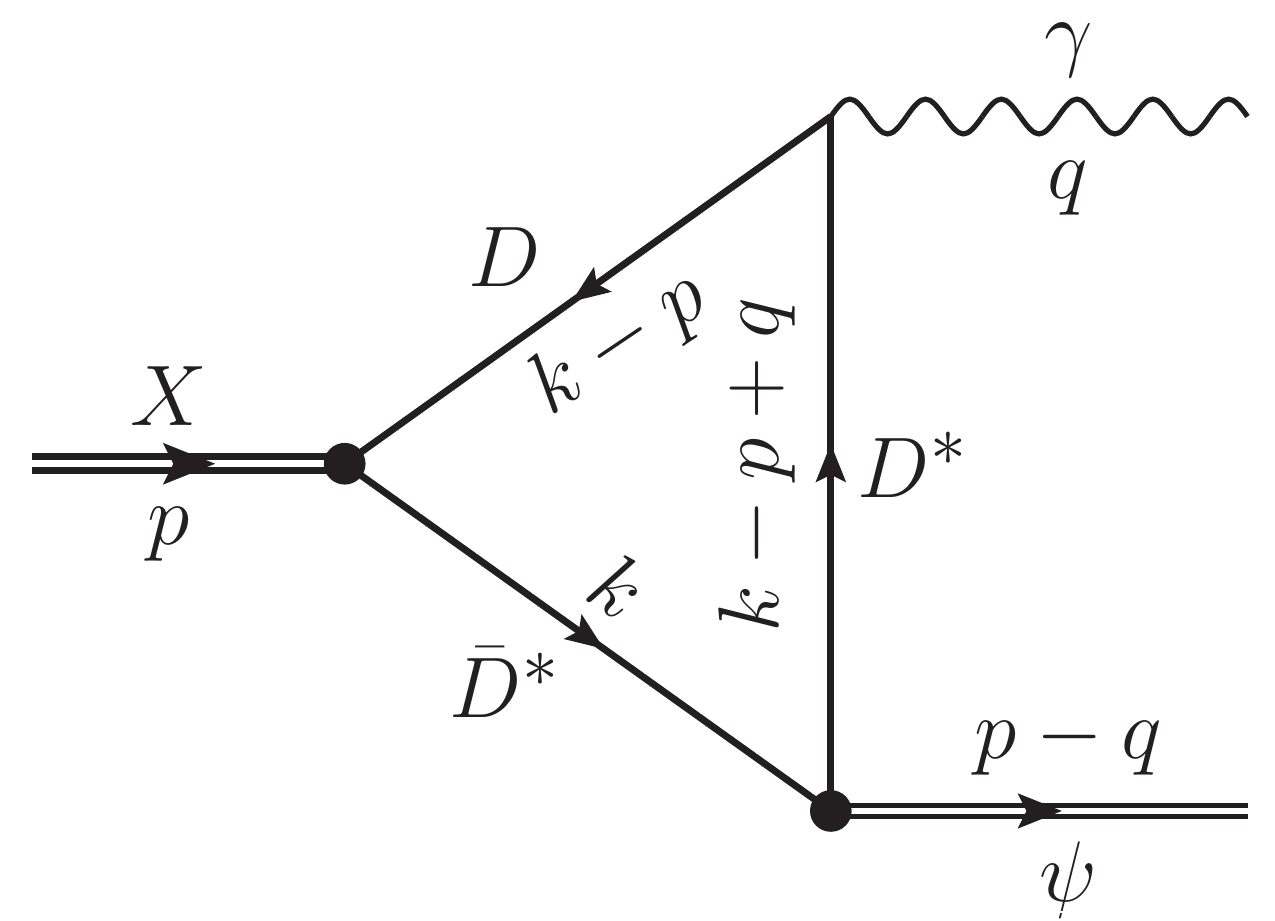,width=0.25\textwidth}&
\raisebox{13mm}{(e)}
~\raisebox{2.5mm}{\epsfig{file=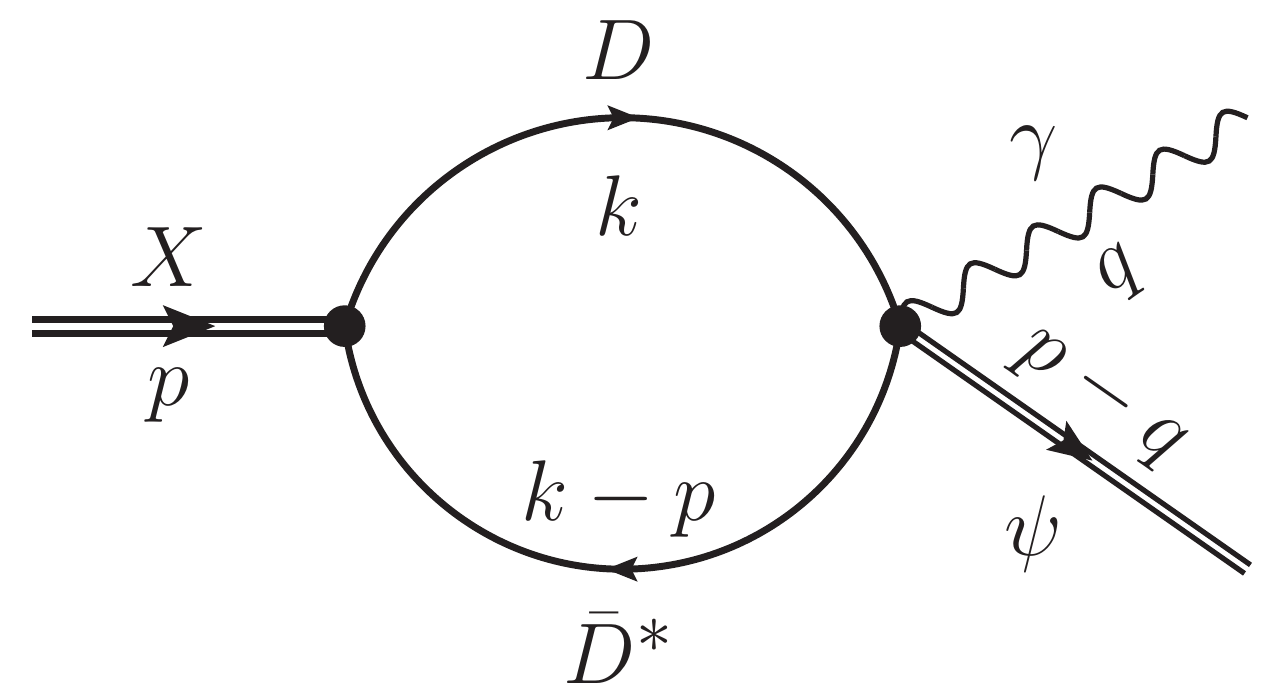,width=0.25\textwidth}}&
\raisebox{13mm}{(f)}
~\raisebox{5.5mm}{\epsfig{file=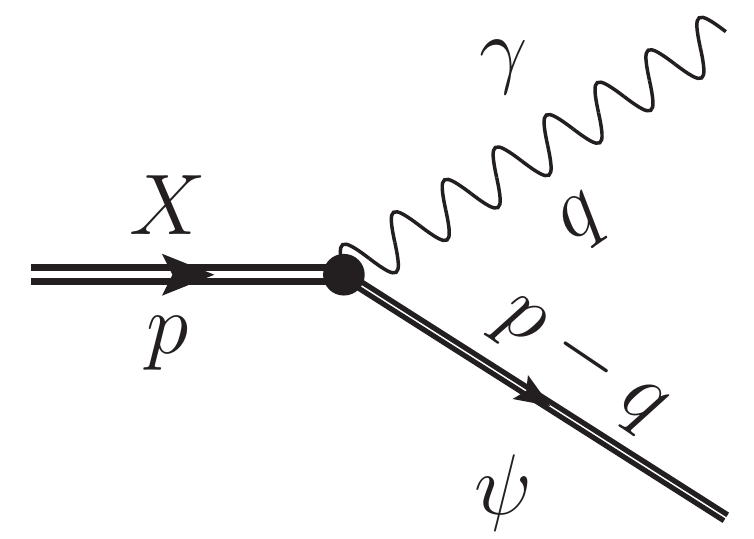,width=0.15\textwidth}}
\end{tabular}
\caption{Graphs (a)-(e) describe the radiative transition $\X\to \pp$ in the mesonic loop mechanism. The charge conjugated loops are not shown but included in the calculation. Graph (f) describes the 
contribution of the compact component of the $X$ wave function.} \label{fig:triangle}
\end{center}
\end{figure*}

The amplitude of the radiative decay $X \to \pp$ (where $\psi=J/\psi$ or $\psi'$) is given by a sum of the graphs depicted in Fig.~\ref{fig:triangle}.
Radiative transition could proceed either via $\bar{c}c$ component of the $\X$ wave function (graph (f)) or via its mesonic component (graphs (a)-(e)), and the latter mechanism dominates if the $\X$ 
is a mesonic molecule. 
 
Assume for a moment that the $\X$ is a pure $D \bar D^*$ molecule and present the vertex $X_\sigma(p)\to D\bar{D}^*_\nu(k)$ (where
$\sigma$ and $\tau$ are the polarisation vector indices of the $\X$ and $D^*$, respectively) in the form
\be
\varGamma_{\sigma\tau}^{(X)}(p,k)=\frac{1}{\sqrt{2}}x_{\rm nr}\sqrt{M_Xm_*m}\,g_{\sigma\tau},
\ee
where $m$, $m_*$ and $M_X$ are the masses of the $D$, $D^*$ and $\X$, respectively, and $x_{\rm nr}$ is the nonrelativistic coupling constant which can be extracted from the $\X$ binding energy 
\cite{Guo:2013zbw}.

The $\psi_{\mu}(p)\to\bar{D}(k_1)D(-k_2)$, $\psi_{\mu}(p)\to \bar{D}^*_{\nu}(k_1)D(-k_2)$ and $\psi_{\mu}(p)\to\bar{D}^*_\alpha(k_1)D^*_\beta(-k_2)$ transition vertices can be written as
\be
V_\nu^{\bar{D}D}(k_1,-k_2)=g_2\sqrt{m_\psi}m (k_1+k_2)_\nu,
\label{psidd}
\ee
\be
V_{\nu\tau}^{\bar{D}D^*}(k_1,-k_2)=2g_2\sqrt{\frac{m_\psi m}{m_*}}\epsilon_{\nu\tau\alpha\beta}k_2^ \alpha k_1^\beta,
\label{psiddstar}
\ee
\be
V_{\nu\alpha\beta}^{\bar{D}^*D^*}(k_1,-k_2)=g_2\sqrt{m_\psi}m_*\Bigl[(k_1+k_2)_{\nu}g_
{\alpha\beta}-(k_1+k_2)_{\beta}g_{\nu\alpha}-
(k_1+k_2)_{\alpha}g_{\mu\beta}\Bigr],
\label{psidsds}
\ee
where heavy quark symmetry relations are used \cite{Colangelo:2003sa,Guo:2009wr,Guo:2010ak}.

Electric couplings of the photon to the $D$ mesons are obtained from gauging the corresponding terms of the interaction Lagrangian. Namely, the electric
vertex $D^\pm(k_1) \to D^\pm(k_2)\gamma_{\rho}(q)$ ($k_1=k_2+q$) reads
\be 
\varGamma^{(e)}_\rho(k_1,k_2)=e(k_1+k_2)_{\rho},\quad q=k_1-k_2, 
\label{el1} 
\ee 
where $e$ is the electric charge. Similarly, the electric vertex $D^{*\pm}_{\nu}(k_1) \to
D^{*\pm}_{\tau}(k_2)\gamma_{\rho}(q)$ ($k_1=k_2+q$) is 
\be
\varGamma^{(e)}_{\nu\tau\rho}(k_1,k_2)=e\Bigl[(k_1+k_2)_{\rho}g_{\nu\tau}-k_{1\tau}g_{\nu\rho}-k_{2\nu}g_{\tau\rho}\Bigr]. 
\label{psipsigam} 
\ee 
It can be easily verified that both above vertices satisfy the appropriate Ward identities. Finally, gauging the vertex (\ref{psiddstar}) yields a four-point vertex $D\bar{D}^*\psi \gamma$ (see graph 
(e) 
in Fig.~\ref{fig:triangle}).

Magnetic transition vertices $D^{*a}_{\mu}(k_1) \to D^{*b}_{\nu}(k_2)\gamma_{\lambda}(q)$ and $D^{*a}_{\mu}(k_1) \to
D^b(k_2)\gamma_{\lambda}(q)$ ($k_1=k_2+q$) follow from a covariant generalisation of the chiral Lagrangian from Refs.~\cite{Amundson:1992yp,Cheng:1992xi} in the form
\be
\varGamma_{\nu\tau\rho}^{(m)ab}(q)=
em_*(q_\tau g_{\nu\rho}-q_\nu g_{\tau\rho})\left(\beta
Q_{ab}-\frac{Q_c}{m_c}\delta_{ab}\right),
\label{DsDsgamm}
\ee
\be
\varGamma_{\nu\rho}^{(m)ab}(q)=
e\sqrt{mm_*}\varepsilon_{\nu\rho\alpha\beta}v^{\alpha}q^\beta
\left(\beta Q_{ab}+\frac{Q_c}{m_c}\delta_{ab}\right),
\label{DsDgamm}
\ee
respectively, where $v_{\mu}$ is the four-velocity of the heavy quark, $Q=\text{diag}(2/3,-1/3)$ is the light quark charge matrix, and $m_c$ and $Q_c=2/3$ are the charmed quark mass and charge, 
respectively. The terms proportional to $Q_c/m_c$ come from the magnetic moment of the $c$-quark and terms proportional to $\beta$ are
from the light quark cloud in the $D$ meson.

Thus, the radiative decay amplitude $X\to\gamma\psi$ in the $D$-meson loop mechanism takes the form
\be
M^{\rm loop}=\varepsilon^\nu(\psi)\varepsilon^\sigma(X)\varepsilon^\rho(\gamma)M_{\nu\sigma\rho}^{\rm loop}, 
\ee 
where 
\be 
M_{\nu\sigma\rho}^{\rm
loop}=\frac{1}{\sqrt{2}}ex_{\rm nr}g_2m\sqrt{M_Xm_\psi}I_{\nu\sigma\rho},
\label{Mampl}
\ee
with $I_{\nu\sigma\rho}$ given by the sum of the contributions from the individual diagrams in Fig.~\ref{fig:triangle}(a)-(e).
The amplitude (\ref{Mampl}) is gauge-invariant which is ensured by the transversality of the magnetic vertices (\ref{DsDsgamm}) and
(\ref{DsDgamm}), as well as by the Ward identities.

The loop integral $I_{\nu\sigma\rho}$ in the amplitude (\ref{Mampl}) is divergent. Thus, one needs to supply the amplitude with a counter term corresponding to the contact $X\gamma\psi$ interaction 
(see graph (f) in Fig.~\ref{fig:triangle}) which takes the form
\be 
M^{\rm cont}=\lambda\varepsilon_{\mu\sigma\rho\nu}\varepsilon^\mu(\psi)\varepsilon^\sigma(X)
\varepsilon^\rho(\gamma)q^\nu 
\label{Mcont} 
\ee 
and which is manifestly gauge-invariant. To renormalise the model one absorbs the divergence of the loop into the bare constant $\lambda$ providing
the contact amplitude (\ref{Mcont}) with a finite strength $\lambda_r$.

Calculations were performed in the $\overline{\rm MS}$~\footnote{The renormalisation scheme MS (from Minimal Subtraction) corresponds to the dimensional regularisation of the loop integrals with a consequent omission of their divergent parts. In the modified $\overline{\rm MS}$ scheme one additionally omits finite constants which alway accompany the divergent part of the integral.} scheme for the three values of the regularisation scale $\mu$: $\mu=M_X/2$, $\mu=M_X$ and $\mu=2M_X$. Given the uncertainties in the value of 
the coupling constant $x_{\rm nr}$ as
well as in the constants $g_2$ and $g_2'$ which define the coupling of the vector mesons $J/\psi$ and $\psi'$ to the $D$ mesons, it is reasonable
to introduce the ratios
\be
r_x\equiv\left|\frac{x_{\rm nr}}{x_{\rm nr}^{(0)}}\right|,\quad
r_g\equiv\left|\frac{g_2}{g_2^{(0)}}\right|,\quad
r_g'\equiv\left|\frac{g_2'}{g_2^{(0)}}\right|,
\label{eq:rcouplings}
\ee
where, as the starting point of the model estimates, the values $|x_{\rm nr}^{(0)}|=0.97$~GeV$^{-1/2}$~\cite{Guo:2013zbw} and
$|g_2^{(0)}|=2$~GeV$^{-3/2}$~\cite{Colangelo:2003sa,Guo:2010ak} can be conveniently adopted.

\begin{table}[t]
\begin{center}
\begin{tabular}{|l|c|c|c|}
\hline
 &$\mu=M_X/2$&$\mu=M_X$&$\mu=2M_X$\\
\hline
$\vphantom{\int_0^1}\varGamma(X\to\gamma J/\psi)$ [keV] &9.7$(r_xr_g)^2$ &23.5$(r_xr_g)^2$ &43.2$(r_xr_g)^2$ \\
\hline
$\vphantom{\int_0^1}\varGamma(X\to\gamma\psi')$ [keV] &3.8$(r_xr_g')^2$ &4.9$(r_xr_g')^2$ &6.0$(r_xr_g')^2$ \\
\hline $\ds
R=\frac{\vphantom{\int_0^1}\varGamma(X\to\gamma\psi')}{\varGamma(X\to\gamma J/\psi)}$
&0.39$(g_2'/g_2)^2$ &0.21$(g_2'/g_2)^2$
&0.14$(g_2'/g_2)^2$ \\
\hline
\end{tabular}
\end{center}
\caption{The calculated radiative decay widths $\varGamma(X\to\gamma\psi)$ for $\psi=J/\psi,\psi'$ and their ratio $R$.} \label{tab:res}
\end{table}

The numerical results for the radiative decay widths and their ratio for the case of $\lambda_r=\lambda_r'=0$ are given in Table~\ref{tab:res}.
We notice that for $g_2'/g_2\simeq 1$ the calculated ratio $R$ is less than unity, however it is much larger than the value obtained in Ref.~\cite{Swanson:2004pp}. Meanwhile, the 
ratio $g_2'/g_2\simeq 3$ is already sufficient to reproduce the experimental ratio (\ref{RLHCb}) even in the purely molecular picture.

Another source of uncertainty in the ratio $R$ is the value of the renormalised contact interaction $\lambda_r^{(\prime)}$. Indeed, the requirement of the final result to be a renormalisation group 
invariant means that the variation of the loop contribution on the parameter $\mu$ is to be compensated by the corresponding variation of the contact
term. The former, as seen from Table ~\ref{tab:res}, is quite large, so that to compensate for it the contribution of the contact term should be at least of the same order of magnitude, that is, 
should be also large. Therefore, the radiative decays appear to be sensitive to the short-range part of the $\X$ wave function.

To estimate the contribution of the contact term model considerations can be used. As the most natural short-range contribution into the
$\X$ wave function is provided by the genuine quark-antiquark state then, to define the ratio $R$, one may take the estimates for the radiative
decay widths of the $2^3P_1$ $\bar{c}c$ charmonium in various quark models. Typical examples of such estimates are collected in Table~\ref{widths}
from which one can see that, in spite of a large spread in the predictions, the values $\varGamma(X(\bar{c}c)\to
\Jp)\simeq 50$~keV and $\varGamma(X(\bar{c}c)\to \p)\simeq100$~keV can be taken as natural ones, that gives the ratio $R\simeq 2$.

Recall now that the relative weight of the quark and mesonic component and, correspondingly, the contributions of the quark and meson loops are defined by the probabilities $Z$ and $1-Z$, 
respectively (see a detailed discussion in Section~\ref{weinapp} above). Thus, if one employs a more realistic model for the $\X$ in which its wave function contains a sizable admixture of the 
charmonium with $Z\simeq 0.4$-$0.5$ (see the results of Section~\ref{dataanalysis}), then the experimental ratio can be reproduced already for $g_2'/g_2\simeq 2$, which is in line with the ratio of 
these couplings found in Ref.~\cite{Dong:2009uf} and referring to the analysis of Ref.~\cite{Colangelo:2003sa}.

Therefore, while the results of the calculations presented in Table~\ref{tab:res} should be regarded only as order-of-magnitude estimates, two important conclusions can be deduced:
\begin{itemize}
\item The $\X$ radiative decays, in particular, the value $R$ for the ratio of the widths, is more sensitive to the short-range part of the $\X$ wave function than to the long-range part. Thus, these 
decays cannot be used to confirm or rule out the molecular picture for the $\X$.
\item The value of the ratio $R$ of the order of unity and, in particular, its experimentally measured value (\ref{RLHCb}) does not contradict the molecular picture for the $\X$, in contrast to 
the prejudice existing in the literature based on early model estimates.
\end{itemize}

\section{Conclusions}

More than 60 ago Weinberg answered the question whether the deuteron was a composite or elementary particle. While a relevant approach was developed and a proof was presented for the deuteron to be a 
proton-neutron molecule, reservations were outlined on the applicability of the approach to other hadronic resonances. To quote Weinberg, ``Nature is doing her best to keep us from learning
whether the elementary particle deserves that title.'' \cite{Weinberg:1965zz} In most cases, the nature indeed is not very accommodating, but the $\X$ case offers a nice exception.

To begin with, the mass of the $\X$ coincides almost exactly with the $D^0 \bar D^{*0}$ threshold, that means that a sizable admixture of the $D \bar D^*$ pairs in the $\X$ wave function is 
unavoidable, 
unless the coupling of the $\X$ to the said pairs is pathologically small. This threshold proximity allows one to develop a low-energy effective field theory which, in turn, provides a 
systematic approach to the calculations of the production and decay processes for the $\X$, investigating in such a way both long-range (molecular) and short-range (compact) component of the wave 
function. Another important advantage of the $\X$ is the existence of quark model predictions for a nearby $2^3P_1$ {\it bona fide} charmonium which constitutes the most natural candidate for the 
short-range part of the $\X$ wave function.
Finally, the experimental situation is very friendly: detailed measurements of the $\X$ production and decays are available allowing one to test various model predictions. All 
this turns the $\X$ from an ``enfant terrible'' of the charmonium spectroscopy into a perfect laboratory for the mesonic molecules studies.
\bigskip

Work of Yu.S.K. was supported by the Ministry of Science and Education of Russian Federation (grant 14.W03.31.0026), work of A.V.N. was supported by 
the Russian Science Foundation (grant 18-12-00226).


\end{document}